\documentclass[a4paper,fleqn,usenatbib]{mnras}
\usepackage{newtxtext,newtxmath}
\usepackage[T1]{fontenc}
\usepackage{ae,aecompl}
\usepackage{xtab,afterpage}
\usepackage[usenames]{color}
\usepackage{graphicx}
\usepackage{caption}
\usepackage{subcaption}
\usepackage{txfonts}
\usepackage{natbib}
\usepackage{amssymb}
\usepackage{textcomp}
\usepackage{float}
\usepackage{gensymb}
\usepackage{eqnarray,amsmath}

\DeclareRobustCommand{\ion}[2]{%
\relax\ifmmode
\ifx\testbx\f@series
{\mathbf{#1\,\mathsc{#2}}}\else
{\mathrm{#1\,\mathsc{#2}}}\fi
\else\textup{#1\,{\mdseries\textsc{#2}}}%
\fi}
\def\rmxaa{Rev. Mexicana Astron. Astrofis.}

\newcommand{\Sm}{$\Sigma_{*}$}

\title[Global and Local Star Formation Main Sequence]{SDSS-IV MaNGA: Effects of Morphology in the global and local Star Formation Main Sequences}

\author[M. Cano-D\'{\i}az et al.]{
M. Cano-D\'{\i}az,$^{1}$\thanks{E-mail: mcano@astro.unam.mx}
V. \'Avila-Reese,$^{2}$ 
S.F. S\'anchez,$^{2}$ 
H. M. Hern\'andez-Toledo,$^{2}$ 
\newauthor{
A. Rodr\'{\i}guez-Puebla,$^{2}$ 
M. Boquien,$^{3}$ 
H Ibarra-Medel,$^{4,2}$
}
\\
$^{1}$ CONACYT Research Fellow - Instituto de Astronom\'{\i}a, Universidad Nacional Aut\'onoma de M\'exico, Apartado Postal 70-264, Mexico D.F., 04510 Mexico \\
$^{2}$ Instituto de Astronom\'{\i}a, Universidad Nacional Aut\'onoma de M\'exico, Apartado Postal 70-264, Mexico D.F., 04510 Mexico
\\
$^{3}$ Centro de Astronom\'{\i}a (CITEVA), Universidad de Antofagasta, Avenida Angamos 601, Antofagasta, Chile
\\
$^{4}$ University of Illinois Urbana-Champaign, Department of Astronomy, 1002 W Green St, Urbana, Illinois, 61801, USA 
}

\date{Accepted XXX. Received YYY; in original form ZZZ}
\pubyear{2018}
\begin{document}
\label{firstpage}
\pagerange{\pageref{firstpage}--\pageref{lastpage}}
\maketitle


\begin{abstract}
                                                                                 We study the global star-formation rate (SFR) vs. stellar mass (M$_*$) correlation, and the spatially-resolved SFR surface density ($\Sigma_{SFR}$) vs. stellar mass surface density (\Sm) correlation, in a sample of  $\sim2,000$ galaxies from the MaNGA MPL-5 survey. We classify galaxies and spatially-resolved areas into star-forming and retired according to their ionization processes. We confirm the existence of a Star-Forming Main Sequence (SFMS) for galaxies and spatially-resolved areas, and show that they have the same nature, with the global as a consequence of the local one. The latter presents a bend below a limit \Sm value, $\approx 3\times  10^7$ M$_\odot$kpc$^{-2}$, which is not physical. Using only star-forming areas (SFAs) above this limit, a slope and a scatter of $\approx1$ and $\approx0.27$ dex are determined. The retired galaxies/areas strongly segregate from their respective SFMS's, by $\sim -1.5$ dex on average. We explore how the global/local SFMS's depend on galaxy morphology, finding that for star-forming galaxies and SFAs, there is a trend to lower values of star-formation activity with earlier morphological types, which is more pronounced for the local SFMS. The morphology not only affects the global SFR due to the diminish of SFAs with earlier types, but also affects the local SF process. Our results suggest that the local SF at all radii is established by some universal mechanism partially modulated by morphology. Morphology seems to be connected to the slow aging and sharp decline of the SF process, and on its own it may depend on other properties as the environment.

\end{abstract}


\section{Introduction} \label{Intro}

The star formation rate (SFR) and stellar mass (M$_*$) are two fundamental properties of galaxies that relate their current evolutionary status (the rate of gas conversion into stars) with their previous evolution. Thus, the SFR--$M_*$ diagram resumes crucial aspects of galaxy evolution. In this diagram clearly emerges a {\it statistical} bimodality \citep{Brinchmann04,Weinmann06}, with a populated region called the Star Formation Main Sequence \citep[SFMS; after][]{Noeske07}, which results in an approximately power-law relation between SFR and $M_*$, with a scatter of $0.2-0.35$ dex and a slope close to but lower than one, both at redshifts $z\sim 0$ \citep[e.g.][]{Salim07,Gavazzi15,Renzini15,catalan15}, and at higher redshifts \citep[e.g.,][see also \citeauthor{Speagle14} \citeyear{Speagle14} and \citeauthor{Rodriguez-Puebla+2017} \citeyear{Rodriguez-Puebla+2017} for compilations of many works, and more references therein] {Noeske07,Daddi07,Santini+2009,Rodighiero+2011,Whitaker+2014,Sparre15,Schreiber+2015,Katsianis15,Tomczak+2016}. On the other hand, there is an increase with cosmic time of the fraction of galaxies that depart from the SFMS, with much lower values of SFR for their masses, occupying a less well defined region in the SFR--$M_*$ diagram that we will identify as the region of retired or quenched galaxies \citep{Stasinska08}. An interesting question is whether the SFMS and the region of galaxies that depart far away from the SFMS are populated by objects that are indeed globally star forming galaxies (SFGs) and galaxies that are retired from the star formation process, which hereafter will be called as retired galaxies (RGs), respectively. Even more, having this physical characterization of galaxies, the SFMS can be established by construction as the SFR vs. $M_*$ correlation of the SFGs \citep{CanoDiaz16,sanchez18a}.

 Morphology is another of the fundamental properties of galaxies that allows to classify them. A clear example of this classification is the scheme proposed by \citet{Hubble1926}. The refinement of this classification scheme has been possible mainly thanks to the improvement in the quality of astronomical observations, along with systematic observations of the sky in the fashion of galaxy surveys \citep[e.g.][]{york00,driver09}. This has helped to identify common physical characteristics among the different morphological groups \citep[e.g.][]{blanton05}. This successful identification within the different types of galaxies, with statistical significance, has eventually helped to shape our understanding of the global galaxy evolution picture \citep[e.g.,][]{Blanton2009}. Such an understanding can be enriched by studying in the above mentioned SFR--$M_*$ diagram the loci of galaxies of different morphological types. 

Among the many phenomena that galaxies exhibit, and that are closely related to their morphology, star formation is one that should be revisited in detail. For instance, globally, it is known that the vast majority of late-type galaxies present a higher specific star formation rate (sSFR), while early-type galaxies present a much lower one \citep{Eales+2017}. This is directly related to the gas content available for forming stars: while late-type galaxies tend to have a high gas fraction, early-type galaxies have a smaller gas reservoir \citep[e.g.][and references therein]{Calette+2017}. However, external and internal phenomena can influence the triggering or stopping of the SF, such as the presence of nuclear activity. 
In general terms, this behavior is visible in the so called color-magnitude diagram, in which there are two main recognizable regions: 1) the blue cloud, mainly formed by late-type, gas-rich galaxies, and 2) the red sequence, mainly formed by early-type, gas poor galaxies \citep[e.g.][]{Baldry2004,bell06,Blanton2009}. The galaxies that populate the blue cloud are mainly the ones that form the SFMS, and those that comprise the red-sequence are mostly located well below the SFMS in the SFR-$M_*$ diagram, in the cloud of the RGs. The relative few galaxies in the transition zone, called the green valley, are typically not easy to classify by their ionization processes \citep{CanoDiaz16}. They tend to be of intermediate morphological types (lenticulars and barred early-type galaxies), and many of them host Active Galactic Nuclei \citep[AGN; see e.g.,][]{Schawinski14,sanchez18b}. The green valley is interpreted as the location of transient galaxies between the sequence of SFGs and the region of RGs. However, it is important to note that it is possible to find galaxies with late-type morphology in the red cloud \citep{Masters+2010,FraserMcKelvie+2016}, and early-type ones in the blue cloud \citep{Schawinski+2009}. For more details about the relation of galaxy morphology and color see for example, \citet{Bamford+2019} and \citet{Skibba+2009}.

More recently, some studies have found that the SFMS relation has a local (spatially-resolved) counterpart in the diagram that relates the SFR surface density ($\Sigma_{SFR}$) with the stellar mass surface density ($\Sigma_{*}$) \citep{Sanchez13,Wuyts13,Magdis16,CanoDiaz16,Maragkoudakis17,Abdurro17,Hsieh17,Pan18,Ellison18,Medling18,Erroz-Ferrer+2019}. This spatially resolved diagram aims to look for the star formation process at a local level within the galaxies (at $\sim$kpc scales). The local SFMS described by star-forming areas (SFAs) in the $\Sigma_{SFR}$-$\Sigma_{*}$ diagram can be roughly approached as a power-law relation but at the low-$\Sigma_{*}$ seems to flatten. The slope of the relation fitted to a power law presents a wide range of values among different authors, but in general is smaller than 1, and the scatter is about 0.3 dex (in $\log\Sigma_{SFR}$). The reason for the different reported parameters may be related to the methodology used to derive the relations (different methods to derive both the SFRs and the stellar masses, the classification scheme used to select star-forming and retired galaxies/regions) as well as to selection effects introduced by the used sample. In addition to the local SFMS, traced by SFAs, the $\Sigma_{SFR}$-$\Sigma_{*}$ diagram is populated by regions of galaxies with very low signatures of star formation, thus, retired areas (RAs) that form a cloud well below the local SFMS.

As in the case of the global SFR--$M_*$ diagram, in the local $\Sigma_{SFR}$-$\Sigma_{*}$ diagram there is also a sparsely populated transition region between the star-forming and retired areas. The evolution between them --as well as for entire galaxies in the case of the SFR--$M_*$ diagram-- is thought to be a fast process, known as star-formation quenching, which explains the lack of objects in the green valley \citep[e.g.,][]{Schawinski14,sanchez18a}. However, slower time scale quenching processes than the one previously reffered to, have been proposed \citep[see e.g.,][]{Smethurst+2015}. The physical processes that drive this quenching is matter of intense debate currently. The joint study of the global SFR--$M_*$ and local  $\Sigma_{SFR}$--$\Sigma_{*}$ diagrams certainly sheds light on this question.  
On top of that, SFGs or SFAs are expected to present an steady decrease of the star formation activity that may produce an evolution across the corresponding SFMS. This process is known as aging \citep{Casado15}. There are several pieces of evidence indicating that the aging maybe related to the morphology of SFGs, both globally \citep[e.g.,][]{catalan15,catalan17} and locally \citep[e.g.,][]{rosa16a}.

The possibility to study the global and local SF activity of large populations of galaxies, so that correlations like the global and local SFMS can be derived, is mainly due to the development of the Integral Field Spectroscopic (IFS) technique and its application to observational programs aimed to build-up large galaxy surveys. Currently, there are three major surveys of this type: the Calar Alto Legacy Integral Field Area Survey \citep[CALIFA][]{Sanchez12}, the Sydney-AAO Multi-object Integral-field spectrograph \citep[SAMI][]{Croom12, Bryant15}, and the Mapping Nearby Galaxies at APO, \citep[MaNGA][]{Bundy15}. The latter is a key project of the Sloan Digital Sky Survey IV (SDSS IV) \citep{Blanton17}. The goal of the project is to observe 10,000 galaxies at its completion, being the one provided with the larger number of objects.

In this paper we use the MaNGA survey for building both the global SFR--$M_*$ and local $\Sigma_{SFR}$-$\Sigma_{*}$ diagrams, and to establish  the sequences of star-forming and retired galaxies and areas, respectively.  As the sample of available MaNGA galaxies is large enough, we explore the possible dependence of these two sequences with the morphology of the galaxies, as proposed by \citet{catalan15,catalan17}, and \citet{rosa16a}.

The paper is divided in the following sections. In Section \ref{Data} we describe the data and the sample of galaxies. In Section \ref{ana} we describe the derivation for the SFR and $M_{*}$ for the whole selected sample, as well as the classification by the main ionization processes scheme for all the galaxies and their spatially-resolved regions. In Sections \ref{Results} and \ref{Disc_ResolRelations} we present our results from global to local scales. And finally in Sections \ref{Disc} and \ref{Conc} we present the discussion and conclusions, respectively. We also provide further details of our analysis and methodology in the Apendixes A and B.

\section{Sample and Data} \label{Data}

We use the dataset provided by SDSS-IV MaNGA survey \citep{Bundy15} obtained at the SDSS 2.5 m telescope at Apache Point Observatory \citep{Gunn06}. We selected a subsample of galaxies from the MaNGA Product Launch 5 (MPL-5), which is part of the SDSS-IV Data Release 14 (DR14). This survey is observing galaxies selected from a parent sample \citep{wake17} that contains galaxies of all the morphological types, redshifts ranging 0.01 $< z <$ 0.15 and stellar masses, $M_{*}$, between $\sim 10^{9}$ and $\sim 10^{12}$ $M_{\odot}$. In addition a small ancillary sample is being observed to extend the MaNGA sample to lower masses (details about this ancillary sample will be described in a future paper by Cano-D\'{\i}az, et al.). Due to this, the final MaNGA sample extends to masses lower than $10^{9}$ $M_{\odot}$, however these are a minimum amount with respect to larger galaxies.

Observations are performed using hexagonal IFUs with different sizes of bundles with fibers of 2$\arcsec$ in diameter each one \citep{Drory15}, which allows to spatially resolve the observed galaxies up to 1.5 or 2.5 effective radii, $R_{e}$. This observational setup has a spectral coverage ranging from 3,600 to 10,300 $\AA$ at a spectral resolution of R $\sim$ 2000 provided by the dual beam BOSS spectrographs \citep{Smee13}. Galaxy observations are performed simultaneously with smaller fiber bundles dedicated to perform sky subtraction and flux calibration \citep{Yan16}. A three point dithering strategy is used to perform all the observations to achieve a complete spatial coverage in the sources \citep[for this and further details about the observing strategy please refer to][]{Law15}.

The basic data reduction was done with the version 2.1.2 of the MaNGA reduction pipeline \citep{Law16}, which delivers sky subtracted, wavelength and flux calibrated data cubes as final data products. For the analysis of the data cubes we make use of the {\sc Pipe3D} pipeline \citep{Pipe3D_I,Pipe3D_II}, which was developed for the analysis of spectra from IFS data. The pipeline delivers information on the stellar populations and the ionized gas of the galaxies, including the kinematics traced by both components. 

The original sample for this work comprises 2737 galaxies available within the MaNGA collaboration by March 2017 (MPL5), all of them analyzed with Pipe3D. This analysis provides spatially resolved, integrated, and characteristic properties of the galaxies, including the stellar masses and SFRs \citep[e.g.,][]{sanchez18a}. 

We combined the IFS dataset with a new and independent morphological classification of the galaxies. This analysis was directly estimated by a fully visual inspection regardless of any other morphological classification that may be available in different databases \citep[e.g., Galaxy Zoo,][]{galzoo}. The classification was carried out by only two classifiers in order to be homogeneous using the SDSS images (DR8), and it was done in various stages: (i) The $gri$-color composed images of all MaNGA galaxies were displayed through a link to the SDSS server. Different zoom and scale options were used to better judge both (1) the morphological details in the inner/outer parts of galaxies and (2) the immediate apparent galaxy environment. Images were judged according to the standard Hubble morphological classification. (ii) A second and more refined evaluation of the morphology is carried out after (1) applying some basic image processing to the $gri$-SDSS images and (2) judging the geometric parameters (ellipticity, position angle, A4 parameter) after an isophotal analysis. At the end of this second stage we tried to isolate as much as possible lenticular galaxies masquerading as ellipticals. (iii) The third and fourth steps carry out a similar analysis than in steps one and two, but incorporating the recently delivered DESI Legacy Survey \citep{Dey2018} images. For further details on the morphological classification methodology we refer the reader to \citet{HernandezToledo2010}. Details of the morphological properties of the MaNGA galaxies will be presented elsewhere (Hernandez-Toledo et al., in prep.).

\begin{figure}
  \centering
    \includegraphics[width=0.35\textwidth, angle=90]{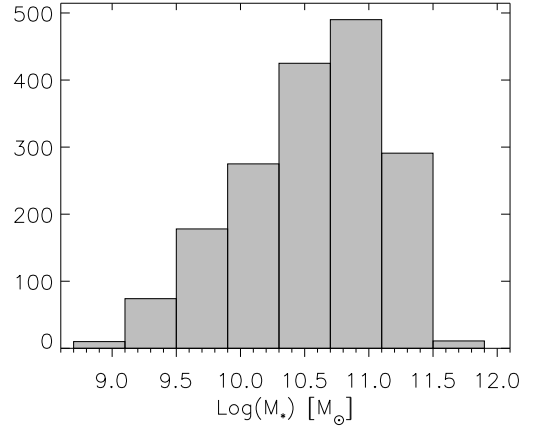}
  \caption{Mass distribution of the final sample of 1,853 galaxies used in this work.}
  \label{MassHistogram}
\end{figure}

Here we present a brief summary for the galaxies analyzed in this article. From our main sample of 2737 galaxies, we discarded those laying outside the stellar mass range $10^{8.5} M_{\odot}< M_{*}$, along with the highly inclined galaxies in order to avoid possible inclination effects in the determination of the SFR. For this purpose, we select only the galaxies with inclinations $i<$ 60\textdegree, since at high inclinations the derivation of the spatially resolved properties may be severely affected. \citep{Ibarra-Medel+2019}. The inclinations were derived using the b/a axis ratio derived by a Sersic fit, reported in the NASA-Sloan Catalogue (NSA)\footnote{The NSA catalogue is accesible through their website: http://nsatlas.org}.  In addition, we discarded the perturbed galaxies, those with unsure morphological classification. After the previous selection we were left with a final sample of 1,754 galaxies. The stellar mass distribution of this final sample of galaxies is presented in Figure \ref{MassHistogram}. Contrary to what it was foreseen by the MaNGA sample selection \citep[e.g.][]{wake17}, the current distribution does not cover the masses in an homogeneous way.

\begin{figure}
  \centering
    \includegraphics[width=0.35\textwidth, angle=90]{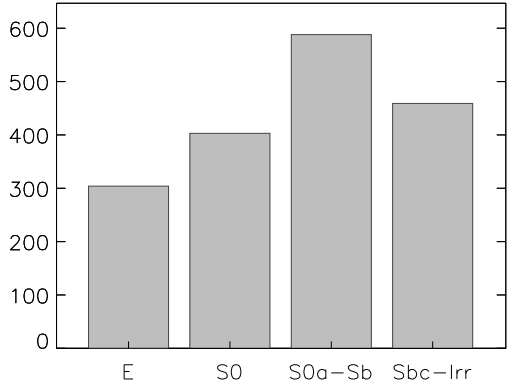}
  \caption{Morphological distribution of the final sample of 1,754 galaxies studied in this work.}
  \label{SampleHistogram}
\end{figure}

 The morphological distribution is presented in Figure \ref{SampleHistogram}, where we show the four morphological groups in which we have divided our sample: (i) Ellipticals (E; 304 galaxies); (ii) Lenticulars (S0; 403 galaxies); (iii) Early-type spirals, which comprises S0a, Sa, Sab and Sb galaxies (S0a-Sb; 588 galaxies); and (iv) Late-type spirals, which comprises Sbc, Sc, Scd, Sd, Sm and Irregulars (Sbc-Irr; 459 galaxies). 

\section{Analysis}
\label{ana}

\subsection{Star Formation Rates and Stellar Masses} \label{SFR}

As indicated before, we use the results from the analysis provided by the {\sc Pipe3D} pipeline \citep{Pipe3D_II}. Both, the SFR and $M_{*}$ for all the galaxies in our sample were derived from these data products using the same methodology already explained in our previous work on the topic \citep{CanoDiaz16}.

A brief summary of the procedure to obtain the quantities required in this study is as follows: (i) the stellar masses, $M_{*}$, were calculated after fitting the spectra with a model of the stellar populations, using the GSD156 single-stellar population (SSP) library \citep{cid-fernandes13}. A spatial binning was adopted in the cubes with two purposes, (a) to increase the S/N to a maximum goal of 50 per $\AA$ in each spaxel and (b) to preserve as much as possible the original shape of the light distribution of the galaxies across the field of view \citep{Pipe3D_II}. Once performed the binning, the pipeline performs a multi-SSP decomposition of the stellar population, and corrects for the binning effect afterwards, by adopting a {\it dezonification} procedure \citep{cid-fernandes13}. This procedure consists of re-scaling the best-fitted stellar population model, derived for each bin, to the flux intensity in each spaxel within the considered bin. Under this assumption all spectra within each bin have similar stellar population properties (in particular same ages, metallicities, dust attenuation and therefore same Mass to light, M/L, ratio), and they differ only in their photometry. This way the {\sc Pipe3D} tool is able to deliver spaxel-wise maps of any stellar property. In particular this software provides stellar mass surface density ($\Sigma_{*}$) maps; (ii) after the stellar population fitting was performed, {\sc Pipe3D} subtracts this contribution to the cubes to obtain spectra of the ionized gas emission lines. These emission lines are then fitted to obtain spatial maps of their main characteristics, such as flux, equivalent widths (EWs), etc. In particular, we used the information of the {\mbox{\rm{H}$\alpha$}} line to transform its luminosity to SFR applying the \citet{Kennicutt98} conversion factor, and the {\mbox{\rm{H}$\beta$}} line to extinction correct the {\mbox{\rm{H}$\alpha$}} luminosity; (iii) we used other lines flux maps ({\mbox{\rm{H}$\beta$}}, {\mbox{\rm [O{\small III}]}}, and {\mbox{\rm [N{\small II}]}}) to perform the classification of galaxies and regions within them based on their dominant ionizing source: star forming, AGNs, retired or undetermined, following the scheme already adopted in previous articles \citep{Sanchez13,CanoDiaz16,laura17,sanchez18a}, as described in the next section. 

We note that the SFR history, in particular the current value of SFR (for example, averaged during the last $\sim 30-100$ Myr of the observed galaxy), can be calculated with the inversion method making use of the full observed spectrum. As showed in \citet{sanchez18b}, despite the differences between the H$_{\alpha}$- and SSP-based SFRs, both present a similar correlation with the stellar mass in what regards the SFMS. For galaxies much below the SFMS, meaning the RGs, the {\mbox{\rm{H}$\alpha$}}-based SFRs correlate tightly with $M_{*}$ while the SSP-based ones present a more scattered trend in the SFR--$M_{*}$ diagram. In fact, the {\mbox{\rm{H}$\alpha$}}-based SFRs for the RGs should be interpreted as an upper limit as other sources of ionization besides young hot stars can be relevant in these galaxies.  In Section \ref{ResultsFullIntegrated} we discuss about this question in more detail.

\begin{figure}
  \centering
   \includegraphics[width=0.35\textwidth, angle=90]{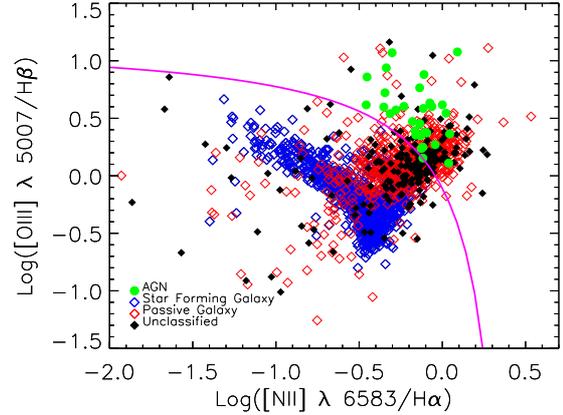}
  \caption{Distribution of the analyzed sample of galaxies along the BPT diagram, color coded according to our ionization classification scheme: blue for SFGs, red for RGs, green for AGNs, and black for unclassified sources. Each symbol corresponds to the line ratios derived from integrated quantities for each galaxy. Magenta solid line represents the \citet{Kewley01} demarcation limit.}
 \label{IntegratedBPT}
\end{figure}

\subsection{Classification by ionization processes} \label{classification}

We classify the sources by the dominant ionizing process. For doing so, we follow the classification scheme presented in \citet{Sanchez13} and adopted in several previous studies \citep[e.g.][]{laura16,CanoDiaz16,Barrera-Ballesteros+2016,laura17,sanchez18a}. This classification makes use of a combination of the EW({\mbox{\rm{H}$\alpha$}}) classification introduced by \citet{CidFernandes11} and the Kewley demarcation limit (KL) \citep{Kewley01} in the Baldwin-Phillips-Terlevich (BPT) diagram \citep{Baldwin81}.

We visually present our classification scheme in a classic BPT diagram in Figure \ref{IntegratedBPT}, using the {\it integrated} light (across the FoV) from all of our 1,754 sources, which means that each symbol corresponds to one galaxy. The galaxies are classified as follows: (i) green symbols account for galaxies for which the [\ion{O}{iii}]/H$\beta$ and  [\ion{N}{ii}]/{\mbox{\rm{H}$\alpha$}} line ratios lie above the KL, and whose EW({\mbox{\rm{H}$\alpha$}}) are \textgreater 6 \AA, meaning that the dominant ionization process in these galaxies comes from the nuclear activity. This group represents only the $\sim 2\%$ of our sample \citep[see also][]{sanchez18a}; (ii) blue symbols represent the galaxies that lie below the KL and whose EW({\mbox{\rm{H}$\alpha$}}) are \textgreater 6 \AA, which means that the ionization processes in these galaxies are dominated by star formation, as it is shown in Figure 3 of \citet{Sanchez14}. This is the biggest group in our sample, as it represents the $\sim 51\%$; (iii) red symbols account for galaxies whose EW({\mbox{\rm{H}$\alpha$}}) are \textless 3 \AA\ regardless of their position in the BPT diagram, meaning that probably the dominant processes in the gas ionization come from old stellar populations \citep{CidFernandes11}. Around $35\%$ of our sample belong to this group; (iv) black symbols represent galaxies whose EW({\mbox{\rm{H}$\alpha$}}) are in between 3 \AA and 6 \AA, which means that their main ionization process remain uncertain being probably in the transition between the star forming regime to the retired one. This group makes a $\sim 12\%$ of our sample. In this plot we can see that, as expected, the star forming galaxies (SFGs; blue symbols) and the AGN dominated galaxies (green symbols) lie below and above the KL drawn with a black solid line, respectively. On the other hand, the passive or retired galaxies (RGs; red symbols) and the unclassified ones (black symbols) lay mostly close to the demarcation limit or beyond it, towards the denominated LINERs location \citep[e.g.][]{Kauffmann2003, Lacerda18}.

 \begin{figure}
  \centering
   \includegraphics[width=0.35\textwidth, angle=90]{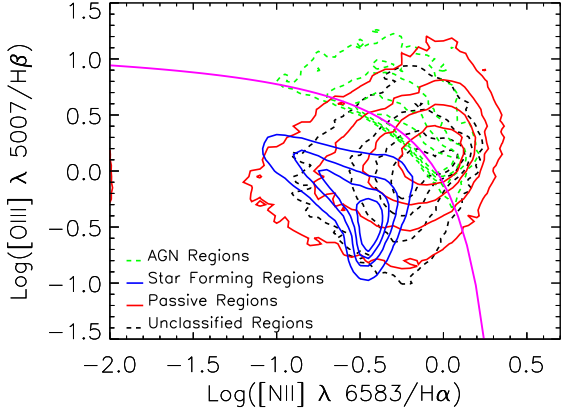}
  \caption{Local BPT diagram showing isocontours of the individual regions of the galaxies used to derive the local SFMS and passive or retired region in blue and red respectively, along with the regions whose ionization classification was undefined and determined to be triggered by nuclear activity in dashed black and green isocontours respectively. The magenta solid line represents the Kewley demarcation limit.}
 \label{SpatiallyResolvedBPT}
\end{figure}

As the goal of this work is to explore both the global SFR-M$_*$ and the local $\Sigma_{SFR}-\Sigma_{*}$ diagrams, then we need to segregate also the local individual regions of each galaxy in our sample and classify them in terms of their ionization type. Using the previously described scheme, in Figure \ref{SpatiallyResolvedBPT} we present a spatially-resolved BPT diagram, as a contour plot. Blue and red isocontours represent the contributions from the star-forming and retired areas (hereafter SFAs and RAs) within the galaxies, respectively, which are the regions that we will focus on for this study. Along with those, we also present the isocontours corresponding to regions whose main ionization source was not possible to be classified (black dashed line), and for those that were identified to be induced by nuclear activity (green dashed line), respectively. For all the four cases, the isocontours presented in the figure correspond to the 80$\%$, 60$\%$, 40$\%$ and 20$\%$ of the total amount of data, while the magenta solid line represents the KL demarcation. This plot, which is clearly consistent with the global BPT diagram (Figure \ref{IntegratedBPT}) shows indeed that the SFAs and RAs are following the same behavior as their global counterparts The total amounts of classified regions used in this plot are reported in Table \ref{LocalFractions} in Section \ref{morph_loc}.

\section{Results} \label{Results}

In this Section we explore the distributions of galaxies within the {\it global} SFR-M$_{*}$ and  $\overline{\Sigma}_{SFR}-\overline{\Sigma}_{*}$ diagrams (where $\overline{\Sigma}_{SFR}$ and $\overline{\Sigma}_{*}$ are the average of $\Sigma_{SFR}$ ans $\Sigma_{*}$ per galaxy), as well as the distribution of areas within galaxies in the (spatially-resolved) {\it local } $\Sigma_{SFR}-\Sigma_{*}$ diagram. We separate the galaxies and areas in star-forming and retired using the criteria explained in the previous Section, and study their correlations in these diagrams.

 \begin{figure}
  \centering
   \includegraphics[width=0.35\textwidth, angle=90]{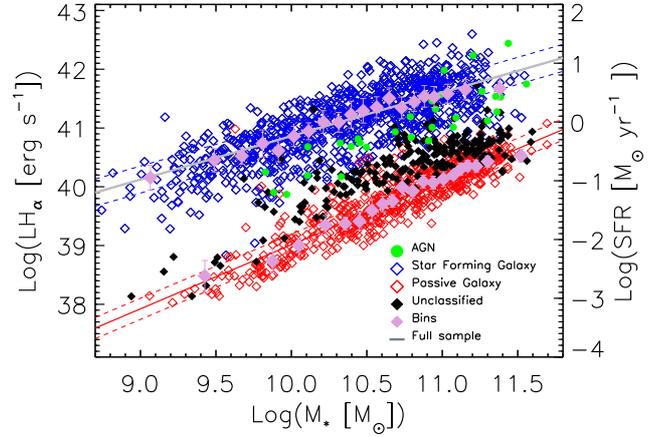}
  \caption{Global SFR-M$_*$ diagram for the analyzed sample of galaxies. Each symbol corresponds to a single galaxy, color coded according to its dominant ionization source as shown in Fig. \ref{IntegratedBPT}.
 Pink symbols represent the binned data used to fit a log-linear relation for both, the SFGs (blue diamnods) and RGs (red diamonds). Blue and red solid lines represent the best fits for each subsample of galaxies, while the corresponding dashed lines show a measure of the scatter around both relations. For comparison the \citet{Renzini15} relation is shown as the solid gray line.}
 \label{FullSampleIntegrated}
\end{figure}

\subsection{Global (integrated) SFR-M$_{*}$ diagram} \label{ResultsFullIntegrated}

Following \citet{CanoDiaz16}, we integrate the local masses and SFRs across the FoV of the MaNGA galaxy datacubes. These FoVs correspond to galactocentric distances $<1.5 R_{e}$ ($<2.5 R_{e}$) for $\sim$70\% ($\sim$30\%) of the objects. The derived distribution in the SFR-M$_{*}$ diagram is shown in Figure \ref{FullSampleIntegrated}. Notice that we also show for comparison the luminosity of H$_{\alpha}$, which was used to derive the SFRs through the \citet{Kennicutt98} relation. Indeed, the reported SFR is just a linear scale of the H$_{\alpha}$ luminosity. 

We plot with different colors the dominant ionization source in each galaxy following the classification scheme described in the previous Section. We observe that galaxies whose ionization is dominated by star formation, the SFGs (represented in blue), are clearly segregated from those dominated by ionization of old stars, the RGs (represented in red). Both populations present clear correlations between L$_{\rm H\alpha}$ and M$_{*}$, with Pearson correlation coefficients larger than $r_c>$0.7, in general. For the particular case of the SFGs, where L$_{\rm H\alpha}$ certainly traces the SFR, its correlation is well represented by a linear trend in logarithmic quantities. The blue solid line is the fit to what we identify as the global SFMS, while the dashed lines show its corresponding standard deviation.
The linear regression in the log-log plane was applied to the means and standard deviations in mass bins, showed with the pink filled symbols and error bars, respectively.
Each bin comprises at least a 5$\%$ of the total amount of analyzed data (for details about the binning methodology, refer to online available Appendix A). Along this article we adopt this binning scheme, followed in previous articles \citep[e.g.][]{Sanchez13,CanoDiaz16,sanchez17a,bb17}. This scheme prevents partially the contribution to the fit of
possible outliers, samples the distribution in a more representative way, and limits the bias due to the particular sample mass distribution (see Fig. \ref{MassHistogram}). If not adopted, the fitting would be dominated by the range of stellar masses with larger number of galaxies (M$_* \sim$10$^{10.5-11}$M$\odot$), with a possible limited constrain to the slope.

 \begin{figure}
  \centering
   \includegraphics[width=0.35\textwidth, angle=90]{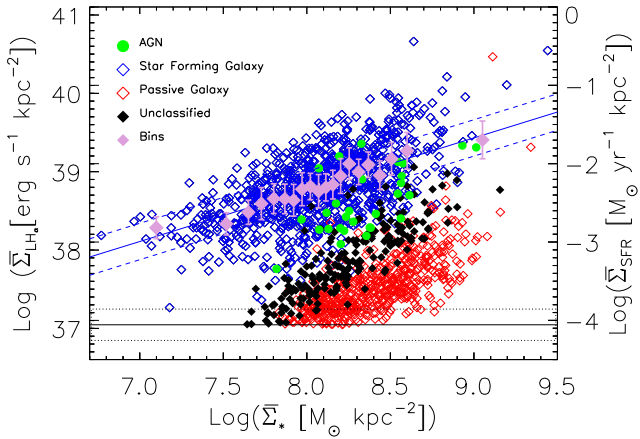}
  \caption{Global intensive $\overline{\Sigma}_{SFR}-\overline{\Sigma}_*$ diagram for the analyzed sample of galaxies. Each symbol corresponds to a single galaxy, color coded according to its dominant ionization source as shown in Fig. \ref{IntegratedBPT}. 
 Pink symbols represent the binned data used to fit a log-linear relation for both, the SFGs (blue diamonds) and RGs (red diamonds). The blue solid line represent the best fit for the star-forming galaxies, while the dashed blue lines show a measure of the scatter around the relation. The black solid horizontal line shows the average detection limit in the $H_{\alpha}$ line flux density for the MaNGA observations, enclosed by dashed black lines that represent its scatter. 
} 
 \label{LocaltoGlobalSFMSandRS}
\end{figure}

\begin{table*}
 \begin{center}
    \begin{tabular}{ l  c  c  c c c c p{5cm}}
    \hline
    \hline
    SFMS & \shortstack{Pearson Correlation \\ Coeff. ($\rho$)} & \shortstack{Confidence Interval \\ $\rho$ 99$\%$} & Slope &  \shortstack{log(SFR)$_{10.5M_{\odot}}^{*}$ \\ log(M$_{\odot}$ yr$^{-1}$)}  & \shortstack{Standard Deviation \\ ($\sigma$)} & \shortstack{No. of galaxies \\ in the relation} \\ \hline
    Full Sample & 0.76 & (0.72, 0.80) & 0.74$\pm 0.01$ & 0.13$\pm 0.05$ & 0.23 & 892 \\ \hline
    E & 0.67 & (0.39, 0.83) & 0.42$\pm 0.01$ & -0.29$\pm 0.14$ & 0.31 & 45 \\ 
    S0 & 0.57 & (0.19, 0.81) & 0.42$\pm 0.02$ & -0.13$\pm 0.22$ & 0.35 & 34\\ 
    S0a-Sb & 0.73 & (0.66, 0.79) & 0.68$\pm 0.01$ & 0.12$\pm 0.07$ & 0.22 & 357 \\ 
    Sbc-Irr & 0.85 & (0.81, 0.88) & 0.79$\pm 0.01$ & 0.20$\pm 0.06$ & 0.19 & 456 \\ \hline
    CALIFA$^{\dag}$ & 0.84 & (0.76, 0.89) & 0.81$\pm 0.02$ & 0.16$\pm 0.21$ & 0.20 \\ 
    RP15$^{\dag \dag}$ & - & - & 0.76$\pm 0.01$ & 0.34$\pm 0.03$ & - \\ \hline
    \hline
RGs Relation & \shortstack{Pearson Correlation \\ Coeff. ($\rho$)} & \shortstack{Confidence Interval \\ $\rho$ 99$\%$} & Slope &  \shortstack{log(SFR)$_{10.5M_{\odot}}^{*}$ \\ log(M$_{\odot}$ yr$^{-1}$)}  & \shortstack{Standard Deviation \\ ($\sigma$)} & \shortstack{No. of galaxies \\ in the relation} \\ \hline
    Full Sample & 0.89 & (0.87, 0.91) & 1.09$\pm 0.01$ & -1.55$\pm 0.06$ & 0.18 & 616 \\ \hline
    E & 0.90 & (0.86, 0.93) & 0.90$\pm 0.01$ & -1.37$\pm 0.10$ & 0.17 & 196 \\ 
    S0 & 0.92 & (0.89, 0.94) & 0.98$\pm 0.01$ & -1.53$\pm 0.08$ & 0.15 & 290 \\ 
    S0a-Sb & 0.82 & (0.72, 0.88) & 1.09$\pm 0.01$ & -1.51$\pm 0.13$ & 0.25 & 129 \\
    Sbc-Irr & - & - & - & - & - & 1 \\ \hline
   CALIFA$^{\dag}$ & 0.85 & (0.77, 0.90) & 0.86$\pm 0.02$ & -1.29$\pm 0.26$ & 0.22 \\ \hline
        \end{tabular}
	\caption{Correlation coefficients and coefficient of the log-linear fits to the SFGs and RGs in the SFR--$M_*$ diagram. The linear regressions are applied to the binned data, where the size of the bins are equivalent to 5$\%$ of the total amount of data for full sample cases and, to 15$\%$ for each of the morphological groups. The $\sigma$ value is the total standard deviation around the fitting. $^{*}$ Normalization coefficient defined at the characteristic stellar mass of 10$^{10.5}$M$_{\odot}$. In the case of the comparison values from CALIFA and RP15, it was calculated using the reported values of the zero point in each work, and summing the uncertainty values for the zero point and the slope. $^{\dag}$ \citet{CanoDiaz16}. $^{\dag \dag}$\citet{Renzini15}.}
	\label{IntegratedFits}
 \end{center}
\end{table*}

A least squares minimization routine (Y vs. X) was used to perform the fitting. The adopted routine (used trough all the paper unless clearly stated) was the LINFIT function implemented in the Interactive Data Language (IDL). This routine takes into account the individual errors per bin, estimated by the population standard deviation of the bins. The results of the fitting are presented in Table \ref{IntegratedFits}, including the slopes and the normalizations defined at the characteristic stellar mass of 10$^{10.5}$M$_{\odot}$, the correlation coefficients, the 99\% confidence intervals of the parameters and the standard deviation in the SFR measured along the best fitted relations.  
For comparison, the linear regression with all the data points (not binning in mass) using a hierarchical 
Bayesian model \citep{Kelly07} is presented in the online available Appendix B. Both fitting approaches, using data averaged in mass bins and using all the data points, are actually similar.

We find that the SFR-M$_*$ correlation for our SFGs is indeed tight with a total standard deviation of $\approx $0.23 dex, and a slope of 0.74 (see Table \ref{IntegratedFits}). We define this correlation as the SFMS.  Our results are in agreement with most of previously published analysis of the SFMS in the local Universe, which were mainly established {\it statistically} in the SFR-$M_*$ diagram but not by defining SFGs with a physical criterion a priori as done here (see the references in the Introduction). In particular, we find similar results to those of \citet{Renzini15}, and those using the CALIFA survey \citep{catalan15,CanoDiaz16,rosa16a}, which also make use of IFS data. For comparison purposes, we list in Table \ref{IntegratedFits} some of these results from the literature. 

The SFMS informs about the shape of the star formation history (SFH). Indeed, it connects the derivative of the in-situ stellar mass growth (i.e, the SFR) with the integral of the SFH (i.e., M$_*$, after taking into account the stellar mass loss and ignoring the contribution of dry mergers). Thus, the slope of the correlation informs about the shape of the SFH, as discussed by \citet{Speagle14} and \citet{sanchez18b}. In particular, a sub-unity slope, as found here, implies that the global SFHs of SFGs are on average shallower than an exponential one.

For the RGs, we find a strong correlation too, for which we do not provide a linear fit. Instead, we have to keep in mind the actual meaning of this relation in the log SFR-log $\Sigma_{*}$ plane. For RGs, the ionization comes from other sources, including post-AGB stars and AGN ionization or rejuvenation in the outer regions \citep[e.g][]{sarzi10,papa13,sign13,Gomes16a,Gomes16b,Belfiore17a}. Therefore, the H$\alpha$-based SFRs for RGs should be considered as an upper limit, as recently demonstrated by \citet{bitsakis18}. Indeed, as the color code clearly indicates, the distribution of RGs located in the cloud segregated from the SFMS have been selected for not being compatible with star-formation, based on their location in the BPT diagram (Fig. \ref{IntegratedBPT}). The fact that the amount of {\mbox{\rm{H}$\alpha$}} luminosity scales with M$_*$ in these galaxies, with a slope compatible with one, may indicate that the ionization somehow scales with the amount of stars. This is in agreement with the fact that old stars are the dominant ionizing source \citep{Binette94,Stasinska08,sign13}. In this regard, for the current work which focuses on the study of the SFMS, the main result from the observed distribution in the log SFR-log M$_*$ plane is that, when the main ionizing source of an entire galaxy is compatible with LINER-like emission, based on its location in the BPT diagram, this galaxy is located well below the SFMS ($>2\sigma$), being easily classified as retired.

As seen in Fig. \ref{FullSampleIntegrated}, galaxies classified a priori by their dominant ionization source cannot be strictly separated into SFGs and RGs by a simple line in the log SFR-log M$_*$ diagram. However, a line that best separate both populations of galaxies can in principle be established. We find that an approximate criterion for separating SFGs from RGs (Salpeter IMF) can be the line that follows the galaxies that have an EW({\mbox{\rm{H}$\alpha$}}) near to 5, which we find is given by:
\begin{equation}
\log(\textstyle\frac{SFR}{M_\odot yr^{-1}}) $=$ -\left(12.20\pm0.15\right) + \left(1.08\pm0.01\right) \log(\frac{M_*}{M_\odot}).
\end{equation}
The details of the derivation of this demarcation line are given in the online available Appendix D, and it is also shown in Figure \ref{SFMSandRSEWHalpha_int} as a gray continuous line. As expected, many of the undefined galaxies by their dominant ionization source (black dots in Fig. \ref{FullSampleIntegrated}) lie around this separation line in the log SFR-log M$_*$ diagram, as well as some SFGs, which means that using this demarcation line cannot avoid contamination of other sources, as well as losing some star-forming ones. We warn that the usage of this demarcation line will not be as accurate as the usage of an spectroscopic selection criteria for the star-forming sources, like the one used in this work (see Section \ref{classification}). 

Finally, we should highlight the location of galaxies whose dominant ionization are AGNs. They are located mostly in the so-called "green valley" \citep[see e.g.,][]{sanchez18a}, the proposed transition region between the SFGs (located at the SFMS) and the RGs (located well below the SFMS). It is assumed that whatever is the mechanism that produces the quenching in galaxies, the fact that there is a depopulated intermediate regime between the two well defined and heavily populated areas (in this diagram or other diagrams, like the color-magnitude one), is an evidence of a fast transition. This is extensively discussed in different articles, making a clear distinction between a slow and smooth decrease of the SFR activity (aging) versus a sharp and abrupt halt of this activity (quenching), as described in e.g., \citet{Schawinski14} and \citet{casa17}. Indeed, the location of AGN hosts in the green valley has been proposed as a possible evidence of a connection between the quenching and the nuclear activity \citep[e.g.,][the latter argues in favor of a gradual suppression of the SF instead of an abrupt one]{Schawinski14,shimi17,Nandra07,Salim07}, although recent studies suggest that they could be co-evolving processes \citep{sanchez18a}.

\begin{figure*}
    \begin{subfigure}[h]{0.39\textwidth}
        \includegraphics[width=0.75\textwidth, angle=90,keepaspectratio]{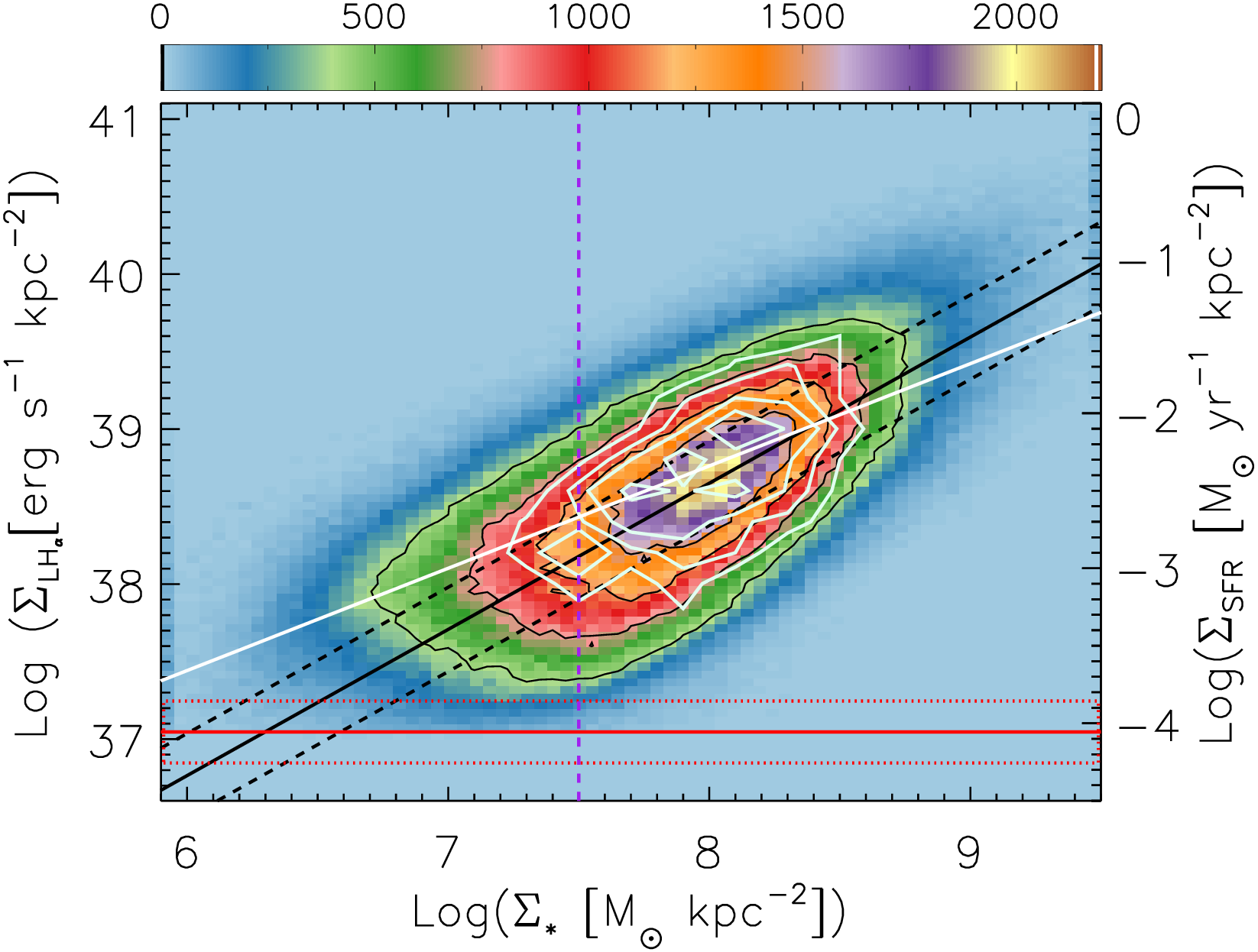}
        \label{SRSFMS}
    \end{subfigure}
    ~\quad
    \begin{subfigure}[h]{0.39\textwidth}
        \includegraphics[width=0.75\textwidth, angle=90,keepaspectratio]{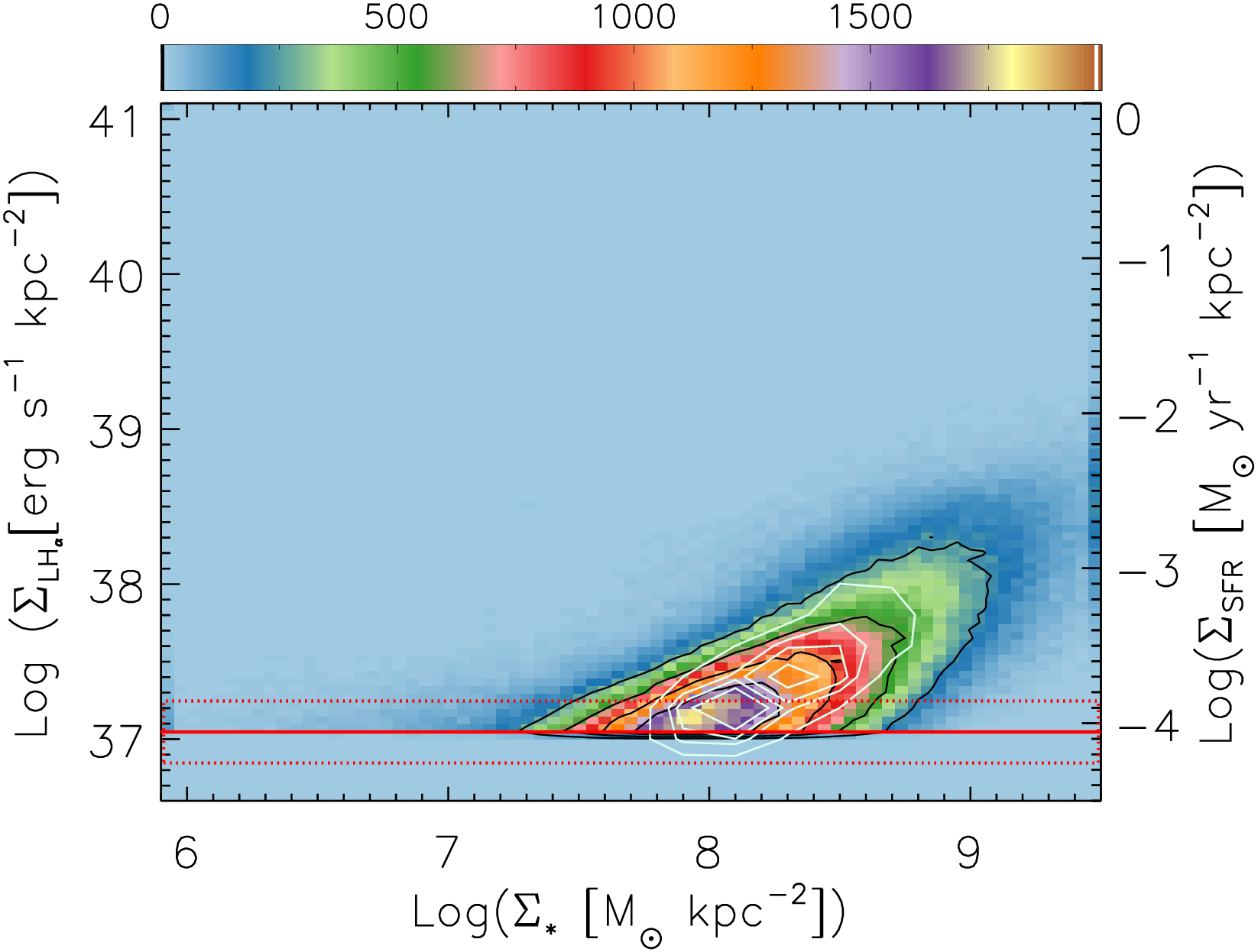}
        \label{SRRS}
    \end{subfigure}
\caption{Local (spatially-resolved) $\Sigma_{\rm SFR}-\Sigma_{*}$ diagrams for SFAs (left panel) and RAs (right panel). The color-image indicates the density of points in the plots. The black isocontours represent the 80$\%$, 60$\%$, 40$\%$ and 20$\%$ of the total amount of data. The red solid horizontal line shows the mean detection limit in the $H_{\alpha}$ line flux density for the MaNGA observations, while the area enclosed by the doted red rectangle is a measure of its scatter. The black solid line in the local SFMS plot, shows the linear fitting done to the data at the right of the $\Sigma_{*}$ cut, which corresponds to the vertical purple dashed line. As a comparison in both panels, we over-plot in white the isocontours of the global $\overline{\Sigma}_{\rm SFR}-\overline{\Sigma}_{*}$ distributions shown in Fig. \ref{LocaltoGlobalSFMSandRS}, and for the right panel the fit of the global relation already presented in the same figure is also plotted for comparison purposes.
}\label{SpatResSFMSandRS}
\end{figure*}

\subsection{Global (integrated) $\overline{\Sigma}_{\rm SFR}-\overline{\Sigma}_{*}$ diagram}
\label{localSR_SFRandRS}

The SFMS described in the previous Section is a relation between two extensive quantities that depends on the selected aperture and may scale with the cosmological distance if a Malmquist bias is involved. In this sense, it is safer to explore the possible physical connection between the two involved galaxy properties using two intensive quantities rather than extensive ones. The most simple way is to analyze the integrated SFR and stellar mass averaged per unit area, i.e., $\overline{\Sigma}_{\rm SFR}$ and $\overline{\Sigma}_{*}$.  Besides that, the relationship between these two global intensive quantities is directly comparable with the same quantities but spatially resolved, that is, the local ${\Sigma}_{\rm SFR}-\Sigma_{*}$ diagram to be presented in the next subsection.

To calculate the integrated $\overline{\Sigma}_{\rm SFR}$ and $\overline{\Sigma}_{*}$ we divide the global $M_{*}$ and SFR of each galaxy by their respective total areas, estimated by co-adding the areas of all the spaxels used to derive each one of these global quantities. It is important to remember here that these quantities are affected by the IFU size of the observations, which allows to have observations of the galaxies only up to 1.5 - 2.5 R$_{e}$. 
Figure \ref{LocaltoGlobalSFMSandRS} shows the distribution of $\overline{\Sigma}_{\rm SFR}$ along $\overline{\Sigma}_{*}$, following the same nomenclature adopted in Fig. \ref{FullSampleIntegrated}. The black solid horizontal line indicates the average detection limit of the $H_{\alpha}$ flux density for the MaNGA observations, which corresponds to $\approx 37$ log(erg s$^{-1}$ kpc$^{-2}$) or  $\approx -4.1$ log($M_{\odot}$ yr$^{-1}$ kpc$^{-2}$) in terms of log($\Sigma_{SFR}$). This limit was derived as the average of the 3$\sigma$ detection limit for all the galaxies, estimated as the mean H$\alpha$ surface brightness detected with a signal-to-noise ratio between 2.5 and 4.5$\sigma$, based on the errors reported by {\sc Pipe3D}. The average detection limit is enclosed by two dotted lines, which correspond to the $\pm \sigma$ dispersion of the distribution of all the individual detection limits in the galaxies of our sample. 

As in the case of the SFR--M$_*$ correlations, we find that the global surface SFR and surface stellar mass density present clear correlations, for both the SFGs and RGs. As in the previous subsection, we calculate the mean and standard deviation in bins, which in this case are in $\Sigma_*$, each containing at least 15$\%$ of the data (pink symbols and error bars in Fig. \ref{LocaltoGlobalSFMSandRS}), and use them to perform a direct linear regression in the log-log diagram. For both the SFGs and RGs, the obtained regressions are tight and with a sub-unity slope. The Pearson correlation coefficients and confidence interval, along with the best fitted parameters, together with the standard deviations along the derived relations are presented in Table \ref{SResolvedFits}.  The fact that the tight relations hold with intensive quantities is a clear evidence of a physical connection between the two analyzed quantities. In Appendix \ref{ApenB} (online available), we present the linear regression with all the data points (no binning in $\Sigma_*$) using the hierarchical Bayesian method by \citet{Kelly07}. This fitting procedure allows to take into account censored data, i.e., upper limits, which are the data below the average H$\alpha$ flux detection limit mentioned above. The slope obtained with this procedure is only slightly steeper than for the binned data (see Table \ref{FitsComparissons}). 

As we already discussed in the previous subsection, the nature of the described relations for both populations of SFGs and RGs is physically different. In particular, only the first one (the surface density SFMS) is certainly informing about the current shape of the SFH in galaxies. Note that this SFMS, $\overline{\Sigma}_{\rm SFR}$ vs. $\overline{\Sigma}_{*}$, is similar to the SFMS in extensive quantities, with standard deviations of $\approx 0.23$ dex, for both.

\begin{table*}
 \begin{center}
    \begin{tabular}{ l c c c c c p{5cm}}
    \hline
    \hline
\shortstack{$\Sigma_{\rm SFR}-\Sigma_{*}$\\ SFGs/SFAs Relations} & \shortstack{Pearson \\ Correlation  \\Coeff. ($\rho$)} & \shortstack{Confidence \\ Interval \\ $\rho$ 99$\%$} & Slope &  \shortstack{log($\Sigma_{SFR}$)$_{8M_{\odot}Kpc^{-2}}$$^{*}$ \\ (M$_{\odot}$ yr$^{-1}$ Kpc$^{-2}$)} & \shortstack{Standard \\ Deviation \\ ($\sigma$)} \\ 
\hline
\shortstack{SFGs Global} & 0.60 & (0.55, 0.66) & 0.66$\pm0.10$ & -2.34$\pm0.86$ & 0.23 \\ 
    \shortstack{SFAs Full sample} & 0.62 & (0.62, 0.62) & 0.94$\pm0.08$ & -2.48$\pm0.69$ & 0.27 \\ \hline
    \shortstack{SFAs E} & 0.71 & (0.70, 0.72) & 1.26$\pm0.07$ & -2.73$\pm0.63$ & 0.31 \\ 
    \shortstack{SFAs S0} & 0.72 & (0.71, 0.73) & 1.13$\pm0.06$ & -2.80$\pm0.56$ & 0.31  \\   
    \shortstack{SFAs S0a-Sb} & 0.68 & (0.68,0.68) & 1.02$\pm0.07$ & -2.60$\pm0.61$ & 0.26  \\ 
    \shortstack{SFAs Sbc-Irr} & 0.62 & (0.62,0.62) & 0.94$\pm0.08$ & -2.36$\pm0.70$ & 0.26   \\ \hline 
    \shortstack{SFAs CALIFA$^{\dag}$} & 0.63 & (0.62, 0.63) & 0.72$\pm0.04$ & -2.19$\pm0.33$ & 0.16 \\ \hline
\hline
  
    \end{tabular}
	\caption{Correlation coefficients and coefficient of the log-linear fits to the SFGs/SFAs and RGs/RAs in the global/local $\Sigma_{SFR}-\Sigma_*$ diagrams. For the global and local correlations, 100\% and 80\% of the data are used, respectively. The linear regressions are applied to the binned data, where the size of the bins are defined to contain 15$\%$ and  1\% of the total amount of data in the selected group for the global and local quantities, respectively. The $\sigma$ value is the total standard deviation around the fitting.    
    $^{*}$ Normalization coefficient defined at the characteristic stellar surface density of 10$^{8}$ M$_{\odot}Kpc^{-2}$. In the case of the comparison value from CALIFA, this value was calculated using the reported value of the zero point in that work, and summing the uncertainty values for the zero point and the slope. 
    $^{\dag}$ \citet{CanoDiaz16}.}
	\label{SResolvedFits}
 \end{center}
\end{table*}

\subsection{Local (spatially-resolved) $\Sigma_{\rm SFR}-\Sigma_{*}$ diagram}\label{ResultsFullSR} 

So far, we have described the connection between the integrated SFR and stellar mass for individual galaxies (extensive quantities), and between their integrated surface SFR's and surface mass densities (intensive quantities).
Previous studies \citep[e.g.,][]{Sanchez13,CanoDiaz16} have described a {\it local} (spatially-resolved) $\Sigma_{\rm SFR}-\Sigma_{*}$ relation for star-forming areas, SFAs, that hold to kpc scales, and that can be referred as the local SFMS. \citet{Hsieh17} explored the dependence of the shape of this local SFMS with galaxy properties. As other fundamental relations regulating the local star-formation activity in galaxies, like the Schmidt-Kennicutt-relation \citep[e.g.][]{kennicutt89,bolatto17}, it is an open question until which scale the SFMS relation is valid, and if the global (extensive or intensive) relation is directly connected to the local one. In this subsection we perform a direct comparison between the local and global correlations, using intensive quantities, i.e., the surface SFR's and mass densities.

To build the local SFMS in the $\Sigma_{\rm SFR}-\Sigma_{*}$ diagram, we classify the individual areas within the galaxies, using the same scheme described in Sec. \ref{classification}, as either star-forming or retired areas (hereafter denoted as SFAs and RAs, respectively), irrespective of the dominant ionization of their host galaxies. 
Therefore, in a galaxy classified as SFG it is possible to find both SFAs (that should dominate) and RAs.
The opposite is expected for RGs, as nicely described in previous studies \citep[e.g.][]{sign13,Gomes16a,Belfiore17a}.
 
Left panel of Figure \ref{SpatResSFMSandRS} shows the local $\Sigma_{\rm SFR}-\Sigma_{*}$ correlation for the SFAs extracted from the full sample of galaxies analyzed here, that is, the local SFMS.  Due to the large number of individual areas (1,265,176), we represent the distribution as density plots rather than individual points, using both, image scale and density iso-contours. Note that the SFAs in this plot are from all the galaxies, that is, in this subsection we study the spatially-resolved correlation for SFAs in a universal sense, {\it independent of the properties of the galaxies to which they belong}.
As in Fig. \ref{LocaltoGlobalSFMSandRS}, the red solid lines shows the average for the sample detection limit in $\Sigma_{\rm SFR}$. Also, the $\pm 2\sigma$ dispersion around this average is shown with the dotted lines.

We calculate the mean and standard deviation of $\log\Sigma_{\rm SFR}$ in bins of $\log\Sigma_{*}$, containing each 1\% of the data (see the means and dispersions in Fig. \ref{MassSFRbins} in Appendix A, and more details in Appendix \ref{ApenB}, both available as online material).
As in the case of the global relation, the means and standard deviations are used for calculating the linear regression and find thus the slope, normalization, and scatter of the local $\Sigma_{\rm SFR}-\Sigma_{*}$ relation. Before proceeding this way, it should be noted that the correlation bends significantly at the low-$\Sigma_{*}$ side, becoming shallower than at higher values of $\Sigma_{*}$ (for a clearer visualization of this, see  Fig. \ref{MassSFRbins}). In a recent study by \citet{Erroz-Ferrer+2019}, the authors have recognized this bending, and characterized it with their fitting methodology. However, our approximation to this bending is that most probably it is caused due to selection effects, i.e., it does not have a physical origin. We have analyzed the histograms of $\log\Sigma_{\rm SFR}$ in the individual $\Sigma_*$ bins and found the following result. At high values of $\Sigma_*$ the histograms present a quite symmetrical and smooth distribution of $\log\Sigma_{\rm SFR}$ but in bins below $\Sigma_*\approx 3\times 10^7$ M$_{\odot}$ kpc$^{-2}$, the distribution becomes more and more asymmetric, with a sharp cut-off at the low-$\Sigma_{\rm SFR}$ end, evidencing clear effects of the detection limits of each galaxy combined with the sample selection particularities and the radial aperture of the observations. Therefore, for the MaNGA galaxies studied here, the mean and standard deviation values of $\log\Sigma_{\rm SFR}$ in bins of $\Sigma_{*}$ below $\approx 3\times 10^7$ M$_{\odot}$ are likely strongly biased by these effects, which we will explore in more detail elsewhere.

We have applied the direct linear regression LINFIT to the data above $\Sigma_{\rm *,lim}\approx 3\times 10^7$ M$_{\odot}$ kpc$^{-2}$. The fit and its scatter are shown in the left pannel of Fig. \ref{SpatResSFMSandRS} and the results are given in Table \ref{SResolvedFits}. In the online Appendix \ref{ApenB}, for comparison we present the results for the linear regression not imposing any limit in $\Sigma_*$. As expected, the obtained slope is much shallower than the one with the limit (0.65 vs. 0.94, respectively). As seen in Fig. \ref{MassSFRbins}, the slope of the fit without imposing a limit is significantly influenced by the data below $\Sigma_{\rm *,lim}$. Since these data are affected by the detection limit and selection effects mentioned above, we caution about using them for inferring the spatially-resolved $\Sigma_{\rm SFR}-\Sigma_{*}$ correlation; a similar problem is expected for other surveys. In the online Appendix \ref{ApenB} we present also a log-linear regression to all the raw data (not binned) above $\Sigma_{\rm *,lim}$ using the hierarchical Bayesian model of \citet{Kelly07} with the option of censoring data, i.e., including upper limits, which here we identify with the data below the average H$\alpha$ flux detection limit. The results are presented in Table \ref{FitsComparissons}. They do not differ significantly to those presented in this subsection, reinforcing the robustness of our results.

To compare directly the spatially resolved diagram with the global one derived using the global intensive quantities for SFGs, we over-plot the latter in Fig. \ref{SpatResSFMSandRS} as the white isocountours.This figure shows that the distribution of individual areas in galaxies coincides roughly with the distribution of the averages of the whole galaxies, though the later one is limited to the region of larger values of the local distribution. This is well understood since the average surface densities of the considered parameters across the optical extension of a galaxy is a mass/flux weighted parameter, and therefore, it is biased towards the local values corresponding to the more dominant central regions (which have larger $\Sigma_{\rm SFR}$ and $\Sigma_{*}$ values).

In general, the global and spatially-resolved SFMSs coincide, suggesting that they are in principle the same relation. If so, we can claim that {\it the SFMS holds from the global galaxy scales down at least to kiloparsec scales.} However, in more detail, the local relation has a steeper slope, close to the fundamental value of one, and it presents a larger scatter than the global (integral) one, 0.27 vs. 0.23 dex. As mentioned above, the global SFMS is by construction slightly biased with respect to the local one, and also the sample selection can introduce some biases when using the properties averaged across the optical extension of the galaxies.  Regarding the slope close to one found for the spatially-resolved regions, regardless the global classification of their host galaxies, it suggest a universal mechanism for the local star-formation process able to keep along the galaxies $\Sigma_{\rm SFR}\propto \Sigma_{*}$. However, the scatter around this relation is not so small (0.27 dex) and it could be due to the existence of secondary effects on this universal mechanism that are regulated by some global property(ies) of the host galaxy. One of these properties that modulate the amplitude (normalization) of the relation is likely the cold gas fraction, which is strongly related to the morphology of the galaxies. In particular, in the next subsection and Section 5 we will see that the SFAs segregate along the local SFMS according to the morphology of the galaxies that host these SFAs. 

Finally, for completeness, in the right panel of Fig. \ref{SpatResSFMSandRS} we present the same spatially-resolved ${\Sigma}_{\rm SFR}- \Sigma_*$ diagram as in the left panel but for the RAs in all galaxies. As seen, the data are strongly affected by the detection limit (red-solid line). 
At least for the highest surface density areas, where the detection limit is not relevant, there is evidence of a correlation between $\Sigma_{LH_\alpha}$ and $\Sigma_*$
A single isocontour plot for all the areas (SFAs and RAs) in the ${\Sigma}_{\rm SFR}- \Sigma_*$ diagram is shown in Appendix \ref{ApenC} (available online). One clearly sees how the SFAs segregate from the RAs.

\subsection{The effects of morphology}\label{morph}

As reported in the previous subsections, the spatially-resolved SFAs of all galaxies follow a correlation between ${\Sigma}_{\rm SFR}$ and $\Sigma_*$, with a slope close to one. At the global level, the SFGs follow also a close correlation, though with at slope $<1$. In the literature it has been reported that both correlations may present secondary trends with some galaxy properties \citep[e.g.][]{LaraLopez+2010,Salim+2014,catalan15,rosa16a,Hsieh17,sanchez18b}. In particular, previous results have indicated that these correlations segregate by galaxy morphology \citep{rosa16a}, as well as the fractions of SFGs/SFAs and RGs/RAs depend on morphology. We explore here these possibilities for the current data. In particular, we are interested in exploring whether the local SFMS segregates or not by 
the morphology of the galaxy that host the SFAs. On its own, morphology is a property that can depend on other properties like the galaxy environment or that correlates with global properties like gas fraction, metallicity, etc.

\subsubsection{Morphological distribution in the SFR-M$_*$ diagram}\label{morph_int}

\begin{figure*}
    \centering   
    \captionsetup[subfigure]{labelformat=empty}
    
  \begin{minipage}{1\linewidth}
  \centering
    \begin{subfigure}[b]{0.45\textwidth}
        \includegraphics[width=0.75\textwidth, angle=90]{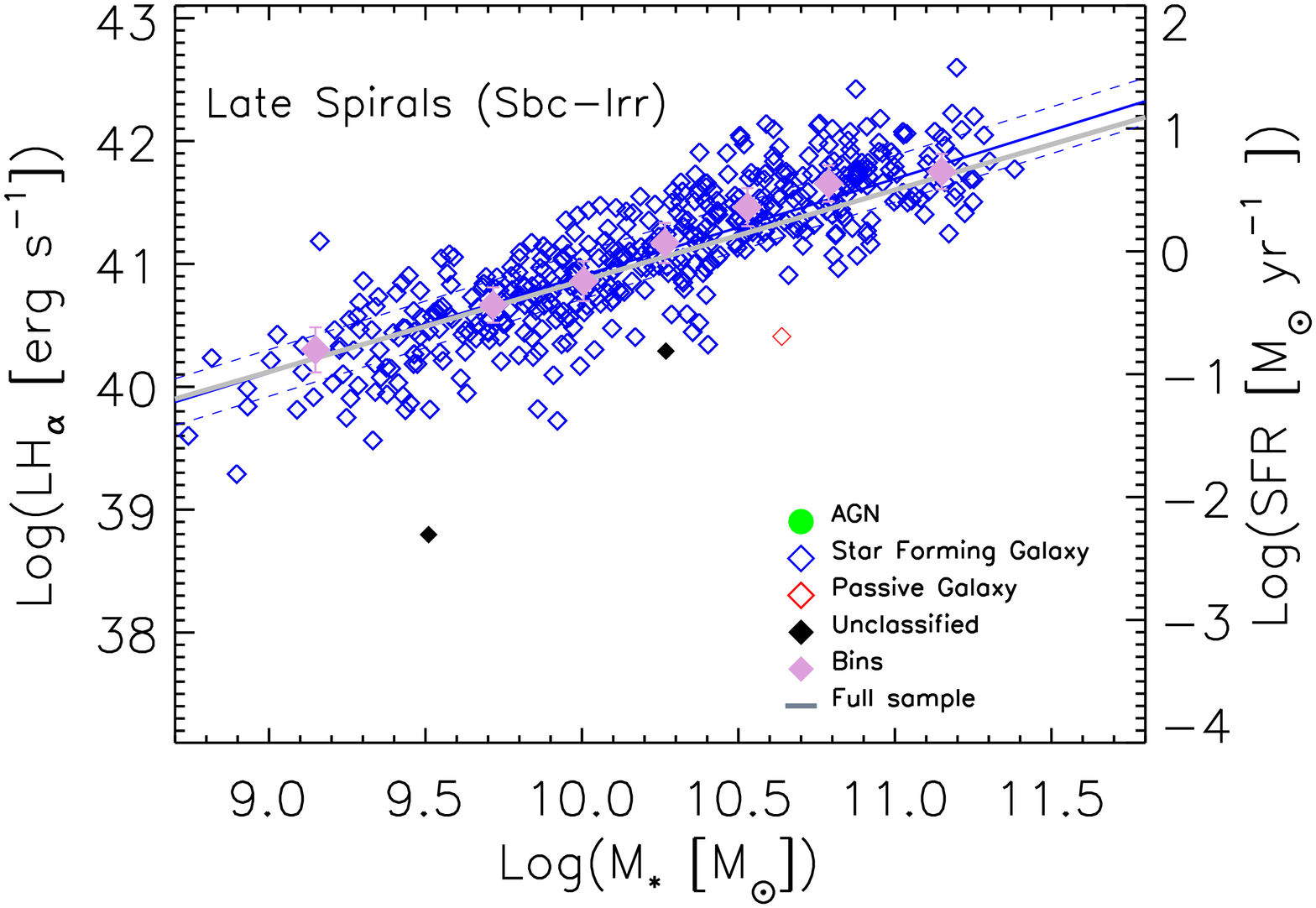}
     \label{fig:INT_Ellipticals}
     \end{subfigure}
     \begin{subfigure}[b]{0.45\textwidth}
        \includegraphics[width=0.75\textwidth, angle=90]{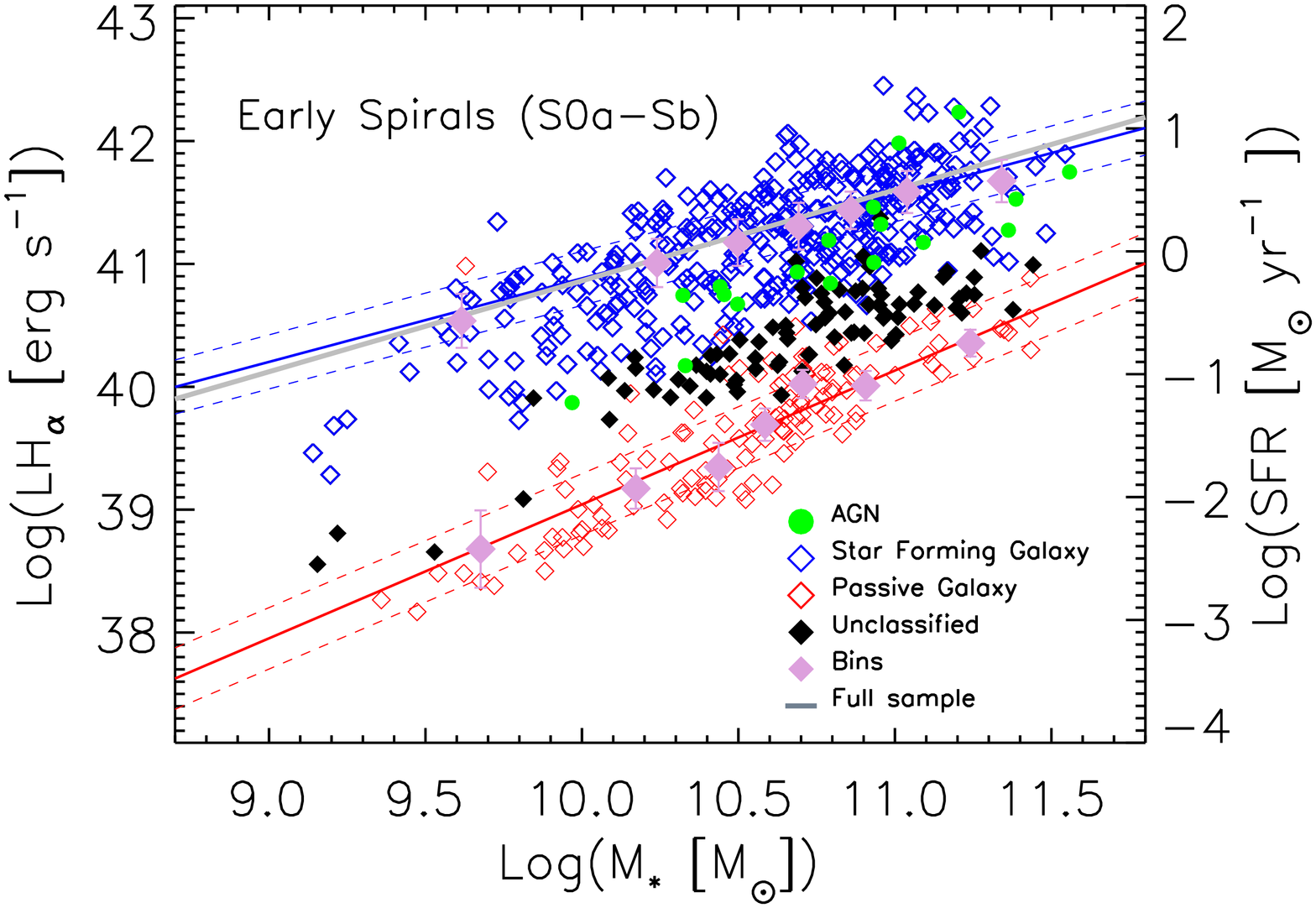}
    \label{fig:INT_Lenticulars}
    \end{subfigure}
 \end{minipage} \par\medskip
  \par\medskip
  \vfill     
 
 \begin{minipage}{1\linewidth}
 \centering
    \begin{subfigure}[b]{0.45\textwidth}
        \includegraphics[width=0.75\textwidth, angle=90]{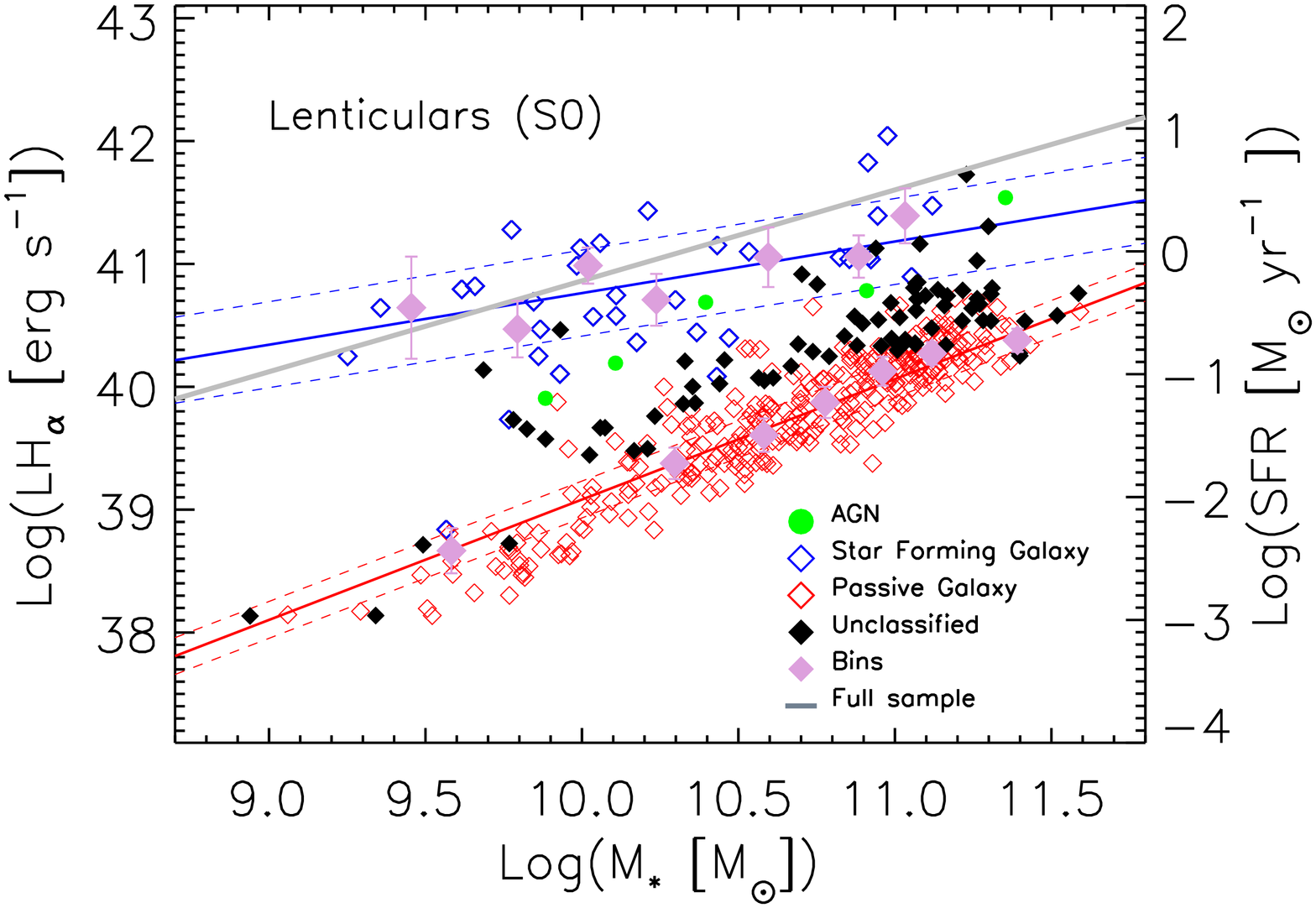}
    \label{fig:INT_Spirals}
    \end{subfigure}
    \begin{subfigure}[b]{0.45\textwidth}
        \includegraphics[width=0.75\textwidth, angle=90]{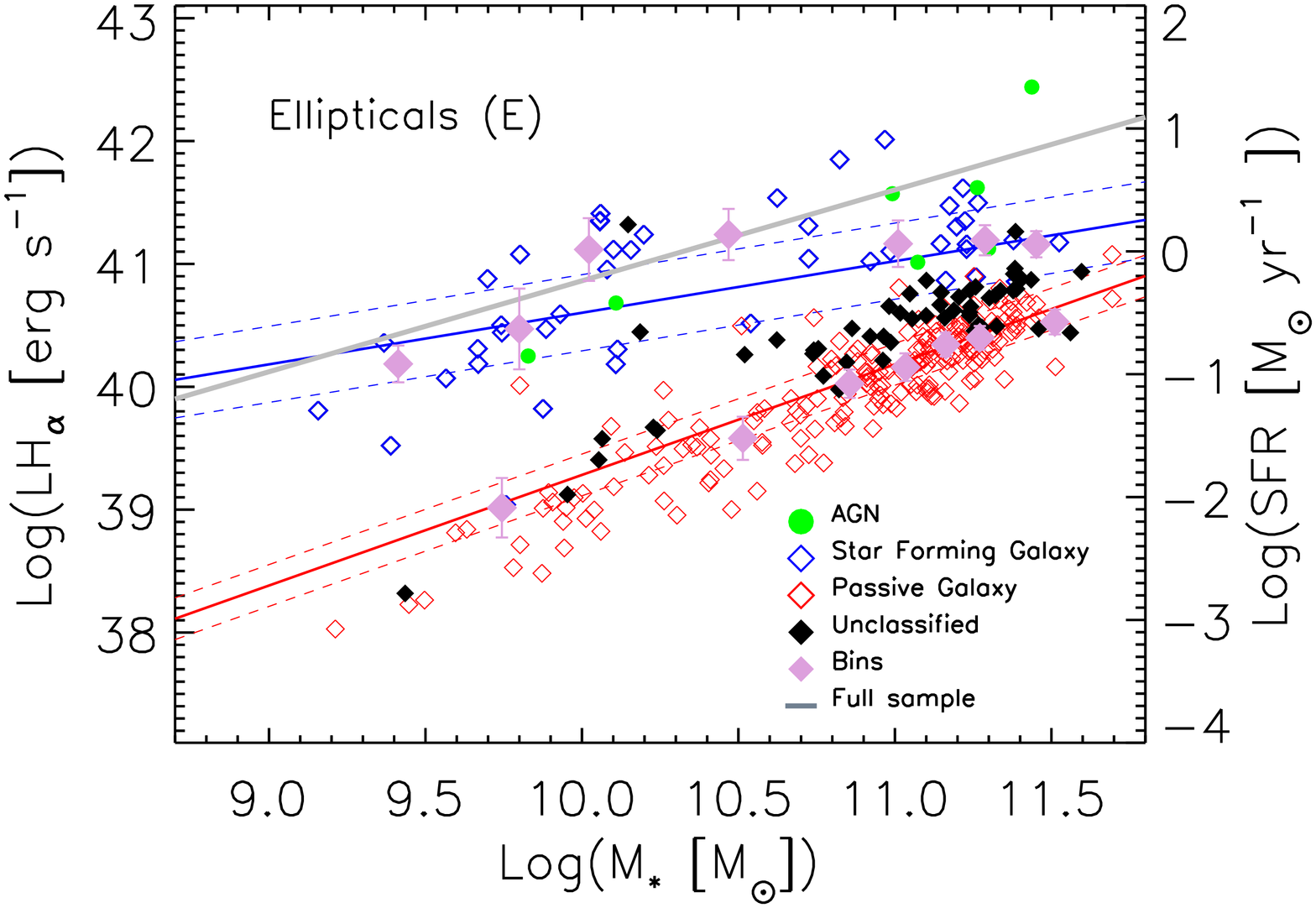}
    \label{fig:INT_Iregulars}
    \end{subfigure}    
       \end{minipage} \par\medskip
  \par\medskip
  \vfill 

\caption{Global SFR-$M_{*}$ diagram for sub-samples segregated by morphology: late types on the top and early types at the bottom. The nomenclature adopted in the different panels is the same as in Figure \ref{FullSampleIntegrated}.}
\label{IntSFMS_Morphology}
\end{figure*}

In Figure \ref{IntSFMS_Morphology}, we present the global SFR-M$_{*}$ diagram for different galaxy morphologies, using the same nomenclature adopted in Figure \ref{FullSampleIntegrated}. Each distribution of points for both the SFGs and RGs in the different morphological groups has been analyzed exactly the same way as it was described in Sec. \ref{ResultsFullIntegrated}, using the same binning scheme and applying a direct linear regression to the logarithmic means and standard deviations, with the only difference that the bins for these relations were derived using 15$\%$ of the total amount of data in each relation. The results from this analysis are listed in Table \ref{IntegratedFits}, including the same parameters described in the quoted Section. Again, alternative fitting results are provided in the online available Appendix B. 
 From Fig \ref{IntSFMS_Morphology}, it is clearly seen that the SFG and RGs present different fractions and fall in slightly different regions in the  SFR-M$_{*}$ diagram according to their morphological types.
 
In Table \ref{GlobalClassification1} we present the fractional distribution by morphology for each one of the groups of galaxies classified according to their global main ionization process, while  Table \ref{GlobalClassification2} presents the fractional distribution by the global main ionization process for each one of the groups classified by morphology (galaxies with undetermined sources of ionization were excluded for these statistics). 
It is important to remark that these fractional distributions refer to the MaNGA sample without applying any correction for selection effects and completeness. In particular, there is an excess of S0/S0a galaxies with respect to volume-complete determinations in the Local Universe (considering an adequate volume correction, this selection effect should be corrected).  Many of the S0/S0a galaxies are actually undetermined (or transitory between the star forming and retired regimes) according to their ionization source.

\begin{table}
\caption{Distributions by morphology in the ionization-based galaxy groups}
    \begin{tabular}{ c c c c c c}
    \hline
    \hline
\shortstack{Ionization \\ group} & \shortstack{\# of \\ galaxies} & E($\%$) & S0($\%$) & S0a-Sb($\%$) & Sbc-Irr($\%$) \\ 
\hline
SFG & 892  & 5.1 & 3.8 & 40.0 & 51.1 \\ 
RG & 616  & 31.8 & 47.1 & 20.9 & 0.2 \\ 
AGN & 30  & 23.3 & 16.7 & 60.0 & 0.0 \\ 
Undetermined & 216  & 25.9 & 34.3 & 38.9 & 0.9 \\ 
\hline  
\hline  
    \end{tabular}
	\label{GlobalClassification1}
\end{table}

\begin{table}
\caption{Distributions by the dominant ionization process in the morphology-based groups}
    \begin{tabular}{ c c c c c c}
    \hline
    \hline
\shortstack{Morphological \\ group} & \shortstack{\# of \\ galaxies} & SFG($\%$) & RG($\%$) & AGN($\%$) & \shortstack{\# of undet.\\ galaxies} \\ 
\hline

E & 248 & 18.2 & 79.0 & 2.8 & 56 \\ 
S0 & 329 & 10.3 & 88.2 & 1.5 & 74 \\ 
S0a-Sb & 504 & 70.8 & 25.6 & 3.6 & 84 \\ 
Sbc-Irr & 457 & 99.8 & 0.2 & 0.0 & 2 \\ 
\hline  
\hline  
    \end{tabular}
{\raggedright \textbf{Note.-} The percentages are with respect to the number of galaxies with a globally determined main ionization processes (second column).}
	\label{GlobalClassification2}
\end{table}

The SFMS discussed above was built with a total of {892} SFGs galaxies, represented  in Fig. \ref{FullSampleIntegrated}, \ref{LocaltoGlobalSFMSandRS} and \ref{IntSFMS_Morphology} with blue symbols. The SFGs comprise {$\sim$51\%} of the full sample, in agreement with the estimations by \citet{sanchez18b} in this regards. According to Table \ref{GlobalClassification1}, the great majority of SFGs are of late types ($\sim 91\%$ in S0a to Irr); only $\sim 9\%$ of the SFGs are of early types (E/S0).   
On the other hand, the RGs (616) comprise {$\sim$35\%} of the full sample, and are represented in Figs. \ref{FullSampleIntegrated}, \ref{LocaltoGlobalSFMSandRS} and \ref{IntSFMS_Morphology} by red symbols. Out of them {$\sim 79\%$} are of early types and $\sim 21\%$ are S0a-Sb galaxies, with only one galaxy of a type later than Sb. 
Finally, according to Table \ref{GlobalClassification1}, $\sim 2\%$ of the full sample are galaxies globally classified as AGNs, being most of them of S0--Sb types, and $\sim 12\%$ remain as undetermined, being they of all morphological types excepting the latest ones. 

So, while the morphological distribution of the SFGs is strongly biased towards late types, in such a way that the SFMS is populated mainly by late-type galaxies, in the case of the RGs, they present a broader morphological distribution, with a non-negligible fraction of them being early-type spirals. 
The fractions of SFGs as a function of morphology reported here are in agreement with previous works that showed that the SFMS is heavily populated by late-type spiral and irregular galaxies, which typically are gas rich, and by consequence, where the star formation is expected to occur. Also, the dependence with morphology of the RGs reported here is a known result from several previous studies \citep[e.g.][]{Weinmann06,James08,Blanton2009,Schawinski14,Casado15,rosa16a}, suggesting the connection between global quenching and morphology.

It is interesting also to know how are the fractions of galaxies for a given morphological group, according to their main ionization process.  According to Table \ref{GlobalClassification2}, $\sim 28\%$ of the early-type galaxies are actually in the regime of active star formation, while $\sim 26\%$ of late-type galaxies are retired. 
Previous studies noticed about the non-neglible fraction of early-type galaxies with star formation \citep[e.g.][and more reference therein]{Thomas+2010,Gomes16b,Lacerna+2016}. This fraction is composed, among other groups, by (i) classical early-type galaxies that have recently suffered a rejuvenation induced by the capture of gas, and (ii) compact blue galaxies that probably acquired the early-type morphology in recent dissipation gas-rich mergers \citep[see for a discussion, e.g.][]{Lacerna+2016}. On the other hand, the existence of late-type galaxies with low levels of star formation (RGs), can be associated to environmental processes (e.g., ram pressure, tidal stripping, harassment, etc.) that devoid the gas from the disks, or by a pure extinction of the star-formation activity due, for example, to the total consumption of the available gas and the lack of further gas accretion, the stabilization of the gaseous disk by the presence of a spheroid, etc. (see subsect. \ref{discussion-1} for a discussion, and references therein). On the other, since our late type subsample comprises S0a-Sa galaxies that have prominent bulges -likely dominated by retired areas-, they could be contributing partially to the fraction of late-type galaxies classified as RGs.

On top of the general segregation in the distribution within the SFR-M$_*$ diagram by morphology (Fig. \ref{IntSFMS_Morphology}), we also see that SFGs present different SFMSs according to their morphologies. The SFMS becomes shallower the earlier the galaxies are (see also Table \ref{IntegratedFits}), and the earlier-type the SFGs are, the lower becomes the rate at which they form stars at a fixed stellar mass. This effect with the morphology was already noticed in previous studies \citep[e.g.][]{catalan15,rosa16a}, and is a consequence of the change of the SFHs by galaxy type \citep[e.g.][]{rgb17}. At a first order this could be due to the lack of gas in earlier type galaxies \citep[e.g.][]{Calette+2017}, but it is not the only cause. Indeed, detailed  analysis of the gas content seems to indicate that the depletion time (and the star-formation efficiency) also change with morphology, connected with the dynamical properties of galaxies \citep[][]{colo18}; a further discussion on this topic will be presented in subsection \ref{Discussion-3}.

\subsubsection{Morphological distribution in the local $\Sigma_{\rm SFR}-\Sigma_*$ diagram}\label{morph_loc}

Knowing that the shape of the global SFMS segregates by morphology, and based on the results of subsection \ref{ResultsFullSR} about the similarity between the global and local SFMSs,
it is worth to understand if this segregation 
holds at local scales. This is relevant since the physical reason for the segregation 
would change based on that. If the local SFMS is the same notwithstanding the morphology of the host galaxies, then all SFAs in galaxies would form stars at a similar rate for a fixed $\Sigma_*$, and therefore the global segregation by  
morphology would indicate that the decline in the global SFR is just due to the presence of a lower fraction of SFAs. We know, indeed, that earlier type SFGs have a lower fraction of SFAs than later types. However, we do not know if those SFAs are forming stars at a lower rate, at a fixed $\Sigma_{*}$ too. 

In this regard \citet{Maragkoudakis17} have done a first study of the morphological dependence of the local SFMS relation using a sample of 31 SFGs with known morphology. Their sample was divided into four groups, close  to our classification. Their results show indeed a segregation with morphology. In particular, they found that the slope grows from the Sa types to the latest ones. However, they also report that the earliest morphological group exhibit the steepest slope (above unity). They attribute this high slope either to the fact that the star-forming regions within these early-type galaxies are as efficient as the ones in later types, or that their sample is too small to be representative of this type of galaxies (11 galaxies). \citet{Medling18} have also presented a study of the local SFMS segregated by morphology, using the SAMI survey data, with a larger sample of galaxies than the one used by \citet{Maragkoudakis17}. They found a systematic difference of the local SFMS relation when segregated by morphology, too. However, they reported that the late-type spirals are the ones describing the steepest slope (with a unitary value) for the local SFMS. They also found that galaxies of different morphological types show a segregation in the $\Sigma_{SFR}$-$\Sigma_{*}$ diagram, as the areas in this diagram coming from SFGs (mostly late-type) have an offset to those coming from clear RGs (mostly early-type).

\begin{figure*}
    \centering   
    \captionsetup[subfigure]{labelformat=empty}
    
  \begin{minipage}{1\linewidth}
  \centering
    \begin{subfigure}[b]{0.45\textwidth}
        \includegraphics[width=0.72\textwidth, angle=90]{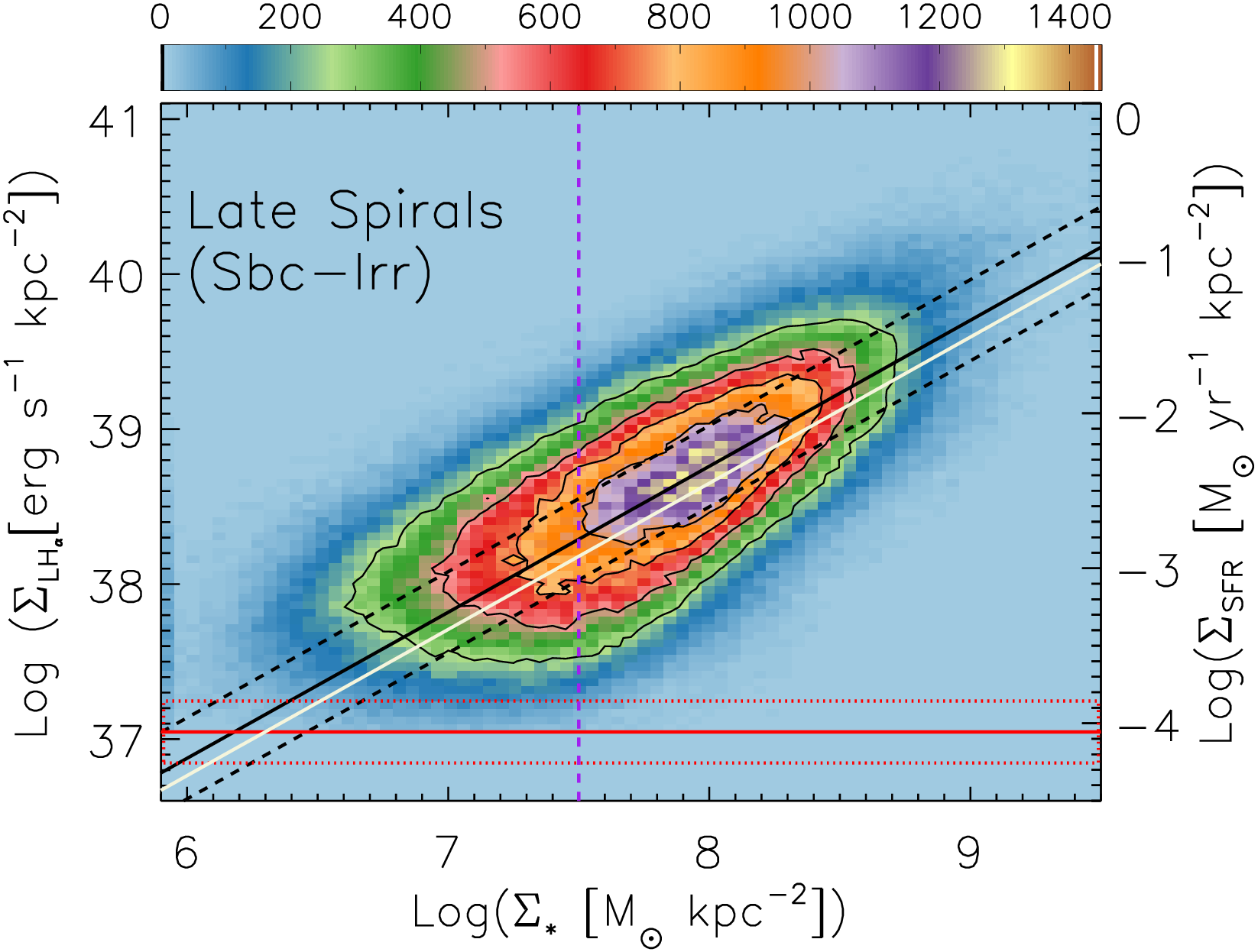}
       \label{fig:SFMS_Ellipticals}
    \end{subfigure}
        \begin{subfigure}[b]{0.45\textwidth}
        \includegraphics[width=0.72\textwidth, angle=90]{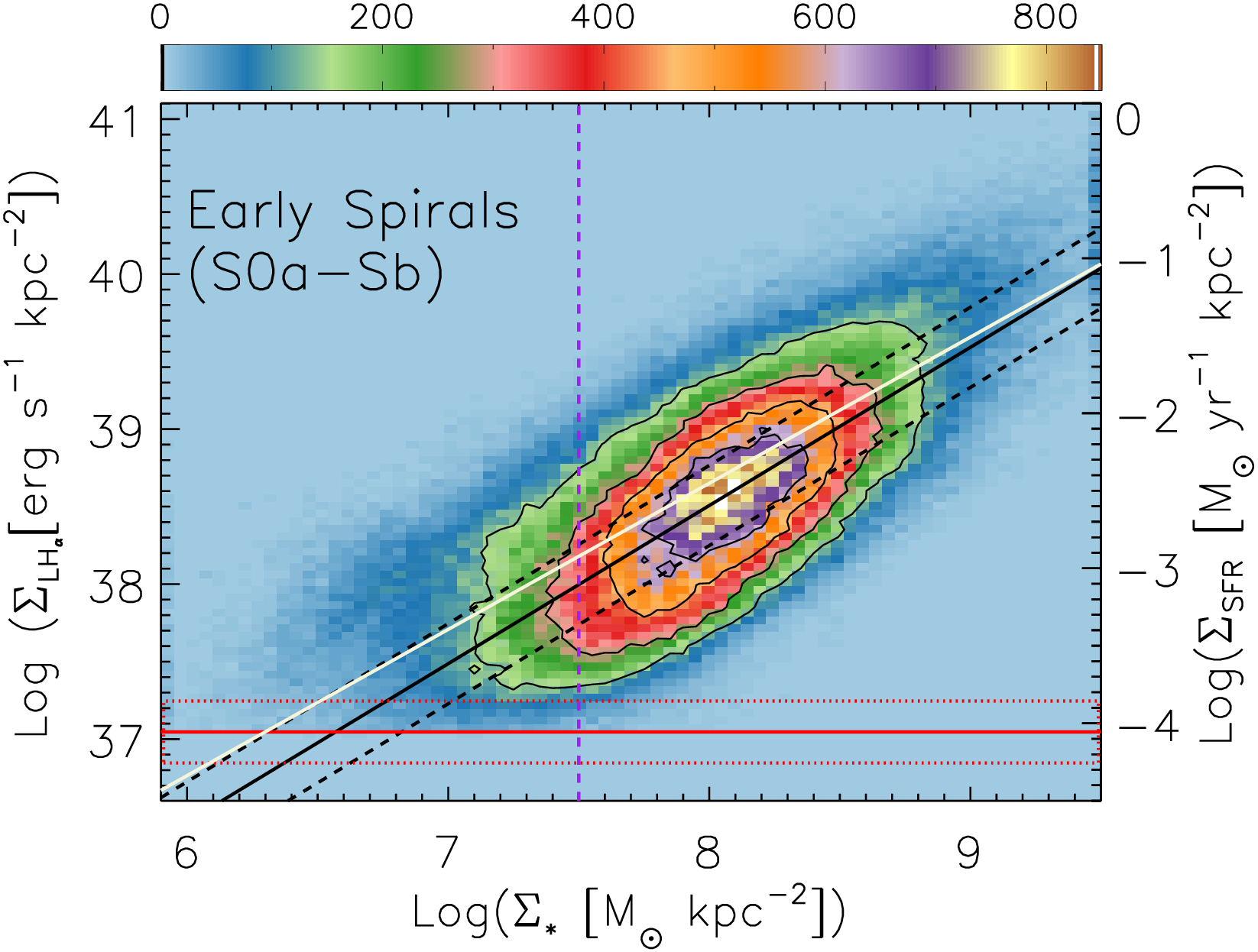}
        \label{fig:SFMS_Lenticulars}
    \end{subfigure}
 \end{minipage} \par\medskip
  \par\medskip
  \vfill            
 \begin{minipage}{1\linewidth}
 \centering
    \begin{subfigure}[b]{0.45\textwidth}
        \includegraphics[width=0.72\textwidth, angle=90]{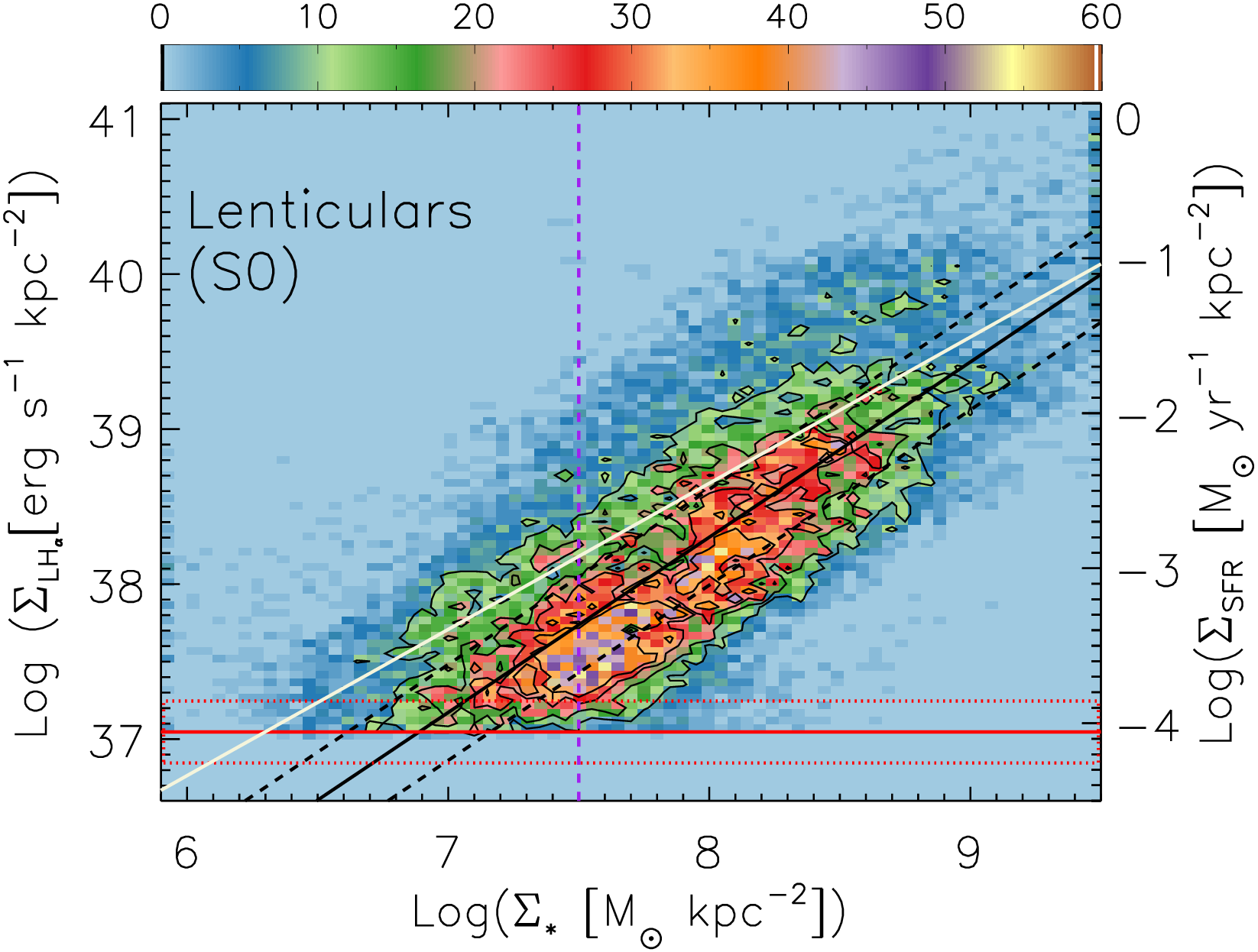}
        \label{fig:SFMS_Spirals}
    \end{subfigure}
    \begin{subfigure}[b]{0.45\textwidth}
        \includegraphics[width=0.72\textwidth, angle=90]{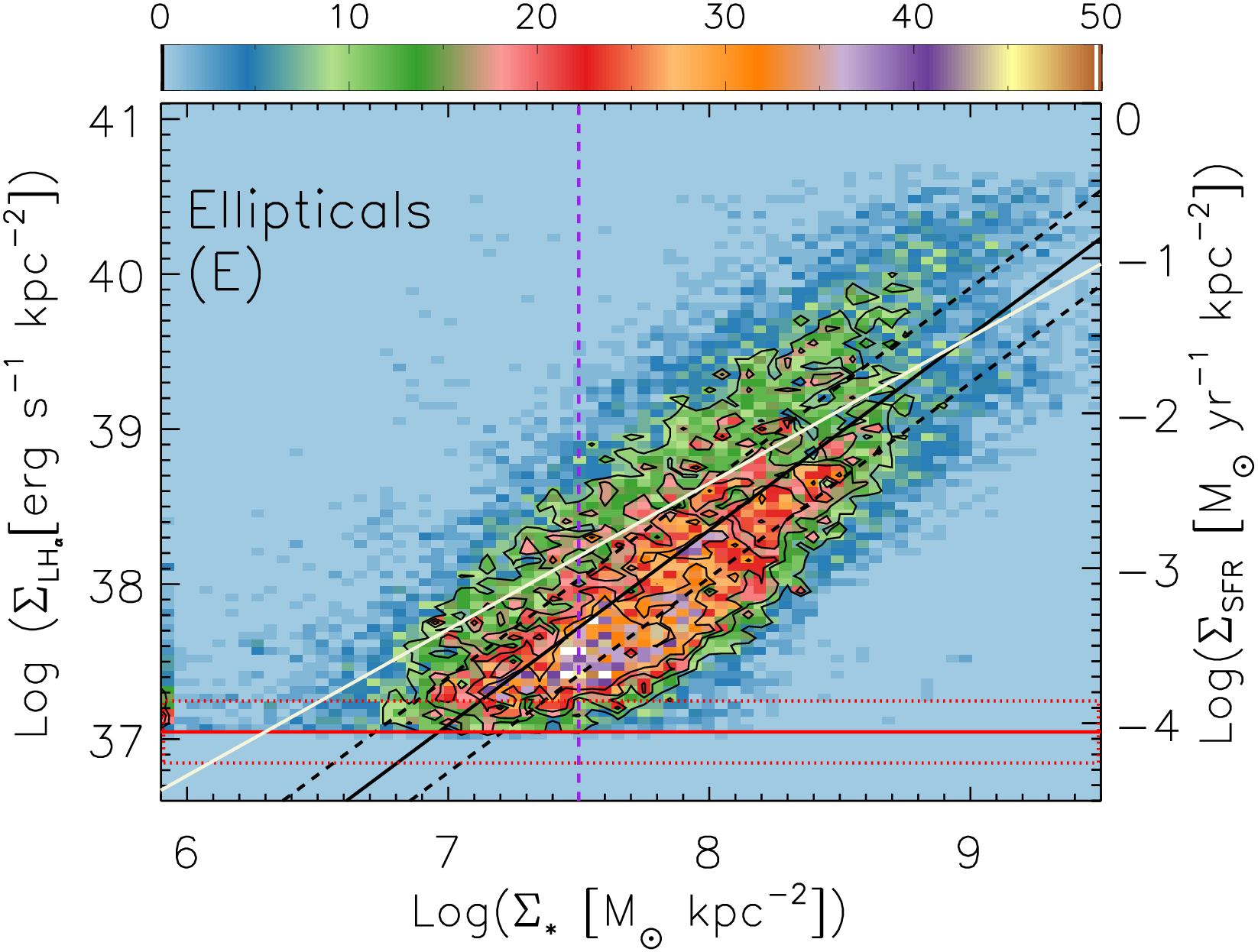}
        \label{fig:SFMS_Irregulars}
    \end{subfigure}
      \end{minipage} \par\medskip
  \par\medskip
  \vfill       
\caption{Local $\Sigma_{\rm SFR}-\Sigma_{*}$ diagram for SFAs of the different galaxies segregated by morphology: Late types at the top, and early types at the bottom. The contours and color-image represent the density of points, following the same nomenclature used in Fig. \ref{LocaltoGlobalSFMSandRS}. The best fitted linear regression is represented as a black solid line, and its dispersion with dashed black lines; the average detection limit is represented with horizontal red solid line, enclosed by the region that represents its 2-$\sigma$ scatter. As a comparison, we show the fit to the full sample of areas with the white solid line, which is the one plotted in Figure \ref{SpatResSFMSandRS} with black color. Finally the cut in $\Sigma_{*}$ used for the fits is represented with the vertical dashed line.}
\label{SpatResSFMS_Morphology}
\end{figure*}

Figure \ref{SpatResSFMS_Morphology} shows the MaNGA local $\Sigma_{\rm SFR}-\Sigma_{*}$ diagrams for SFAs segregated by the morphology of their host galaxies. Like in the case of Fig. \ref{LocaltoGlobalSFMSandRS}, we show a density plot due to the large number of SFAs in each panel. Following the procedure outlined in Sec. \ref{ResultsFullSR}, we analyze the correlation between the two involved parameters. The result of this analysis is listed in Table \ref{SResolvedFits}. For completeness, we  present in Appendix \ref{Sigma_RAs} (available online) the same analysis but for the RAs, whose distributions for different morphologies of the host galaxies are shown in Fig. \ref{SpatResRS_Mophology}. 

 As in the case of the global SFMS, the local one presents a segregation according to the morphology of the galaxy hosting the areas; this segregation is even stronger than in the global SFMS. As seen in Table \ref{SResolvedFits}, the $\Sigma_{SFR}$ values at the characteristic stellar mass surface density of $\Sigma_*\approx $10$^8$ M$_\odot/    $kpc$^{-2}$ decrease continuously from late to early morphological types of the host galaxies. 
Therefore, it is not only that the global star-formation activity (parametrized by the integrated SFR) is either quenched or it presents a steady decrease in earlier-type galaxies because they have a lower number of SFAs, but {\it for similar values of $\Sigma_*$, the SFAs in earlier-type galaxies form stars at lower rates, too.}

\begin{table*}
\caption{Distribution of the areas classified by its ionization process along the global ionization galaxy groups and morphological types }
    \begin{tabular}{ c c c c c c c c c c}
    \hline
    \hline
\shortstack{Ionization \\ group} & \shortstack{\# of \\ areas} & SFG($\%$) & RG($\%$) & AGN($\%$) & \shortstack{\# of areas in \\ undet. galaxies} & E($\%$) & S0($\%$) & S0a-Sb($\%$) & Sbc-Irr($\%$) \\
\hline

SFAs & 1,233,871& 98.7 & 0.5 & 0.8 & 31,305 & 1.9 & 2.1 & 33.9 & 62.1\\ 
RAs & 330,505 & 12.8 & 85.7 & 1.5 & 83,622 & 28.4 & 38.8 & 30.8 & 2.0 \\ 
AGNAs & 202,841 & 86.3 & 8.7 & 5.0 & 27,699 & 11.9 & 10.0 & 38.9 & 39.2 \\ 
\hline  
\hline  
    \end{tabular}
    
{\raggedright \textbf{Note.-} The percentages along the global ionization galaxy groups are excluding galaxies with undetermined main ionization processes.}
\label{LocalFractions}
\end{table*}

The global SFMS is populated by galaxies whose overall main source of ionization is classified as SF. 
It could be that these SFGs are not actively forming stars everywhere, consequently having both retired and star forming areas. In the same way, it could be that not all the SFAs that comprise the local SFMS are contained in SFGs. In Table \ref{LocalFractions} we present the fractional distributions of the star-forming, retired, and AGN-type areas along the different galaxy groups classified both by the global main ionization process and the morphology of the host galaxies. For the 1,265,176 areas classified as SFAs, $\sim 98\%$ of them are in SFGs (for this statistics, galaxies with undetermined global classification were excluded).
Therefore, the vast majority of the SFAs are actually contained in SFGs. Moreover, as discussed in the previous subsection, the global SFMS is populated mostly by late-type galaxies ($\sim$ 91\%). In principle, we should also expect to find a similar contribution from SFAs within these type of galaxies to the local SFMS. Indeed, we find that the vast majority ($\sim$96\%) of the individual regions comprising the local SFMS are from late-type galaxies (see Table \ref{LocalFractions}). 
In fact, finding that $\sim 96\%$ of the total SFAs come from late-type galaxies, explains that the fit to the local SFMS for late-type galaxies is very similar to the one derived for the full sample (see values in Table \ref{SResolvedFits}, and compare the white line with the black lines in the upper panels of Fig. \ref{SpatResSFMS_Morphology}).

In the case of RAs the situation is different: $\sim 13\%$ of them are in SFGs and 1.5\% in AGN galaxies (note that there is a non-negligible number of RAs that are in galaxies with undetermined global ionization process).
Unlike the local SFMS, which is strongly dominated by SFAs contained in SFGs with a very limited contribution from RGs, in the case of the cloud of RAs, a non-negligible fraction of these areas are not in RGs but in SFGs (or in galaxies with undefined classification). An interesting question is that of the late-type galaxies containig a non-negligible fraction of RAs. These galaxies might be in the process of their global star-formation shut down. We find that, while the majority of the RAs are in early-type galaxies ($\sim 67\%$, see Table \ref{LocalFractions}), {$\sim$33\%} of RAs are in {late-type galaxies. 
Thus, the vast majority of SFGs with some RAs correspond actually to late-type galaxies. On the other hand, recall that $\sim 21\%$ of the late-type galaxies are classified as RGs. Therefore, a significant fraction of the RAs are in late-type galaxies that can be either star-forming or already retired galaxies.  A more detailed exploration of the star-forming/late-type galaxies that could be initiating or finalizing their global quenching process will be presented elsewhere.

\begin{figure*}
  \begin{minipage}{1\linewidth}
    \begin{subfigure}[b]{0.5\textwidth}
        \includegraphics[width=0.8\textwidth, angle=90]{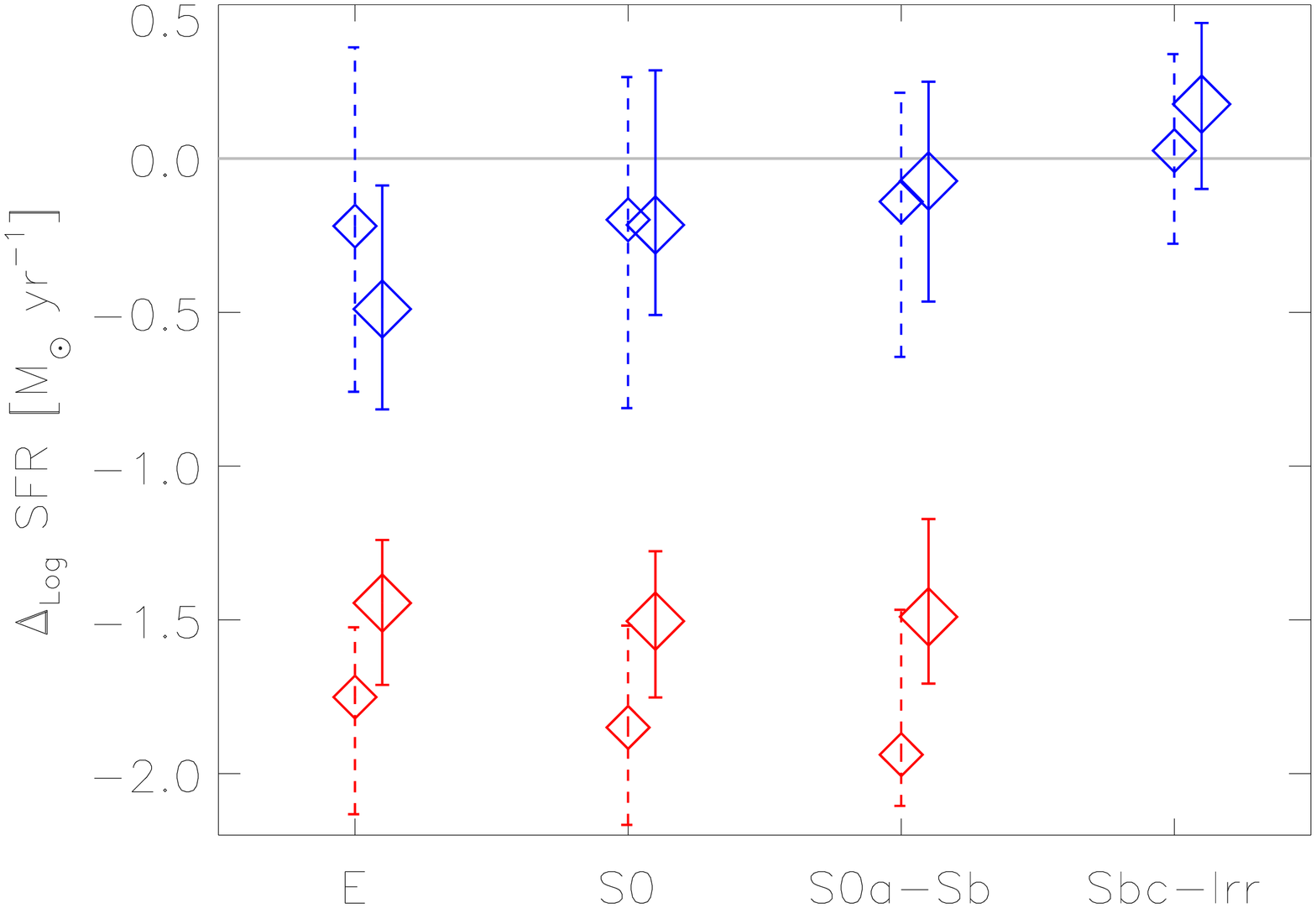}
        \caption{}
        \label{sfms_distances_INT_V1}
    \end{subfigure}
    \begin{subfigure}[b]{0.5\textwidth}
        \includegraphics[width=0.8\textwidth, angle=90]{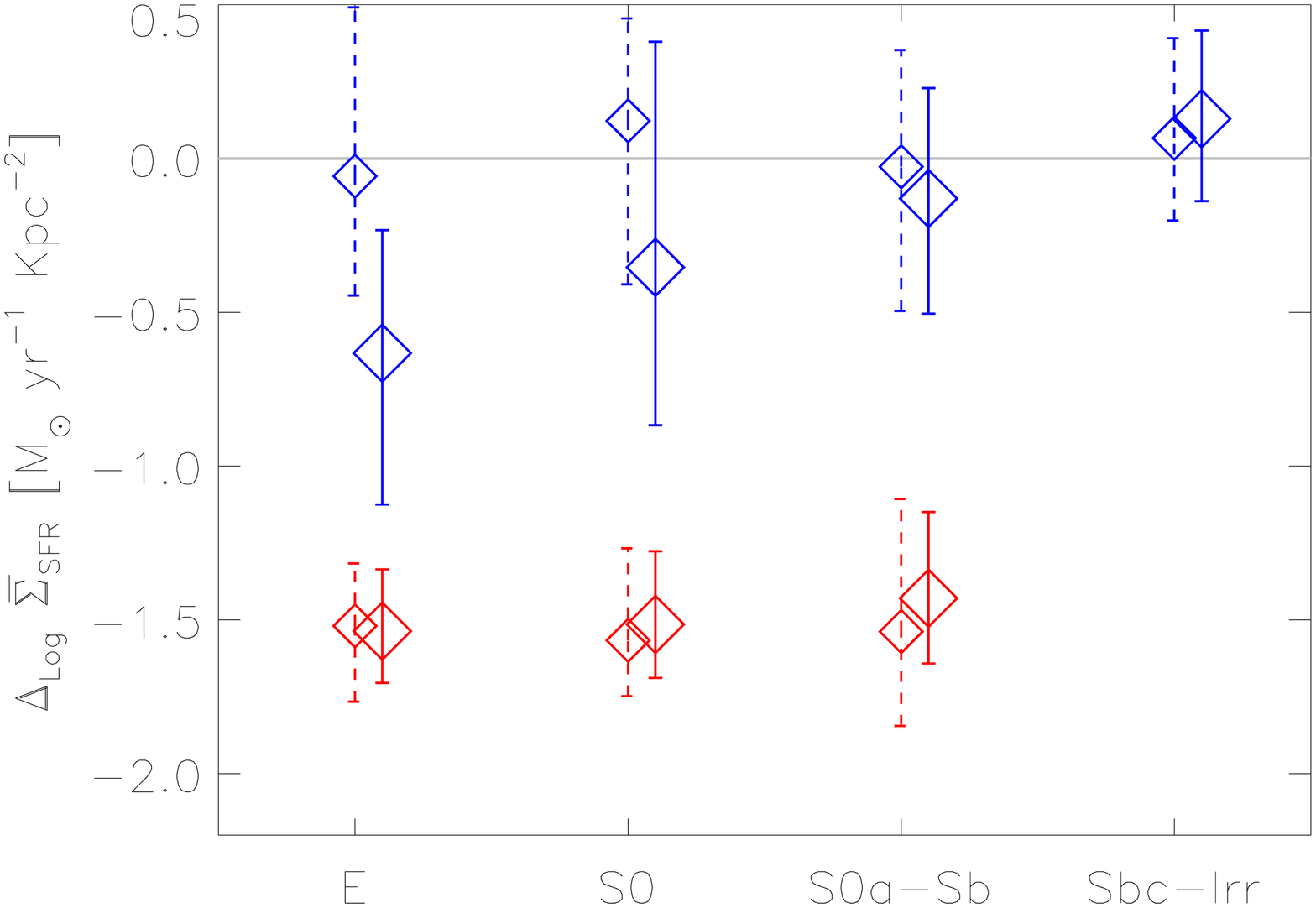}
        \caption{}
        \label{sfms_distances_globaltolocal_V1}
    \end{subfigure}
 \end{minipage} \par\medskip
  \par\medskip
  \vfill

 \begin{minipage}{1\linewidth}
    \begin{subfigure}[b]{0.48\textwidth}
        \includegraphics[width=0.8\textwidth, angle=90]{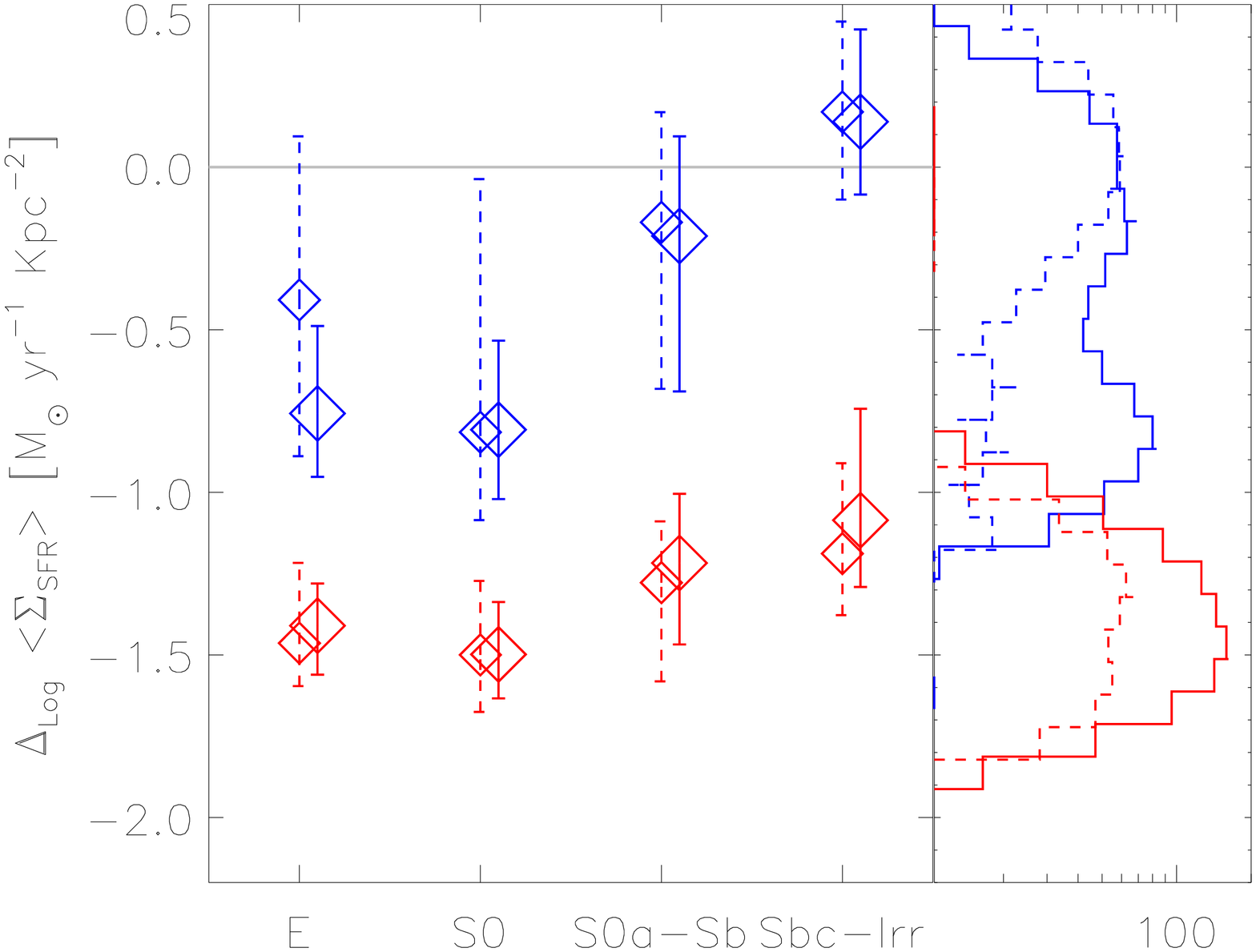}
        \caption{}\label{sfms_distances_V2}
    \end{subfigure}
    \begin{subfigure}[b]{0.48\textwidth}
        \includegraphics[width=0.8\textwidth, angle=90]{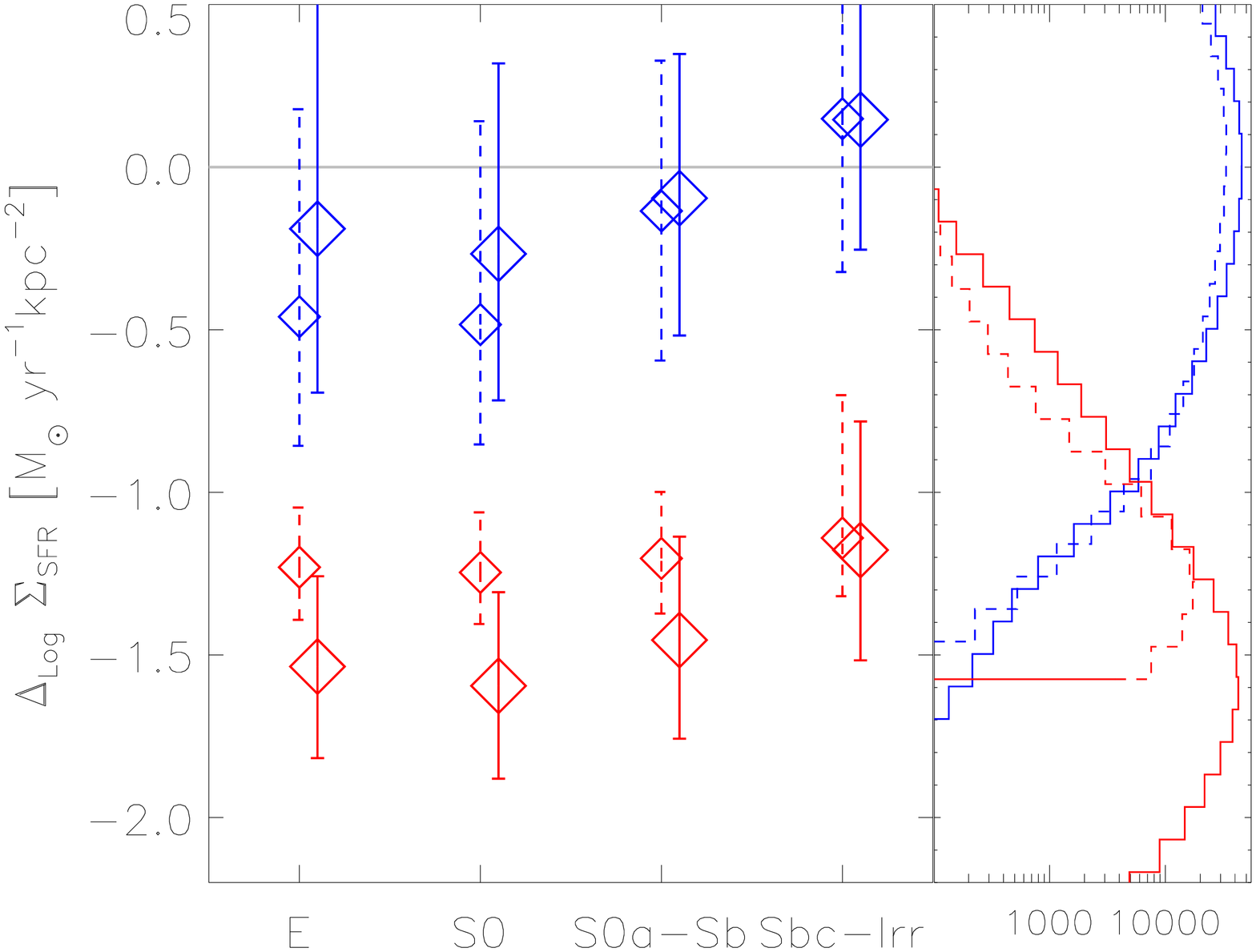}
        \caption{}\label{sfms_distances_points_V1}
    \end{subfigure}
      \end{minipage} \par\medskip
  \par\medskip
  \vfill      
\caption{Top plots: medians of the residuals from the global SFMS for (a) integrated extensive quantities, and for (b) average intensive quantities. Bottom plots: medians of the (c) averaged residuals from the local SFMS and (d) the medians of the individual residuals, also from the local SFMS. In all cases the blue and red colors represent the SFGs/SFAs and RGs/RAs respectively. In plots (a), (b) and (c) the smaller diamonds and dashed uncertainty bars represent the medians of the data below 10.5 M$_{\odot}$, while the bigger diamonds with continous uncertainty bars represent the medians of the data above this value. In the case of plot (d) the threshold value that separates both symbols is 8 M$_{\odot}$ kpc$^{-2}$. Finally, in plots (c) and (d), the distribution of the residuals per mass bin is shown without any distinction per galaxy morphology. The data below and above their corresponding threshold value is represented with dashed and continuous lines respectively.
}
\label{DistancesTofits}
\end{figure*}

\section{Residuals from the SFMS's: $\Delta$SFR and $\Delta\Sigma_{\rm SFR}$}\label{Disc_ResolRelations}

In the previous Section we have explored the distribution of galaxies and regions in the global (SFR-M$_*$) and local ($\Sigma_{\rm SFR}-\Sigma_*$) diagrams, respectively, both for the full sample and for subsamples segregated by morphology. We found that RGs are dominated by RAs, reflecting a global quenching in the star formation, clearly connected with early-type morphologies. On the other hand, SFGs are dominated by SFAs, connected with late-type morphologies. However, a non negligible fraction of RAs are found in SFGs, indicating that the star-formation activity is not homogeneously distributed in these galaxies, and that a fraction of their optical extension is already quenched. Finally, we have found evidence of an aging process in the SFGs (SFAs) as the morphology is earlier: for similar stellar masses (surface densities), the SFRs (surface density SFRs) are lower the earlier the galaxy type.

To quantify and visualize more clearly the different effects mentioned above we explore here the trends of the residuals from both the global and local mean SFMSs with the morphological type, and partially also with M$_*$ and $\Sigma_*$.  The residuals for a given galaxy are defined as:
$\Delta$SFR $\equiv$ SFR$-$SFMS$_{\rm global}$(M$_*$),
$\Delta\overline{\Sigma_{\rm SFR}} \equiv \overline{\Sigma}_{\rm SFR}-$SFMS$_{\rm global}$($\overline{\Sigma}_*$), and $\langle\Delta\Sigma_{\rm SFR}\rangle \equiv \langle\Sigma_{\rm SFR}-{\rm SFMS}_{\rm local}$($\Sigma_*$)$\rangle$; the averaging in the last case refers to the areas {\it within a given galaxy}, and it can be calculated for either the star forming or retired ones, i.e., for each galaxy we can have a value of $\langle\Delta\Sigma_{\rm SFR}\rangle$ for SFAs and another one for RAs.
The residual for each area is defined as:
$\Delta\Sigma_{\rm SFR} \equiv \Sigma_{\rm SFR}-{\rm SFMS}_{\rm local}$($\Sigma_*$). 
Note that the $\langle\Delta\Sigma_{\rm SFR}\rangle$ and $\Delta\Sigma_{\rm SFR}$ are with respect to the local SFMS; they were derived using only those points that are above the $\Sigma_{\rm *,lim}$ limit introduced for the SFAs relation ($>3\times 10^7$ M$_{\odot}$ kpc$^{-2}$). The same limit was adopted for the RAs for this analysis.

The panels of Figure \ref{DistancesTofits} show the residuals with respect to the following relations:

\begin{itemize}
\item (a) residuals from the global extensive SFMS, $\Delta$SFR. 
\item (b) residuals from the global intensive SFMS, $\Delta\overline{\Sigma_{\rm SFR}}$.
\item (c) the average of the residuals $\Delta\Sigma_{\rm SFR}$ from the local SFMS {\it within each galaxy}, $<\Delta\Sigma_{\rm SFR}>$. 
\item (d) residuals from the local SFMS from regions in all galaxies, $\Delta\Sigma_{\rm SFR}$.
\end{itemize}
All panels have a similar layout: the blue color represents the SFGs/SFAs, while the red color represent the RGs/RAs; in panels (a)-(c), the smaller and bigger diamonds refer to galaxies less and more massive than $\log$(M$_*$ / M$_{\odot}$)= 10.5,  while in panel (d), these symbols refer to areas below and above $\log$($\Sigma_*$/ M$_{\odot}$kpc$^{-2}$)=8. 
In panels (c) and (d), we show also the distribution of the residuals of SFAs and RAs in the form of histograms, independent of the morphology of the host galaxies, but separated into the groups above defined of massive/high-$\Sigma_*$ (continuous lines) and less massive/low-$\Sigma_*$ (dashed lines) galaxies/areas, respectively.  Note that in panel (c), most of galaxies appear two times, one for the average of the deviations $<\Delta\Sigma_{\rm SFR}>$ corresponding to SFAs, and other corresponding to the RAs. 

The residuals of the SFGs from the global extensive and intensive SFMS, $\Delta$SFR and $\Delta\overline{\Sigma_{\rm SFR}}$, respectively (blue symbols), follow some trend with morphology: the residuals are slightly above the mean relations for the Sbc-Irr galaxies, while for the E/S0 galaxies, their star formation activities lie on average clearly below the corresponding SFMS, with S0 and S0a-Sb galaxies falling in between. This is another way to present the results found in the previous Section, where we showed that the mean relations in the SFR-M$_*$ diagram of SFGs described by earlier-type galaxies are below those described by later-type galaxies (Fig. \ref{IntSFMS_Morphology}). For the global SFMS, the differences in the residuals with mass are small, especially for the dominant population of late-type galaxies; this just shows that the $\log$ SFR-$\log$ M$_*$ correlation of SFGs is well fitted by a linear relation. In the case of E/S0 types (a minority of the SFG population), the star formation activity is not only on average below the corresponding SFMS, but some segregation with mass is evident in the direction that more massive early-type galaxies deviate more from their SFMS than the less massive ones. 
The residuals of the RGs from the global SFMS (red symbols) are well separated from those of the SFGs, highlighting the well known bimodal distribution of SFGs and RGs. For the latter, the early-type (E/S0) galaxies are dominating and there are not Sbc-Irr galaxies. The residuals of RGs do not depend on the morphological type.

The residuals of the SFAs from the spatially-resolved SFMS, either averaged per galaxy (panel c) or the individual ones (panel d), correlate significantly with the morphology of their host galaxy, showing that the star-forming regime is segregated at the local level by the morphology of the host galaxy, with earlier-type galaxies having lower local SFRs. Note that in panel c, the mean residual, $<\Delta\Sigma_{\rm SFR}>$, refers to the average of all the residuals within a galaxy, and for each galaxy, there are typically two mean residuals, one for the SFAs (blue symbols) and other for the RAs (red symbols). The host galaxy mass actually does not seems to affect too much the deviation of the mean of the residuals of SFAs from the local SFMS; it is rather the morphology which affects the level of SFR of the SFAs. Only for the earliest types, there is some dependence on mass in the direction that SFAs in the more massive galaxies, deviate more from the local SFMS. This shows that the SFHs of local SFAs in massive early-type galaxies decay faster than similar areas but in less massive galaxies.  

In panel (d), the large/small symbols refer to the high/low stellar mass density of the areas, respectively. For the SFAs (blue symbols), their residuals do not seem to depend too much on $\Sigma_*$ for late-type galaxies, while for early-type galaxies, a weak difference with $\Sigma_*$ emerges in the direction that SFAs of lower $\Sigma_*$ deviate on average more from the local SFMS than those of higher  $\Sigma_*$. On the other hand the overall distribution of SFAs around the SFMS looks quite symmetric, both for $\Sigma_*$ values higher and lower than $10^8$ M$_{\odot}$ kpc$^{-2}$. A cut-off at the low end of the residuals is seen for the low-$\Sigma_*$ SFAs (dotted line); this is due to selection effects, we do not consider SFAs with $\Sigma_*$ values lower than $3\times 10^7$ M$_{\odot}$ kpc$^{-2}$. This effect is much more evident for the RAs of low $\Sigma_*$ values (red dotted lines). In general, for the RAs (red symbols), both the residuals averaged per galaxy and the individual ones also show some dependence on the morphology of the host galaxy: in late-type galaxies the RAs have on average higher levels of star formation than in early-type galaxies. This segregation highlights our previous claims, indicating that while the star-formation quenching has a global effect, it seems to be indeed a process driven locally, with a strong connection with the morphology of the galaxies. 

Finally, we should mention that \citet{Thorp19} have a similar study of the residuals using local intensive quantities, following the methodology previously introduced by \citet{Ellison18}, but for a sample of only post-merger MaNGA galaxies. In this study, they identify that the positive offsets of the residuals could be a result of the interaction. In our study we do not explore the possible effect of the environment in our relations, but is a variable that definitely should be introduced in future analysis.

\section{Discussion} \label{Disc}
		
The relationship between the SFR and the stellar mass (or between the SFR and stellar surface densities) actually connects the current value of the SFH (e.g., at the considered redshift, $z\sim 0$ here) and its integral along the past.
Therefore, the fact that galaxies (or their local areas) present a bimodal distribution in the SFR-M$_*$ (or $\Sigma_{\rm SFR}$-$\Sigma_*$) diagram indicates that they present at least two families of SFHs: one following a smooth power-law or exponential decay \citep[e.g.,][]{Speagle14}, forming stars in a steady way (SFGs), and those that depart from that relation towards a more abrupt decay, presenting then a quenching in the SFR (RGs).} This was demonstrated clearly by recent results that explore the cosmological evolution of the SFR of current star-forming and retired galaxies using fossil record methods \citep[e.g.][]{sanchez18b}. In a more general way, some previous works have explored the SFH of galaxies, for example, as a function of mass. In particular, since the pioneering studies by \citet{panter03,panter07} and \citet{heavens04}, who applied the fossil record method to a large survey of local SDSS galaxies, it is clearly established that more massive galaxies are more frequently quenched and present a sharper SFH than less massive ones, a trend known as downsizing \citep[e.g.,][]{Thomas+2005,Cimatti+2006,Fontanot+2009}. More recent results have shown a clear connection with morphology, too \citep[e.g.][]{rgb17,lopfer18}, showing that the quenching, although it becomes a global process, it has a local origin, and in general it gains strength from the center to the outer regions \citep{rosa16a,Belfiore17a,lopfer18,sanchez18a}. Previous results have already shown the presence of retired/quenched areas in galaxies that are globally star-forming \citep[e.g.,][]{sign13}, and the presence of star-forming areas in globally retired ones \citep{Gomes16b}. On top of that, we know that SFHs in SFGs are not all the same, presenting a decline toward lower values that depends on both the mass and the morphology \citep[e.g.][and references therein]{rgb17}. Therefore, it is not surprising that even for SFGs there is a segregation of their location in the SFR-M$_*$ diagram with the morphology, too \citep[see also][]{catalan15,rosa16a}.

\begin{figure}
        \includegraphics[width=0.85\columnwidth, angle=90,keepaspectratio]{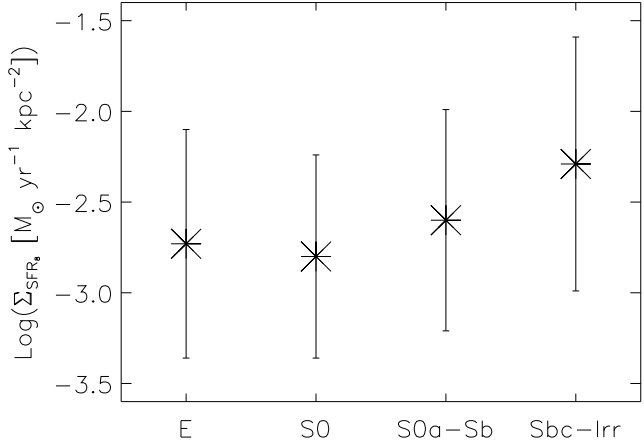}
\caption{Segregation of the local SFMS by morphology. The segregation is measured for $\Sigma_{SFR}$ at the characteristic stellar mass surface density of 10$^{8}$M$_\odot$ kpc$^{-2}$.}
\label{morph_shift}
\end{figure}


Our results show a connection between the global SFMS (and the location of a galaxy in the SFR-M$_*$ diagram) with the local SFMS (and the location of the different areas of these galaxies in the ${\Sigma_{\rm SFR}-\Sigma_*}$ diagram). SFGs make the stronger contribution to the SFAs, the population that defines the SFMS. Therefore, in these galaxies the overall SFH has to be similar to all the different regions that comprise it. However, as indicated before, as the morphology is of earlier type, the fraction of RAs in SFGs increases, and therefore, there is a decline of the global SFR as earlier is the type, with a very a weak dependence on mass (see Fig. \ref{DistancesTofits}). However, this is not the only reason for the decline in the star-formation activity. As shown in Secs. \ref{morph_loc} and \ref{Disc_ResolRelations}, the local star-formation activity has also a decline as earlier is the host galaxy for similar values of $\Sigma_*$ (panels c and d of Fig. \ref{DistancesTofits}). To reinforce this claim, we present in Figure \ref{morph_shift} the mean $\Sigma_{\rm SFR}$ and standard deviations at a fixed characteristic mass surface density of $\Sigma_*$=10$^{8}$M$_\odot$ kpc$^{-2}$ for the different morphological types of the host galaxies (note that the characteristic $\Sigma_*$ is that one used for reporting the normalization of the correlations in Table \ref{SResolvedFits}. For a fixed  $\Sigma_*$, we see a decline of the star-formation in the SFAs for earlier morphological types of the host galaxy. In other words, the SFAs of earlier-type galaxies form stars on average at a lower rate for their stellar-mass surface densities than those of later types, and consequently, overall, earlier SFGs have lower SFRs than the later ones.

On the other hand, RGs are dominated by RAs, with a very small contribution to the population of SFAs. Therefore, once a galaxy is in the cloud of RGs it seems to cease its star-formation activity globally. Again, this is connected with the morphology, with most of RGs being of earlier types and the dominant population of RAs being hosted by this type of galaxies too.

In summary, our results show that the morphology is partially connected to the 
decline in the SFR of galaxies, and therefore to the definition of the shape of the SFH, both globally and locally. Such a decline can be either rapid and abrupt (quenching) or rather slow and associated to the intrinsic SFH of star-forming galaxies (aging). Next, we will discuss both cases as well as some mechanisms that might explain the SFH-morphology connection.

\subsection{The morphology-quenching connection}
\label{discussion-1}

It is broadly accepted that the massive morphological transformation of galaxies (from mostly pure disks to a fraction of them with bulges or completely dominated by the spheroidal component) happened since $z\sim 2$ \citep[e.g.,][]{wolf05,Bruce+2012,Buitrago+2013,Mortlock+2013,Avila-Reese+2014,Margalef-Bentabol+2016}. This epoch roughly coincides with the peak of the global star formation history, suggesting some connection between the quenching of star formation and morphology. \citet{martig09} proposed that the growth of a stellar spheroid can stabilize the gas disk, and quench SF by preventing the fragmentation of bound gas clumps (morphological quenching). More recently, \citet[][see also \citealp{Tacchella+2016} and \citealp{Avila-Reese+2018}]{Dekel+2014} have proposed a quenching process, based on the dissipative shrinkage of the high-redshift gaseous disk (compaction), the triggering of intense SF in the dense centre with the consequent rapid gas consumption and outflows that produce an inside-out quenching. The remaining dense inner stellar structure can have the properties of a bulge.    

Interesting enough, $z\sim 2$ is also the epoch of the peak of the luminosity function of quasars, suggesting that the AGN feedback should play a relevant role in the quenching process of galaxies \citep[][]{lipari94,sanders96,hopkins09,Hopkins+2010}, process that is also expected to happen the inside out. The AGN/quenching connection is nowadays under debate \citep[e.g.][]{sanchez18b}. 
For massive galaxies, a natural complementary mechanism to avoid cold gas infall onto the galaxies is the gas virial shock heating, which is effective when the halo becomes more massive than $\sim 10^{12}$ M$_\odot$ \citep[halo quenching; e.g.,][]{White+1991, Birnboim+2003, Keres+2005,Dekel+2006}, leading this to a dominion of retired galaxies at high masses. Other quenching mechanisms are related to the environment \citep[e.g.,][]{Peng+2010}, and they are less dependent on mass.\footnote{Note, however, that the environment probably plays a more general role in the overall evolution of galaxies, including their morphological transformation.} Actually, all the previously mentioned mechanisms have different time scales and depend on the mass regime, the environment, and the cosmological epoch; according to the situation, some of them are expected to be more dominant than the other ones. 

The dependence with morphology of the global and local star-formation properties of galaxies found here (Figs. \ref{IntSFMS_Morphology}, \ref{SpatResSFMS_Morphology}, and \ref{DistancesTofits}), and the gaps in the residual distributions observed in the latter figure (they suggest a fast transition from star-forming/late-type galaxies to the retired/early-type ones), support partially the compaction and morphological quenching scenarios. In the former scenario, galaxies should pass violent star-formation events associated with the compaction of their gas disks at early times, in such a way that most of the gas reservoir is consumed, in particular in the central regions.  
However, other mechanisms are required for avoiding further gas accretion or star formation (for example the halo, environmental, and morphological quenching mechanisms). 
It is interesting to note that the surface mass densities of the RAs in the latest-type galaxies are higher on average than those of the SFAs in galaxies also of the latest types. Indeed, the small fraction of late-type (Sbc-Irr) retired galaxies contain areas with the largest values of $\Sigma_*$. This is because a late-type galaxy that quenched, transformed its gas into stars, increasing then $\Sigma_*$. However, the fact that these surface densities are even higher on average than those in the early-type retired galaxies (this notably applies for the Sbc-Irr retired galaxies), suggests a process that strongly increases the local surface densities of the galaxy along with the quenching in SF.

Our results regarding the morphological quenching agree with those of \citet{colo18}. In particular, it is notable that at a given $\Sigma_*$, the areas contained in earlier-type galaxies present lower SFR than in the late-type ones (panels c and d of Fig. \ref{DistancesTofits} and Fig. \ref{morph_shift}). However, as \citet{colo18} show, in all the cases it is observed also a decrease of the gas content in those areas with low SF rates, and that seems to be the primary reason of why there is no SF. Therefore, gas is expelled or consumed somehow, although the efficiency may be affected by the extra disc stabilization introduced by the stellar spheroid.

In \citet{sanchez18a} it is shown that indeed the lack of gas is the dominant reason why RGs do not form stars, in agreement with quenching scenarios where gas is efficiently consumed or expelled. At a given stellar mass, these galaxies have a much lower fraction of molecular gas than the star-forming ones. But the results were inconclusive regarding the reason of this lack of gas. Although the effects of an AGN seems to be appealing, with a large fraction of them being found in the intermediate region between SFGs and RGs (the so-called green valley), the different time scales between the quenching ($\sim$ Gyr), and the AGN duty-cycle ($\sim$100 Myr) does not match. Either AGNs are recurrent events, or all we can claim is that there is a co-evolution between AGNs and quenching. However, it could be also that the AGN quenching mechanism combines with other mechanisms for maintaining the galaxy quenched, for example, the halo (or mass) quenching mechanism mentioned above, which works for halos that overcome $\sim 10^{12}$ M$_\odot$ after $z\sim 2$.

Related to the environment, possible quenching mechanisms happen when a galaxy becomes satellite in a structure larger than this galaxy: the accretion of gas it is ceased and the galaxy enters into a regime of starvation or strangulation \citep[e.g.,][and references therein]{Larson+1980,Balogh+2016,Peng15}; as the satellite moves to inner orbits, its cold gas can be stripped by ram pressure and tidal forces producing this a quenching of its SF \citep[e.g.,][]{Gunn+1972,Aragon-Salamanca+1993,Moore+1998,Bekki+2002, Bellhouse+2019}.  
The time scale of quenching since the galaxies become satellites has been found to be relatively large, from $\sim 2$ to $4$ Gyr \citep[see e.g.,][and more references therein]{Wetzel+2013}. In our study, we did not explore the environment of galaxies, but according to the morphology-environment correlation \citep{Dressler1980,Whitmore+1993,Dressler+1997,Cappellari+2011}, early-type galaxies are more probable to be in denser environments being satellites. Therefore, the environmental quenching could also explain the dependence of the global and local SFRs with morphology found here, in particular for low masses, where the halo and AGN quenching mechanisms are not feasible.  
It should be also mentioned that the morphology-environment relation seems to change with redshift \citep{Butcher+1978,Dressler+1994} in such a way that the morphology of galaxies might has been modulated over cosmic time mainly by the environment, and once the morphology is established, then the morphological quenching could affect the SFH as discussed above.

\subsection{The morphology-aging connection}
\label{discussion-2}

As indicated before, the morphology is not only connected with the quenching, but it does affect also the aging. Earlier type SFGs and the SFAs hosted by early-type galaxies are less actively forming stars for a given stellar-mass surface density than later-types or areas hosted by them. This is a very interesting result since it indicates that the dynamical stage, characterized by the morphology, of a galaxy has an influence not only on the global but also on the local SFHs. Indeed, this was somehow already shown by \citet{sanchez18a}, where it was found that the star-formation efficiency of SFGs is slightly larger than that of RGs. That is, for the same amount of molecular gas, the latter form stars at a slightly lower rate. Since most of RGs are early-type galaxies too, this means that these galaxies form stars at lower rates, as we have shown here. Therefore, it is not only the lack of gas but there is also a real decrease in the efficiency of gas transformation into stars.

In this regards, the proposal by \citet{martig09} is relevant. It may well be that the dynamical stage of the galaxy prevents the gas to collapse and form new stars in early-type galaxies, dominated by disordered orbits, with the gas not nicely settled in a thin disk. This is an appealing scenario, that should be confirmed by a detailed orbit analysis. However, recent results by  \citet{zhu18} have shown that indeed the fraction of stars in cold/warm/hot orbits is clearly related to stellar mass and morphology in galaxies. We will explore that scenario in future studies.

\subsection{The global and local SFMS}
\label{Discussion-3}

The connection between the extensive global SFMS and the local one was speculated by \citet{CanoDiaz16}. A pure numerical argument indicates that the local relation may be the driver for the global one (being the later just the integral of the former). Other authors have attempted to connect both relations \citep[e.g.][]{Hsieh17}.

However, we consider that our direct comparison between the intensive global SFMS and the local one clearly demonstrates that they are indeed nearly the same relation, likely modulated by the morphological type of the galaxies or other global properties that correlate with morphology (e.g., gas fraction, stellar mass, gas-phase metallicity, environment). The prevalence of the global relation as a consequence of the local one modulated by some global property is simple to understand: if all SFAs in a galaxy follow a tight $\Sigma_{\rm SFR}-\Sigma_*$ relation, whose normalization depends on a global property, then the average of them must lie in a global $<{\Sigma_{\rm SFR}}>-<{\Sigma_*}>$ correlation with a close shape to the local one. We have shown in  Fig. \ref{DistancesTofits} that the scatter (residuals) around the local $\Sigma_{\rm SFR}-\Sigma_*$ relation of all the SFAs (the local SFMS) segregates by morphology, implying that the scatter around the global relation segregates also by morphology. 
The fact that SFAs from very different galaxies present the same correlation between $\Sigma_{\rm SFR}$ and $\Sigma_*$, though with a segregation by morphology, suggests that the SF process along these galaxies is driven by an universal mechanism that is modulated by the global morphology of the host galaxy, either directly or indirectly, i.e., by some other property that correlates with morphology. 

Being established that the local relation determines the global one, it is important to understand why the local SFAs deviate from the average in galaxies with different morphologies. This decline could be partially related to a decrease of the gas content, as seems to be the case for the quenching mechanisms discussed in subsect. \ref{discussion-1}. However, there are some pieces of evidence that for similar gas fractions, the star-formation efficiency, SFE, of local regions also changes with the morphology of their host galaxies. For example, \citet{colo18} found  that there is a dependence of the local (2-3 kpc scales) depletion time (SFE$^{-1}$), and the morphology and dynamical stage of the host galaxies. Thus, a real change in the efficiency seems to be partially behind the observed SFR decline as earlier are the host galaxies.
A possible explanation for this would be that in the regions of earlier type galaxies, a fraction of the gas clouds are more stable, unable to collapse to form stars \citep[as suggested by][]{martig09}. Another possibility is that SF happens only in the disc of galaxies, and as a larger fraction of gas is orbiting out of the disc, the SFR declines for a fixed stellar mass surface density. In both cases, other fundamental relations regulating the SF activity, like the Schmidt-Kennicutt law, should present a segregation by morphology. 

Some authors have claimed that the gas-phase metallicity, $Z_g$, is a third parameter in the global/local SFMS of galaxies \citep{LaraLopez+2010,Salim+2014}, as a consequence of a possible M-Z-SFR relation \citep[e.g.,][]{ Mannucci+2010}. On the contrary, other authors have not found enough evidence of this in their analyzed galaxy samples \citep{Sanchez13,Izotov+2014,Kashino+2016,bb17,Sanchez+19a}.
Recent explorations indicate that $\Sigma_{\rm SFR}$ depends strongly on $\Sigma_*$ and $\Sigma_{gas}$, with metallicity being only as relatively important as stellar ages or integrated stellar masses \citep{Biprateep+2019}.
It is beyond of the scope of this paper to study this question. However, if metallicity is a third parameter, and since it correlates with morphology \citep[e.g.,][]{GonzalezDelgado+2014}, then our finding that the SFMS segregates by morphology could be partially induced by this primary dependence. At this point, it is important to note that the \citet{Kennicutt98} relation used here to calculate the SFR from the H$\alpha$ luminosity (Sect. \ref{SFR}) is strictly valid only for the solar metallicity. In fact, accurate photoionization models show that the lower metallicity, the higher is the H$\alpha$ luminosity for a given SFR (e.g., Boquien in prep.). This implies that the SFR inferred from a given  observed H$\alpha$ luminosity would be higher as the metallicity is lower, though significant differences are expected only for extremely low or high metallicities.
This is valid not only for H$\alpha$, but for any other indicator of the SFR that relies on the measured UV-photons. Therefore, part of the segregation by morphology in the SFMS of galaxies or their regions could be removed when adopting a metallicity-dependent H$\alpha$-luminosity tracer of the SFR. The latter could also partially give rise to the apparent segregation of the mass-metallicity relation by SFR. We will explore that possibility in forthcoming studies.

In summary, our results suggest that the local star-forming regions in galaxies follow an universal process of SF (e.g., the self-regulated, quiescent disc regime) modulated by the morphology of the host galaxy or by more fundamental properties that correlate with morphology \citep[e.g.,][]{Biprateep+2019}.
We discussed that these properties, besides metallicity, which is under debate, could be the gas fraction and/or the galaxy depletion time. This stresses out the need to study large samples of galaxies with spatially resolved spectroscopic information together with spatially resolved information of the different phases of the cold gas at the same scales. Only with all of this information we will be able to constrain the star-formation laws spatially resolved in galaxies. In this regards, efforts like the EDGE-CALIFA survey \citep{bolatto17} are extremely important pioneering studies. We also discussed that the environment might be behind the morphological transformation, which along with the environmental and morphological quenching processes, could ultimately be the key driver of the SFH of galaxies. Elsewhere, we will perform the study presented in this paper but for different environments.

\section {Conclusions} \label{Conc}

In this paper, we have studied the global and local (spatially-resolved) SFRs of 1754 MaNGA galaxies from MPL-5, selected to be less inclined than 60 degrees, and containing in total 2,126,380 resolved areas. We first classified the galaxies and each spatially-resolved area of them into two groups according to the dominant ionizing process: star-forming galaxies or areas, SFGs/SFAs, and retired galaxies or areas, RGs/RAs. According to our classification method, the ionization source could not be defined for 12.31\% of the galaxies and 10.18\% of the local areas; we excluded then these undetermined galaxies/areas from our analysis.  For the SFGs/RGs, we established their respective correlations in the global extensive SFR-M$_*$ and intensive $\bar{\Sigma}_{\rm SFR}-\bar{\Sigma}_{*}$ diagrams, and for the SFAs/RAs, their correlations in the local ${\Sigma}_{\rm SFR}-{\Sigma}_{*}$ diagram. Our main focus was on the correlations for the SFGs/SFAs. Not surprisingly, these correlations are tight and constitute a physically-based (global or local) Star Forming Main Sequence, SFMS. Finally, using the morphological characterization that we have for all the studied galaxies, we explored how the different correlations segregate by morphology. Our main conclusions are as follow:

\begin{itemize}
\item Both star-forming and retired galaxies, and star-forming and retired areas present a strong segregation (bimodality) in the above mentioned SF diagrams, with a very small fraction of galaxies/areas in between, where actually lie most of our undetermined cases. The slope of the extensive SFMS (SFR vs. M$_*$) is $\approx 0.75$, while the slopes of the intensive global ($\bar{\Sigma}_{\rm SFR}-\bar{\Sigma}_{*}$) and local (${\Sigma}_{\rm SFR}$-${\Sigma}_{*}$) SFMS are $\approx 0.65$ and $0.94$, respectively. In fact, the local SFMS bends significantly at the low-$\Sigma_*$ side; this bending is mainly due the H$_\alpha$ flux detection limit combined with the survey observational constraints. We find that the local SFMS is affected by these effects at values below $\Sigma_{*}\approx 3\times 10^7$ M$_\odot$ kpc$^{-2}$, so that the log-linear fit is performed to the data above this value. In general, we find that the global and local SFMS's  occupy a similar region in the intensive star-formation diagram, so that the global SFMS is tightly related to the local one. This leads us to the conclusion that the SFMS of galaxies is mainly established at small scales ($\ga 1$ kpc$^2$ for the MaNGA resolution).

\item The vast majority ($\sim 96\%$) of the SFAs are contained in galaxies classified as SFGs; most of SFGs are of late types ($\sim 91\%$, from S0a to Irr). The global SFMS is then mostly populated by late-type galaxies, though it contains also a small fraction of early-type (E/S0) galaxies, which were classified as SFGs ($\sim 9\%$). As the RAs, a non-negligible fraction of them are not in RGs but in SFGs, which are typically of late types.  On the other hand, we find that a non-negligible fraction of late-type galaxies ($\sim 21\%$) are classified as RGs. Therefore, a significant fraction of the RAs are in late-type galaxies that can be either yet star-forming or already retired galaxies. 

\item The SFGs present some segregation by morphology from the mean SMFS: the earlier the morphological type, the lower is on average the SFR, even for galaxies of similar masses. The RGs strongly deviate from the SFMS, on average by $-1.5$ dex, without a clear dependence on morphology.

\item The SFAs also segregate in the local SFMS by the morphology of their host galaxies; this segregation is even stronger than in the case of the global SFMS. Therefore, the morphology does not only affect the global star-formation process, due to the suppression in the number of SFAs as earlier is the morphology, but also the local one, forming stars at lower rates at the $\sim$kpc-scales. 

\end{itemize}

Our results show that the local (spatially-resolved) SFMS has a slope close to one, and a scatter that clearly segregates by the morphology of the galaxies hosting the SFAs. This suggests that local star formation (at scales $\ga 1$ kpc$^2$) is established by some universal process that is modulated partially by the global morphology of the galaxy (or other global properties that correlate with morphology like gas fraction, SFE, environment, etc.).
The decline in the local ${\Sigma}_{\rm SFR}$ as earlier-type are the galaxies seem to be primarily induced by a decrease in the amount of cold gas, but also due to a decline in the SFE (from what it is reported in the literature). The reason for that could be a combination of gas expelling, consumption and heating, and some dynamical processes that stabilize the disc against molecular clouds formation. To explore these possibilities in detail one needs to combine the current IFU dataset with equally large datasets of spatially resolved derivations of the molecular and atomic gas content.
A direct exploration of the effects of environment on the global and local SFR of our galaxies is also important, and it will be the topic of a future study.

\section*{Acknowledgements}

The authors acknowledge the thoughtful and detailed Reviewer report, which helped to improve the presentation of this paper. 
We also wish to thank the comments given  by Jorge Barrera-Ballesteros, Karen Masters, and Rog\'eiro Riffel about this work.

We are grateful for the support of the CONACYT grant CB-285080 and funding from the PAPIIT-DGAPA-IA101217 (UNAM) project.

This project makes use of the MaNGA-Pipe3D dataproducts. We thank the IA-UNAM MaNGA team for creating this catalogue, and the CONACyT-180125 project for supporting them.

MCD acknowledges support from UC MEXUS-CONACYT grant CN-17-128. ARP and VAR acknowledges support from UNAM PAPIIT grant IA104118 and from the CONACyT `Ciencia Basica' grant 285721. MB acknowledges FONDECYT regular grant 1170618.

Funding for the Sloan Digital Sky Survey IV has been provided by the Alfred P. Sloan Foundation, the U.S. Department of Energy Office of Science, and the Participating Institutions. SDSS-IV acknowledges support and resources from the Center for High-Performance Computing at
the University of Utah. The SDSS web site is www.sdss.org.

SDSS-IV is managed by the Astrophysical Research Consortium for the 
Participating Institutions of the SDSS Collaboration including the 
Brazilian Participation Group, the Carnegie Institution for Science, 
Carnegie Mellon University, the Chilean Participation Group, the French Participation Group, Harvard-Smithsonian Center for Astrophysics, 
Instituto de Astrof\'isica de Canarias, The Johns Hopkins University, 
Kavli Institute for the Physics and Mathematics of the Universe (IPMU) / 
University of Tokyo, Lawrence Berkeley National Laboratory, 
Leibniz Institut f\"ur Astrophysik Potsdam (AIP),  
Max-Planck-Institut f\"ur Astronomie (MPIA Heidelberg), 
Max-Planck-Institut f\"ur Astrophysik (MPA Garching), 
Max-Planck-Institut f\"ur Extraterrestrische Physik (MPE), 
National Astronomical Observatories of China, New Mexico State University, 
New York University, University of Notre Dame, 
Observat\'ario Nacional / MCTI, The Ohio State University, 
Pennsylvania State University, Shanghai Astronomical Observatory, 
United Kingdom Participation Group,
Universidad Nacional Aut\'onoma de M\'exico, University of Arizona, 
University of Colorado Boulder, University of Oxford, University of Portsmouth, 
University of Utah, University of Virginia, University of Washington, University of Wisconsin, 
Vanderbilt University, and Yale University.

\bibliographystyle{mnras}

\appendix

\section {Binning procedure} \label{ApenA}

\begin{figure*}
    \begin{subfigure}[b]{0.39\textwidth}
        \includegraphics[width=0.75\textwidth, angle=90]{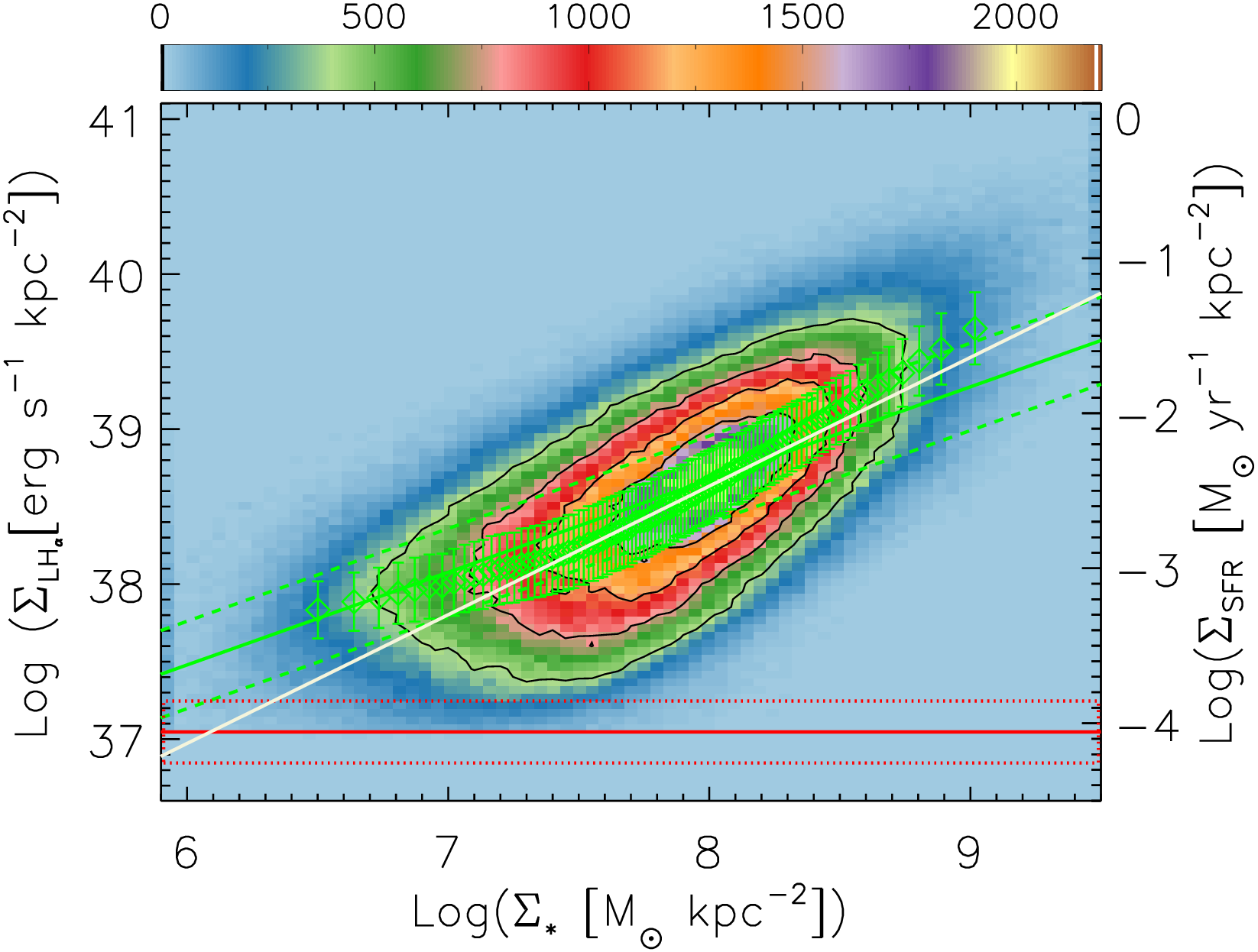}
        \label{fig:SFMS_Massbins}
    \end{subfigure}
        ~\quad
        \begin{subfigure}[b]{0.39\textwidth}
        \includegraphics[width=0.75\textwidth, angle=90]{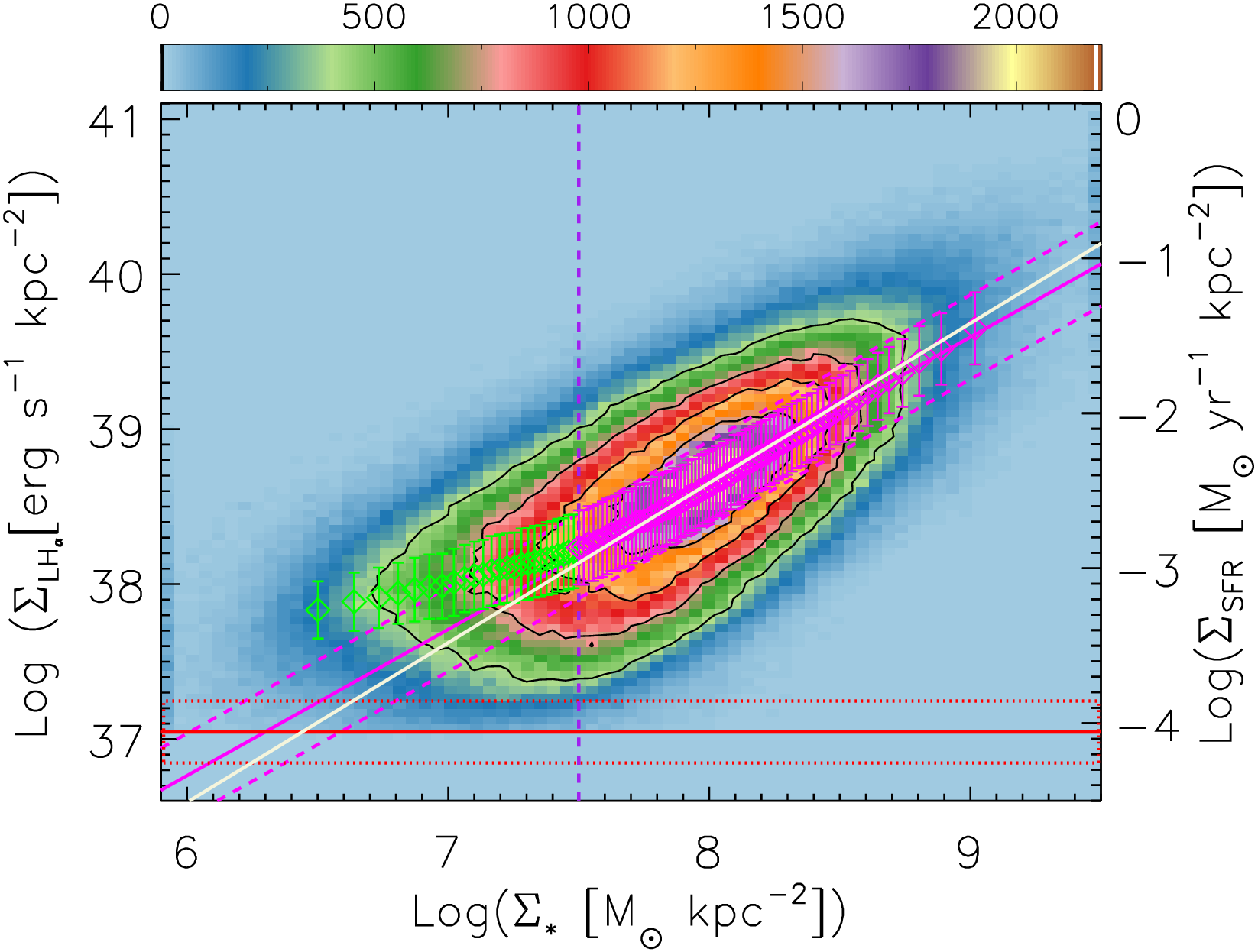}
        \label{fig:RS_Massbins}
    \end{subfigure}

 \caption{Spatially-resolved $\Sigma_{\rm SFR}-\Sigma_{*}$ diagram for SFAs, as Fig. \ref{FullSampleIntegrated}. In the left panel the $\Sigma_{*}$-binned data (mean and standard deviation) are shown in green, while in the right panel the same $\Sigma_{*}$-binned data are shown but we highlight in magenta the data above the $\Sigma_{*}$ cut (purple dashed vertical line) introduced to avoid the selection effects discussed in the main text. The green and magenta solid lines are the fittings corresponding to each case (without and with the cut, respectively), along with their dispersion (dashed lines). For each case we also show in white the corresponding fits applied directly to all the data (no binning) and taking into account the data below the mean detection limit in $\Sigma_{LH_{\alpha}}$ as non-detections. The censored fitting method by \citet{Kelly07} is used for this purpose. The results of the different fits are given in Table \ref{FitsComparissons}.}
\label{MassSFRbins}
\end{figure*}

 For reasons explained in subsection \ref{ResultsFullSR}, the different relations presented in Section \ref{Results} were obtained from linear regressions applied not to all the data points but to the logarithmic means and standard deviations of the ordinates calculated in variable-size bins of the abscissa. 
The binning method used is quite similar as the one used in \citet{CanoDiaz16}. The bins are constructed from the data points in either the SFR vs. $M_{*}$ or $\Sigma_{*}$ vs. $\Sigma_{SFR}$ logarithmic diagrams, along the respective abscissa coordinate ($M_{*}$ or $\Sigma_{*}$), where the size of each bin is chosen to contain 5$\%$ of the total amount of data in the case of the global extensive relations (Figures \ref{FullSampleIntegrated} and \ref{IntSFMS_Morphology}), 15$\%$ for the global intensive relation (Figure \ref{LocaltoGlobalSFMSandRS}), and 1$\%$ in the case of the local relations (Figures \ref{SpatResSFMSandRS}, \ref{SpatResSFMS_Morphology} and \ref{SpatResRS_Mophology}). The half-value of each bin is used as the abscissa value.  The mean of the ordinate (SFR or $\Sigma_{SFR}$) logarithmic values in each bin is used as the corresponding ordinate value, and the respective standard deviation is assigned as a measure of its uncertainty (error bar). The plots in Figure \ref{MassSFRbins} show the bins derived with this method for all the SFAs in our sampple.

The different linear regressions in the log-log diagrams presented in Section \ref{Results} were applied to the binned data as described above.

\section {Alternative Fittings} \label{ApenB}

In sections \ref{ResultsFullIntegrated} and \ref{morph_int}, we presented the global SFR--$M_*$ correlations for SFGs and RGs for the full sample, as well as for four sub-samples divided by the different galaxy morphologies.  These correlations were established using the binning method described in Appendix A. In Table \ref{AppendixIntegratedFits} we present complementary results to the ones presented in Table \ref{IntegratedFits}. These linear regressions are applied to all the data points (instead of the binned data) in each case. The same ordinary least squares routine (Y vs. X) used in sections \ref{ResultsFullIntegrated} and \ref{morph_int} is applied here. The fits presented in \ref{AppendixIntegratedFits} can be compared directly with those reported in \citet{CanoDiaz16} for the CALIFA galaxies, since in that paper the linear regressions were applied to all the (unbinned) data. 

\begin{table*}
    \begin{tabular}{ l c c c}
    \hline
    \hline
SFR-M$_*$ Relation & Slope &  \shortstack{Zero Point \\ log(M$_{\odot}$ yr$^{-1}$)}  & \shortstack{Standard Deviation \\ ($\sigma$)} \\ \hline
    SFGs Full Sample & 0.78$\pm 0.01$ & -8.06$\pm 0.04$ & 0.23 \\ 
    SFGs Ellipticals & 0.60$\pm 0.01$ & -6.49$\pm 0.12$ & 0.31 \\ 
    SFGs S0 & 0.66$\pm 0.02$ & -7.06$\pm 0.20$ & 0.35 \\ 
    SFGs S0a-Sb & 0.84$\pm 0.01$ & -8.85$\pm 0.07$ & 0.22 \\ 
    SFGs Sbc-Irr & 0.88$\pm 0.01$ & -8.98$\pm 0.05$ & 0.19 \\ \hline
    SFGs with CALIFA$^{\dag}$ & 0.81$\pm 0.02$ & -8.34$\pm 0.19$ & 0.20 \\ 
    SFGs RP15$^{\dag \dag}$ & 0.76$\pm 0.01$ & -7.64$\pm 0.02$ & - \\ \hline\hline
    RGs Full Sample & 1.10$\pm 0.01$ & -13.01$\pm 0.05$ & 0.18 \\ 
    RGs Ellipticals & 1.05$\pm 0.01$ & -12.56$\pm 0.09$ & 0.17 \\ 
    RGs S0 & 1.14$\pm 0.01$ & -13.53$\pm 0.07$ & 0.15 \\ 
    RGs S0a-Sb & 1.09$\pm 0.01$ & -12.95$\pm 0.12$ & 0.24 \\ \hline
    RGs with CALIFA$^{\dag}$ & 0.86$\pm 0.02$ & -10.32$\pm 0.24$ & 0.22 \\ \hline
\end{tabular}
	\caption{Correlation coefficients and coefficients of the log-lineal fits to the SFGs and RGs in the global SFR--$M_*$ diagram. The linear regressions are applied to all the data (no mass binning) corresponding to the full samples or to the different subsamples by morphology. The $\sigma$ value is the total standard deviation around the fitting.
$^{\dag}$ \citet{CanoDiaz16}. $^{\dag \dag}$\citet{Renzini15}.}
	\label{AppendixIntegratedFits}
\end{table*}

In Section \ref{ResultsFullSR} we presented the fittings for the SFAs in the local $\Sigma_{SFR}-\Sigma_{*}$ diagram. The fits were performed to the binned data (see Appendix A and Figure \ref{MassSFRbins} for details of the binning method), after a constant cut for the low $\Sigma_{*}$ bins was impossed to avoid the bending at these values. The value of the cut was selected after inspecting the individual $\Sigma_{*}$ histograms and realizing that above $\approx$ 7.5 M$_{\odot}$ kpc$^{-2}$ they become mostly unaffected by the detection limit in $\Sigma_{LH_{\alpha}}$. This cut is shown with a vertical dotted line in the right panel of Figure \ref{FitsComparissons}. 

The bending below the cut ends with the linear approximation that we have been assuming for the SFAs, but as we have explained in Section \ref{ResultsFullSR}, this bending is likely due to selection effects, so that data only above it were taken for the fit. In this Appendix we explore alternative fits that show how the fit changes when not imposing this cut and when the fits are performed to all the data instead of to the binned data. 

In Table \ref{FitsComparissons} we present alternative fits for the full sample of local SFAs in the $\Sigma_{SFR}-\Sigma_{*}$ diagram,and the four different morphology subsamples, using i) a hierarchical Bayesian method, that takes into account the censored data or non-detections within a data set (data below the detection limit), along with the trustable points (data above the detection limit). The method is described in \citet{Kelly07} and it is used on all the SFAs above the 7.5 M$_{\odot}$ kpc$^{-2}$ cut, ii) The same Bayesian approach, but without the 7.5 M$_{\odot}$ kpc$^{-2}$ cut. iii) The ordinary least squares method over the binned data with the cut, and iv) the ordinary least squares method over the binned data without the cut.

\begin{table*}
 \begin{center}
    \begin{tabular}{ l c c c c}
    \hline
    \hline
$\Sigma_{\rm SFR}-\Sigma_*$ Relation & Slope &  \shortstack{Zero Point \\ log(M$_{\odot}$ yr$^{-1}$ Kpc$^{-2}$)} & \shortstack{Standard Deviation \\ ($\sigma$)} & \shortstack{$\Sigma_{*}$ Cut \\ (M$_{\odot}$ yr$^{-1}$ Kpc$^{-2}$)} \\ \hline
    \shortstack{SFG Global unbinned censored fit} & 0.85$\pm0.04$ & -9.20$\pm0.30$ & 0.23 & -\\
    \shortstack{SFG Global binned data fit} & 0.66$\pm0.10$ & -7.62$\pm0.76$ & 0.23 & -\\ 
    \shortstack{SFAs Full Sample unbinned censored fit} & 1.03$\pm0.01$ & -10.69$\pm0.01$ & 0.28 & 7.5\\ 
    \shortstack{SFAs Full Sample unbinned censored fit} & 0.83$\pm0.01$ & -9.11$\pm0.01$ & 0.28 & -\\ 
    \shortstack{SFAs Full Sample binned data fit} & 0.94$\pm0.08$ & -10.00$\pm0.61$ & 0.27 & 7.5\\
    \shortstack{SFAs Full Sample binned data fit} & 0.60$\pm0.03$ & -7.21$\pm0.27$ & 0.28 & -\\
    \shortstack{SFAs Ellipticals unbinned censored fit} & 1.44$\pm0.02$ & -14.37$\pm0.12$ & 0.32 & 7.5\\
    \shortstack{SFAs Ellipticals unbinned censored fit} & 1.04$\pm0.01$ & -11.05$\pm0.05$ & 0.33 & -\\
    \shortstack{SFAs Ellipticals binned data fit} & 1.26$\pm0.07$ & -12.81$\pm0.56$ & 0.31 & 7.5\\
    \shortstack{SFAs Ellipticals binned data fit} & 0.78$\pm0.03$ & -8.93$\pm0.26$ & 032 & -\\
    \shortstack{SFAs S0 unbinned censored fit} & 1.23$\pm0.01$ & -12.63$\pm0.07$ & 0.31 & 7.5\\ 
    \shortstack{SFAs S0 unbinned censored fit} & 0.97$\pm0.01$ & -10.50$\pm0.04$ & 0.31 & -\\ 
    \shortstack{SFAs S0 binned data fit} & 1.13$\pm0.06$ & -11.84$\pm0.50$ & 0.31 & 7.5\\ 
    \shortstack{SFAs S0 binned data fit} & 0.84$\pm0.03$ & -9.46$\pm0.26$ & 0.30 & -\\ 
    \shortstack{SFAs S0a-Sb unbinned censored fit} & 1.38$\pm0.01$ & -13.85$\pm0.08$ & 0.32 & 7.5\\
    \shortstack{SFAs S0a-Sb unbinned censored fit} & 0.87$\pm0.01$ & -9.54$\pm0.01$ & 0.27 & -\\
    \shortstack{SFAs S0a-Sb binned data fit} & 1.02$\pm0.07$ & -10.76$\pm0.54$ & 0.26 & 7.5 \\ 
    \shortstack{SFAs S0a-Sb binned data fit} & 0.71$\pm0.04$ & -8.20$\pm0.30$ & 0.27 & - \\ 
    \shortstack{SFAs Sbc-Irr unbinned censored fit} & 1.01$\pm0.01$ & -10.44$\pm0.01$ & 0.26 & 7.5\\
    \shortstack{SFAs Sbc-Irr unbinned censored fit} & 0.77$\pm0.01$ & -8.49$\pm0.01$ & 0.27 & -\\   
    \shortstack{SFAs Sbc-Irr binned data fit} & 0.94$\pm0.08$ & -9.88$\pm0.62$ & 0.26 & 7.5\\ 
    \shortstack{SFAs Sbc-Irr binned data fit} & 0.62$\pm0.03$ & -7.33$\pm0.25$ & 0.27 & -\\ 
    \hline    
    \shortstack{SFAs CALIFA$^{\dag}$ $\Sigma{*}$ binned data fit} & 0.72$\pm0.04$ & -7.95$\pm0.29$ & 0.16 & -\\
\hline\hline
    \end{tabular}
	\caption{Correlation coefficients and coefficients of the log-linear fits to the SFGs/SFAs and RGs/RAs in the global/local $\Sigma_{SFR}-\Sigma_*$ diagrams for all galaxies and for galaxies in different morphological groups. For the global correlations, 100\% of the data are used, while for the local ones, 80\% of the data in the corresponding samples are used. The $\sigma$ value is the total standard deviation around the fitting.
    $^{\dag}$ \citet{CanoDiaz16}.}
	\label{FitsComparissons}
 \end{center}
\end{table*}

\section{Distribution of SFAs and RAs in the $\Sigma_{\rm SFR}-\Sigma_{*}$ diagram}\label{ApenC}

\begin{figure}
  \centering
   \includegraphics[width=0.35\textwidth, angle=90]{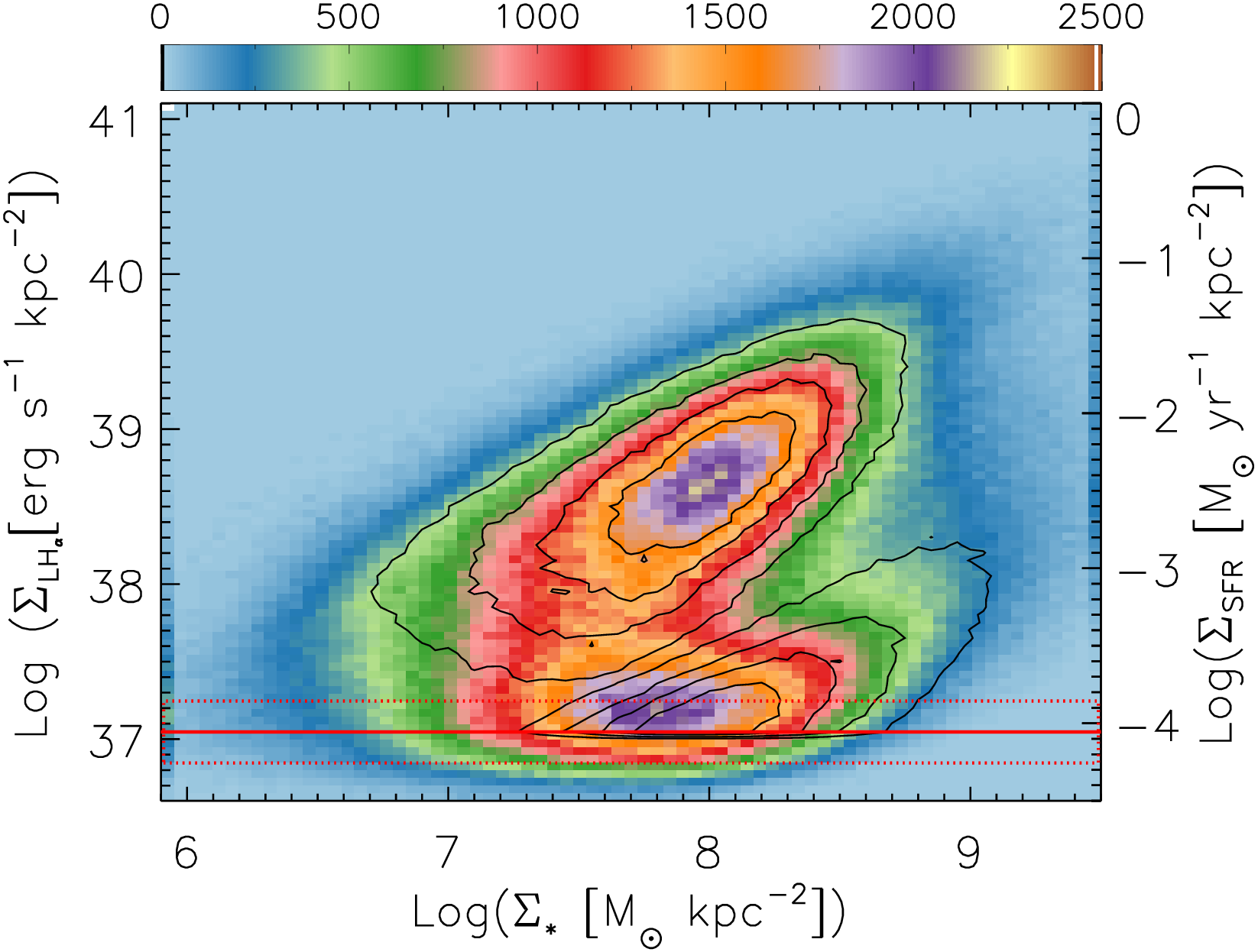}
  \caption{Spatially-resolved $\Sigma_{SFR}-\Sigma_{*}$ diagram for the entirety of areas from our MaNGA galaxies sample. The color-image represent the density of points. The average detection limit in H$\alpha$ is shown with the red horizontal line along with its 2-$\sigma$ dispersion, represented by the dotted red rectangle. The black isocontours are the same already reported in Figure \ref{SpatResSFMSandRS}, which trace the 20, 40, 60, and 80\% of the SFAs an RAs.}
 \label{MasVsSFR}
\end{figure}

To highlight the differences between the two populations in the $\Sigma_{SFR}-\Sigma_{*}$ diagram, meaning the star-forming and the retired areas, we plotted all the areas of all galaxies with detected H$\alpha$ emission, without any segregation based on the EW(H$\alpha$) nor their location in the BPT diagram. The local SFMS and retired regions in Fig. \ref{MasVsSFR}, are highlighted as black iso-contours enclosing the same number of areas as in Figure \ref{SpatResSFMSandRS}. In this figure we can clearly see the segregation between the SFAs and RAs.

\section{A criterion to separate star-forming and retired galaxies/areas based on the EW(H$\alpha$)}\label{EWHalpha}

We present here a simple criterion to separate galaxies (spatially-resolved regions) into star-forming and retired in the SFR--$M_*$ ($\Sigma_{SFR}-\Sigma_{*}$) diagram. These criteria consist of a line in the corresponding log-log diagrams and are based on the characteristic EW(H$\alpha$) at which the galaxies or regions transit from star forming to retired.  

\begin{figure}
  \centering
   \includegraphics[width=0.35\textwidth, angle=90]{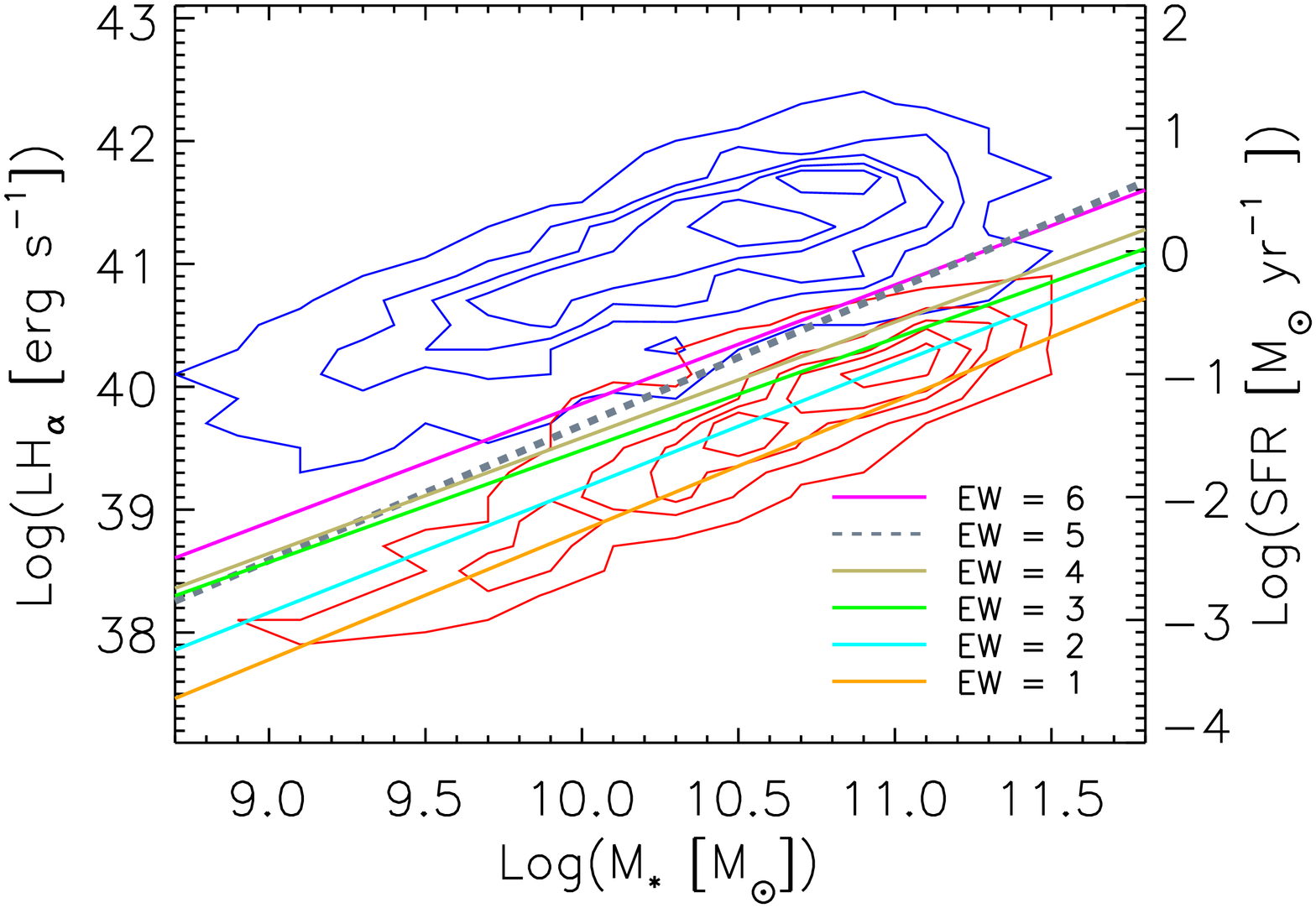}
  \caption{Isocountour density maps of the star-forming (blue lines) and retired (red lines) galaxies, as Figure \ref{FullSampleIntegrated}).  The isocontours trace the 20, 40, 60, 80 and 100\% of the data. Over plotted are the linear fits to galaxies with six different values of EW(H$\alpha$) indicated with the labels. The fit corresponding to EW(H$\alpha$)$\approx 5$ (dotted line) separates better star-forming and retired galaxies.}
 \label{SFMSandRSEWHalpha_int}
\end{figure}

\begin{figure}
  \centering
   \includegraphics[width=0.35\textwidth, angle=90]{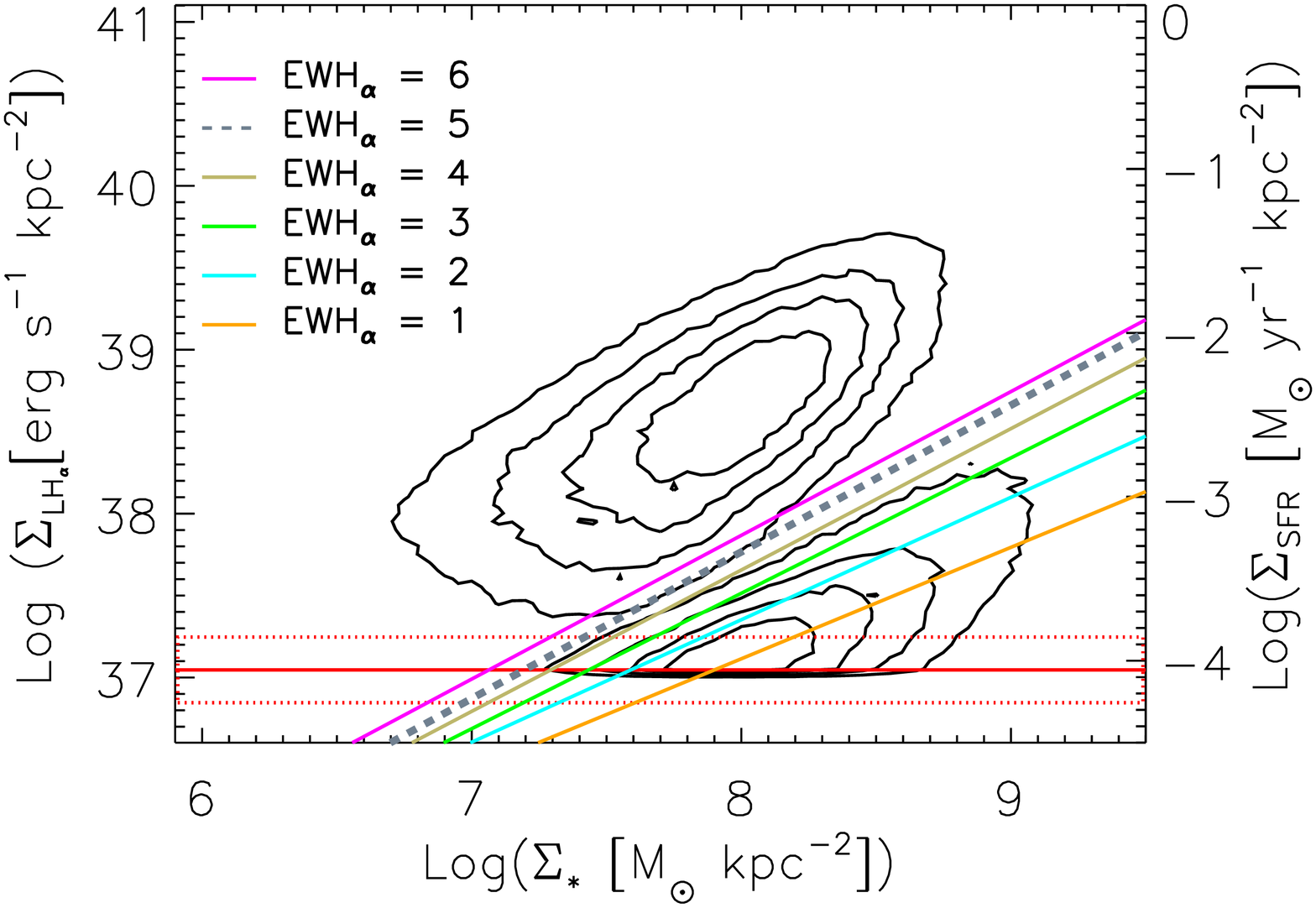}
  \caption{As Figure \ref{SFMSandRSEWHalpha_int} but for the spatially-resolved star-forming and retired areas. The fit corresponding to areas with EW(H$\alpha$)$\approx 5$ also is the one that best separates SFAs and RAs.}
 \label{SFMSandRSEWHalpha}
\end{figure}

Figure \ref{SFMSandRSEWHalpha_int}, like in Figure \ref{FullSampleIntegrated}, shows the isocontours of the SFGs (blue lines) and RGs (red lines) in the global SFR--$M_*$ diagram. A similar diagram but for the spatially-resolved areas in all galaxies, $\Sigma_{SFR}$ vs. $\Sigma_{*}$, is shown in Fig. \ref{SFMSandRSEWHalpha}, which is similar to Figure \ref{SpatResSFMSandRS}. 
In both figures the contours are overplotted with colored lines representing linear fits to the galaxies/areas in the respective diagrams with values of a given EW(H$\alpha$), regardless of their position in the BPT diagram. Galaxies or areas with the following EW(H$\alpha$) values were used:  6, 5, 4, 3, 2 and 1 within a range of $\pm\epsilon$ around these values. For the global case, $\epsilon0.3$ for the EW$(H\alpha)$ values of 1, 2, 3 and 4, and $\epsilon=0.5$ for the other two values, 5 and 6; for these last two values, a narrower band of EW$(H\alpha)$ did not allow to have enough galaxies to derive the fits. For the local case, $\epsilon=0.1$ for all the EW$(H\alpha)$ values. For the global case, the fitting to each one of the groups of constant EW(H$\alpha$) vales was performed to the individual points, while for the local case, it was performed to the binned data. The binning was achieved in the same way that it has been done thorough this paper (see Appendix A).

The results of these exercises, are consistent in both cases. When comparing Figure \ref{SFMSandRSEWHalpha_int} and \ref{SFMSandRSEWHalpha}, it can be easily deduced that one or maybe two of these EW$(H\alpha)_{cte}$ lines could work as a proxy for segregating the SFGs/SFAs to the rest in these diagrams, which could be useful in the case of working with data that do not contain the information needed to classify the sources/regions based on the BPT diagram and the EW$H\alpha$ value. For this purpose we propose the constant line of EW$(H\alpha)_{cte}$ = 5 (dashed gray lines in Figures \ref{SFMSandRSEWHalpha_int} and \ref{SFMSandRSEWHalpha}) as the most suitable one. For the global case, this demarcation line can be expressed as:
\begin{equation}
\log(\textstyle\frac{SFR}{M_\odot yr^{-1}}) $=$ -\left(12.20\pm0.15\right) + \left(1.08\pm0.01\right)\log(\frac{M_*}{M_\odot}).
\end{equation}
For the local case, the demarcation line is:
\begin{equation}
\log(\textstyle\frac{\Sigma_{SFR}}{M_\odot yr^{-1} kpc^{-2}}) $=$ -\left(10.49\pm0.18\right) + \left(0.90\pm0.02\right)\log(\frac{\Sigma_*}{M_\odot kpc^{-2}}).
\end{equation}
Note that our determinations are assuming a Salpeter initial mass function. 

 Figure \ref{SFMSandRSEWHalpha} illustrates clearly that the SFAs that defines the spatially-resolved SFMS, correspond to ionized regions clearly with EW(H$\alpha$)$>3$\AA, in all the cases, and in most of them to regions with a value larger than 6\AA. On the other hand, the RAs have all an EW(H$\alpha$)$<$3\AA, with their peak density corresponding to regions with an equivalent width of $\sim$1\AA. This result agrees with the thresholds presented by \citet{CidFernandes10} and \citet{Sanchez13}, for selecting RGs (or RAs within galaxies), $<$3\AA, and SFAs, $>$6\AA, respectively, supporting the idea that the EW(H$\alpha$) is a good discriminator between both ionizations, as recently explored by \citet{Lacerda18}. The distribution of EWs for SFAs is a consequence of the tight relation found between this parameter and the sSFR, described in \citet{Sanchez13} and recently confirmed by \citet{Belfiore17a}. For RAs, the range of EWs is compatible with the scenario in which this emission, of low intensity, diffuse, and ubiquously distributed across the optical extension of galaxies, is the consequence of ionization by old-stars (e.g., post-AGBs or HOLMES), as proposed by \citet{Binette94} and \citet{Stasinska08}, and recently found in both early-type galaxies \citep{sarzi10} and RAs of later-type galaxies (e.g., Sigh et al. 2013). In all these galaxies/areas the EW(H$\alpha$) is between 1-2\AA  \citep[e.g.][]{Gomes16b}. As already shown by \citet{Lacerda18}, selecting ionized areas based on the EW clearly defines their location in the classical diagnostic diagram, being a more powerful selection method than using only those diagrams. Indeed this result was already shown by \citet{CanoDiaz16} for the properties of the ionized gas integrated across the optical extension of galaxies, and more recently, was adopted as a classification scheme by \citet{SanchezMenguiano18} and \citet{sanchez18a}, in their selection of star-forming and AGN ionized regions in galaxies respectively. The low EW(H$\alpha$) of the ionized gas in the retired galaxies/areas and their location in the BPT diagram \citep{Lacerda18, CanoDiaz16}, support the idea that this ionization is not dominated by the presence of young stars, and in purity its H$\alpha$ luminosity cannot be transformed to a SFR. In a conservative way the estimated SFR for those areas should be considered as an upper-limit, taking into a possible contamination by ionization due to a low level of star-formation. Only under this interpretation it could be included in the current SFR-M$_*$ diagram.


\section{Local $\Sigma_{\rm SFR}-\Sigma_{*}$ diagram for RAs segregated by morphology}\label{Sigma_RAs}
\begin{figure*}
    \centering   
    \captionsetup[subfigure]{labelformat=empty}
    
  \begin{minipage}{1\linewidth}
  \centering
    \begin{subfigure}[b]{0.45\textwidth}
        \includegraphics[width=0.72\textwidth, angle=90]{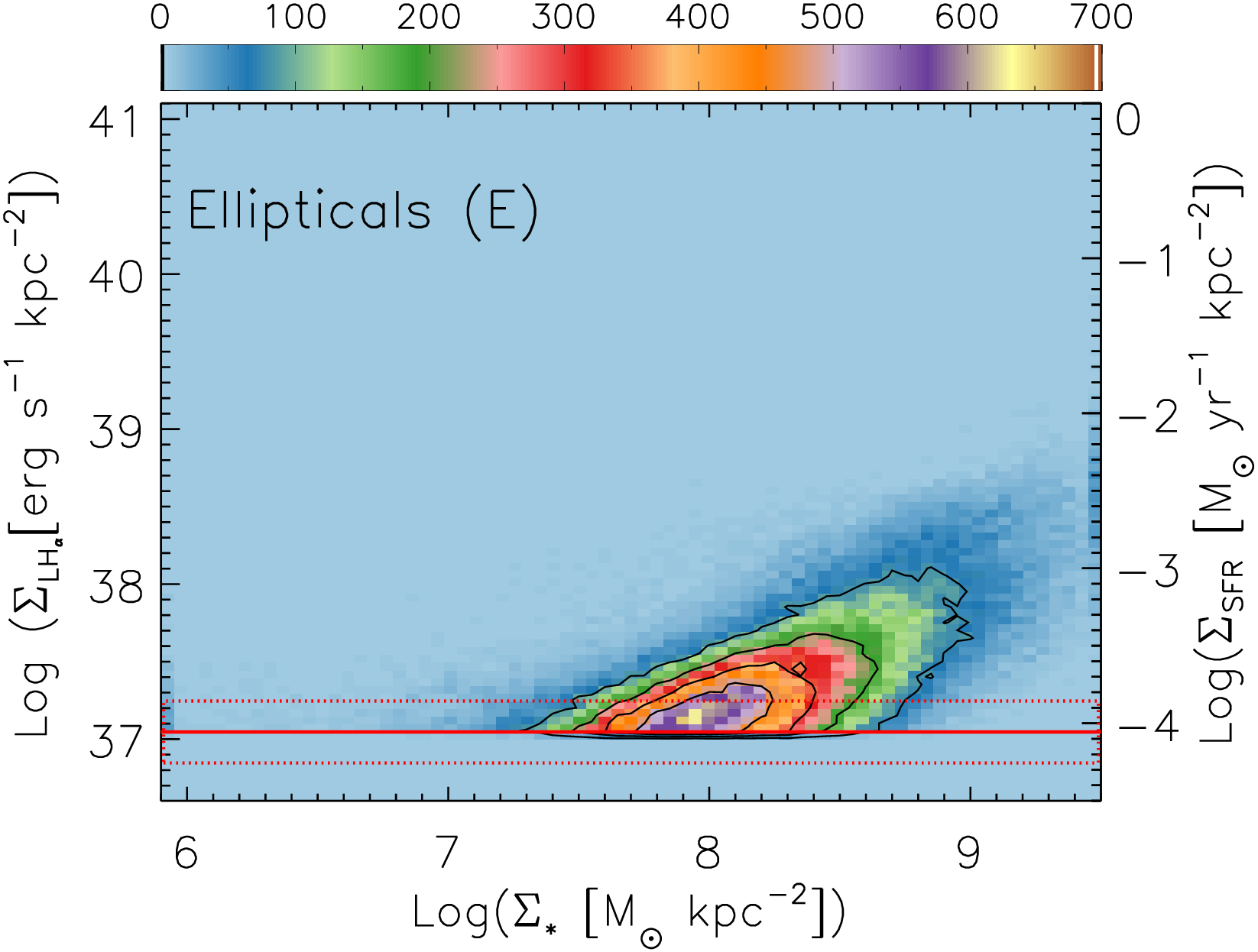}
        \label{fig:SFMS_loc_Ellipticals}
    \end{subfigure}
        \begin{subfigure}[b]{0.45\textwidth}
        \includegraphics[width=0.72\textwidth, angle=90]{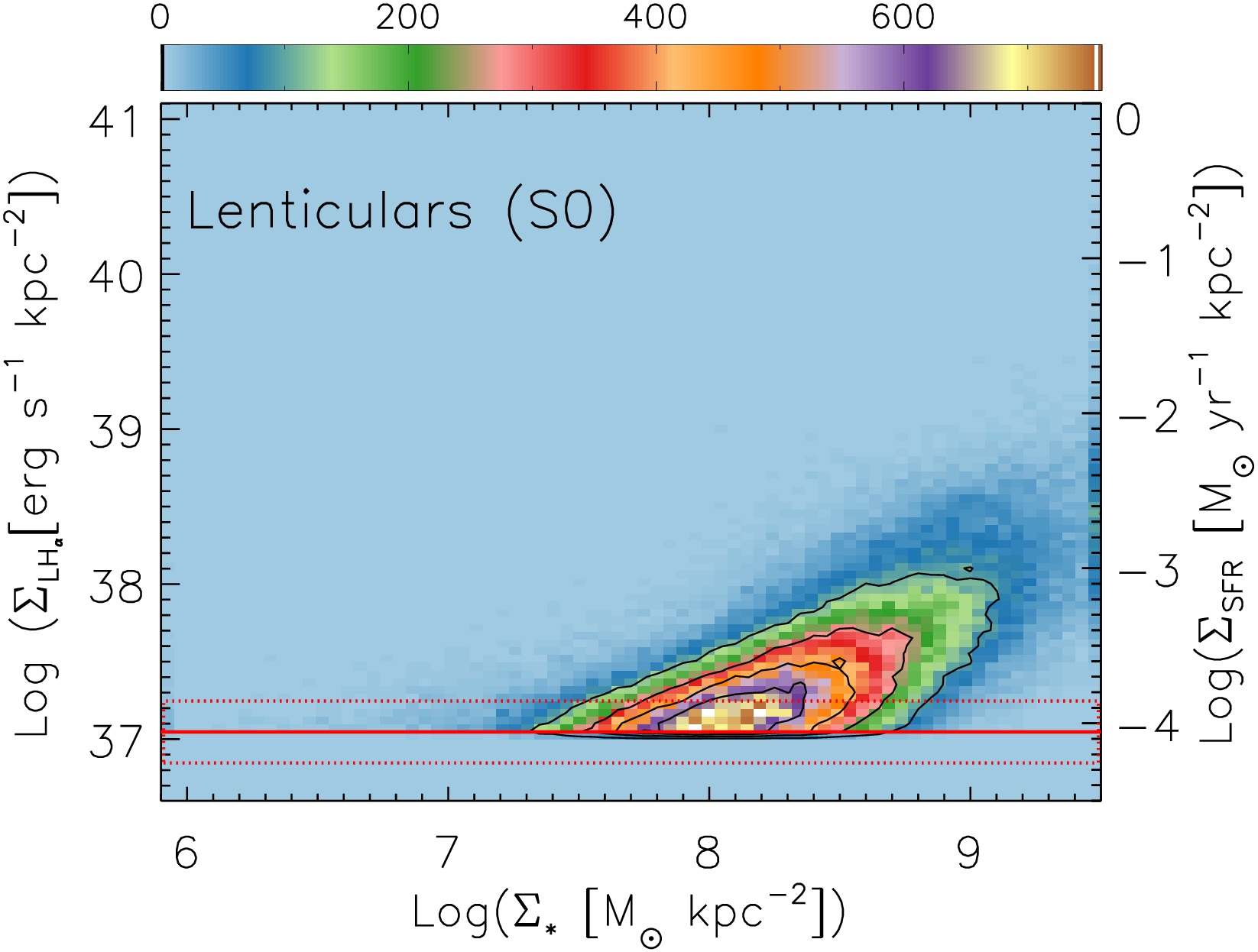}
        \label{fig:SFMS_loc_Lenticulars}
    \end{subfigure}
 \end{minipage} \par\medskip
  \par\medskip
  \vfill          
 \begin{minipage}{1\linewidth}
 \centering
    \begin{subfigure}[b]{0.45\textwidth}
        \includegraphics[width=0.72\textwidth, angle=90]{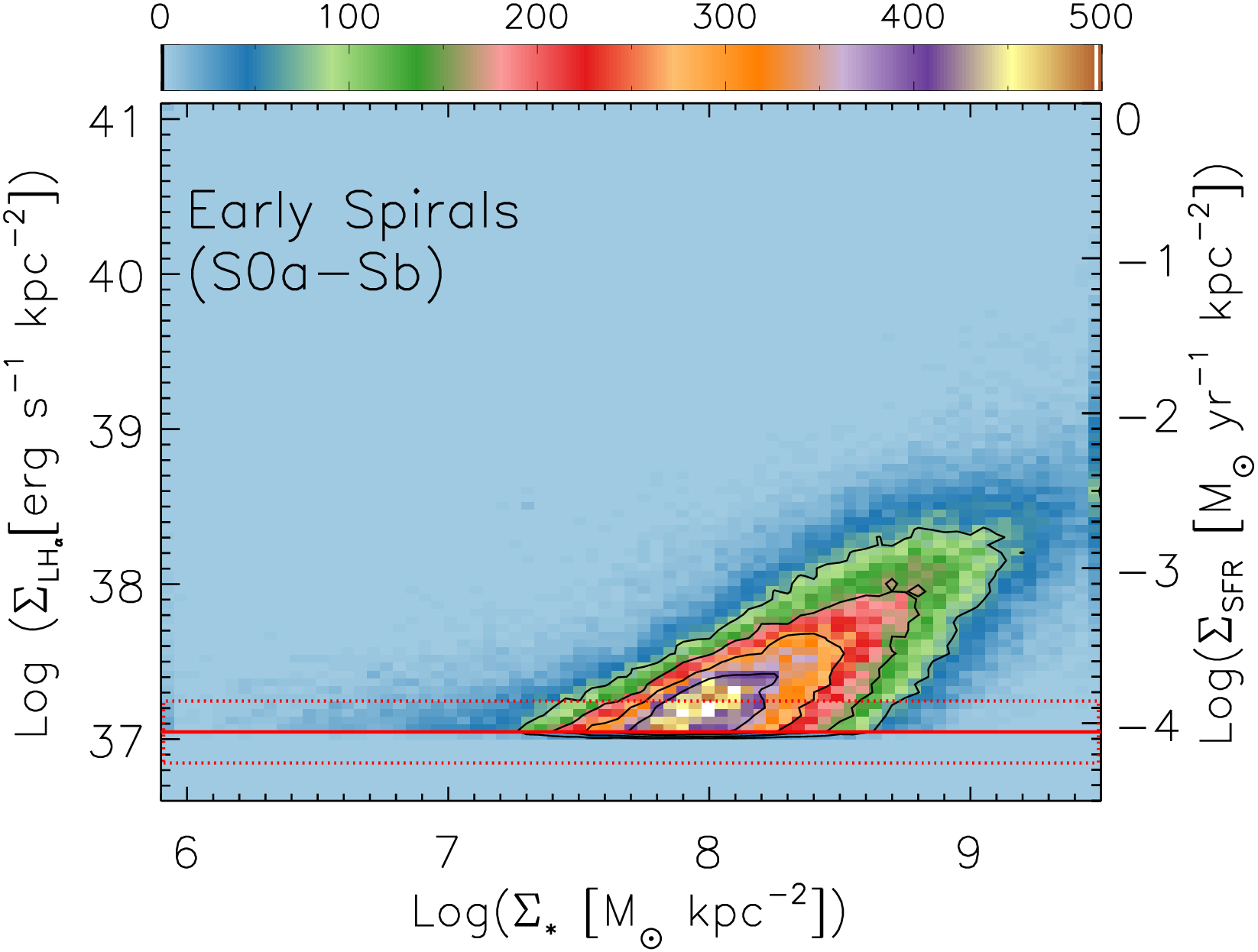}
        \label{fig:SFMS_loc_SpiralsSa}
    \end{subfigure}
    \begin{subfigure}[b]{0.45\textwidth}
        \includegraphics[width=0.72\textwidth, angle=90]{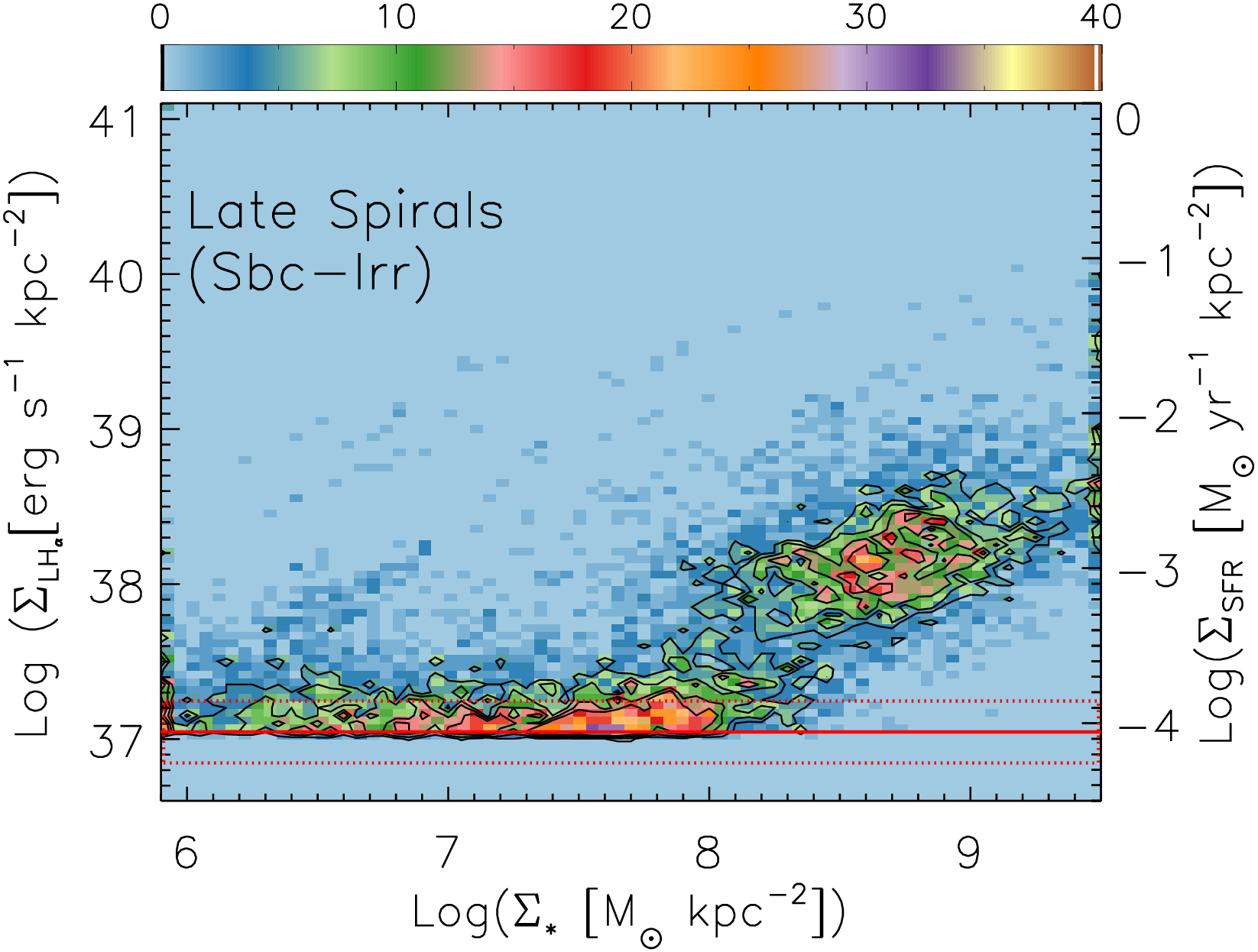}
        \label{fig:SFMS_loc_SpiralsSbc}
    \end{subfigure}
      \end{minipage} \par\medskip
  \par\medskip
  \vfill       
\caption{Spatially-resolved $\Sigma_{\rm SFR}-\Sigma_{*}$ diagram for the retired areas of the different galaxies segregated by morphology: Ellipticals (top left), Lenticulars (top right), Early Spirals (bottom left) and Late Spirals (bottom right). The countours and color-image represent the density of points, following the same nomenclature used in Fig. \ref{LocaltoGlobalSFMSandRS}, as well as the average detection limit is represented as an horizontal red solid line, and its associated scatter.}
\label{SpatResRS_Mophology}
\end{figure*}


In Fig. \ref{SpatResSFMS_Morphology} we show the distribution in the local $\Sigma_{\rm SFR}-\Sigma_{*}$ diagram of the SFAs segregated by the morphology of the host galaxies, together with the best fitted linear regression between these two parameters. This corresponds to the local SFMS derived for each morphological type. For completeness we show in Fig. \ref{SpatResRS_Mophology} the same distributions for the RAs of galaxies, segregated by morphology. The Irr/BCGs have been excluded due to the low number of retired areas detected in those galaxies. The parameters of the best fitted linear regressions shown in the figure are reported in Table \ref{SResolvedFits} for completeness.

\bsp	
\label{lastpage}
\end{document}